\documentclass[pre,onecolumn,superscriptaddress]{article}

\usepackage{authblk}
\usepackage{pdfsync}
\usepackage{boldline}
\usepackage{longtable}
\usepackage{setspace}
\usepackage[inline]{enumitem}

\usepackage[version=3]{acro}
\usepackage[numbers,sort&compress]{natbib}
\usepackage{anysize}
\marginsize{2cm}{2cm}{1.cm}{1.cm}
\usepackage{graphicx}
\usepackage{pdfpages}
\usepackage{epsfig}
\usepackage{verbatim} 
\usepackage{bm}
\usepackage{etoolbox}
\usepackage{multirow}
\usepackage{yfonts}
\usepackage[colorlinks]{hyperref}
\usepackage{fancyhdr}
\usepackage{relsize}
\usepackage{newpxtext,newpxmath}
\usepackage{booktabs,siunitx}
\usepackage{array,booktabs,ragged2e}
\usepackage{enumitem}   
\usepackage{empheq}
\usepackage[font=small,labelfont=bf]{caption}
\newcommand{\mycaption}[2]{\caption[#1]{\textbf{#1}. #2}}
\usepackage[title,toc,titletoc,page]{appendix}
\usepackage[capitalise]{cleveref}

\usepackage{xr-hyper} 
\usepackage{xcite} 
\externalcitedocument{supplementary}
\usepackage[most]{tcolorbox}
\usepackage{xcolor}

\usepackage{eurosym}
\providecommand{\be}{\begin{equation}}
            \providecommand{\ee}{\end{equation}}
\providecommand{\bsp}{\begin{split}}
            \providecommand{\esp}{\end{split}}
\providecommand{\bea}{\begin{eqnarray}}
            \providecommand{\eea}{\end{eqnarray}}
\providecommand{\beas}{\begin{eqnarray*}}
            \providecommand{\eeas}{\end{eqnarray*}}
\providecommand{\iexp}{\varsigma}

\providecommand{\iexpu}{{\iexp_{\scalebox{0.6}{${\rm up}$} }}}
\providecommand{\iexpl}{{\iexp_{\scalebox{0.6}{${\rm low}$}}}}

\providecommand{\betac}{\beta_{\rm c}}
\providecommand{\rmd}{{\rm d}}

\providecommand{\jacs}{{\mathscr J}}

\providecommand{\Tc}{{T_{\rm c}}}
\providecommand{\bw}{\begin{widetext}}
            \providecommand{\ew}{\end{widetext}}
\providecommand{\ra}{\rightarrow}
\providecommand{\bali}{\begin{align}}
            \providecommand{\eali}{\end{align}}

\providecommand{\bsi}{{\bm \sigma}}

\providecommand{\bS}{{\bm S}}
\providecommand{\bJ}{\textit{\textbf J}}
\providecommand{\cM}{{\cal M}}
\providecommand{\cMp}{{\cal M}'}
\providecommand{\pa}{\partial}

\providecommand{\blangle}{\boldsymbol{\langle}}
\providecommand{\brangle}{\boldsymbol{\rangle}}
\providecommand{\rmiterations}{M}

\providecommand{\rmL}{{\rm L}}

\providecommand{\rmR}{{\rm R}}
\providecommand{\rmd}{{\rm d}}

\providecommand{\bE}{{\mathbb{E}}}
\providecommand{\qJ}{L}
\providecommand{\qJs}{K}
\providecommand{\lambdas}{{\overline{\lambda}}}

\providecommand{\dJ}{{\rmd \bJ \,}}

\providecommand{\uarr}{{\uparrow}}
\providecommand{\darr}{{\downarrow}}

\providecommand{\olambdaast}{{\overline{\lambda}_\star}}

\providecommand{\mytitle}{Order parameter for non-mean-field spin glasses}

\providecommand{\revision}[1]{#1}

\NewDocumentCommand{\up}{som}{%
      \IfBooleanTF{#1}
      {\upext{#3}}
      {#3\IfNoValueTF{#2}{\mathord}{#2}\uparrow}%
}
\NewDocumentCommand{\upext}{m}{%
      \mleft.\kern-\nulldelimiterspace#1\mright\uparrow
}

\definecolor{link_color}{RGB}{110,38,14}

\DeclareAcronym{OP}{short = OP, long = order parameter}
\DeclareAcronym{SG}{short = SG, long = spin glass,     long-plural-form = {spin glasses}}
\DeclareAcronym{RG}{short = RG, long = renormalization group}
\DeclareAcronym{MF}{short = MF, long = mean-field}
\DeclareAcronym{HEA}{short = HEA, long = hierarchical Edwards-Anderson model}
\DeclareAcronym{HT}{short = HT, long = high temperature}
\DeclareAcronym{GS}{short = GS, long = ground state}
\DeclareAcronym{DHM}{short = DHM, long =  Dyson's hierarchical model}
\DeclareAcronym{FD}{short = FD, long = fixed distribution}
\DeclareAcronym{FM}{short = FM, long = ferromagnetic}
\DeclareAcronym{PDF}{short = PDF, long = probability density function }
\DeclareAcronym{CDF}{short = CDF, long = cumulative distribution  function }
\DeclareAcronym{L}{short = L, long = left} 
\DeclareAcronym{LHS}{short = LHS, long =  left-hand side}
\DeclareAcronym{MC}{short = MC, long = Monte Carlo}
\DeclareAcronym{pHEA}{short = pHEA, long = hierarchical Edwards-Anderson model with power-law interaction decay}
\DeclareAcronym{cHEA}{short = cHEA, long = hierarchical Edwards-Anderson model with fixed average coordination number}
\DeclareAcronym{R}{short = R, long = right}
\DeclareAcronym{RHS}{short = RHS, long = right-hand side}

\hypersetup{colorlinks=True,
	linkcolor=blue,
	anchorcolor=black,
	citecolor=red,
	urlcolor=black,
  filecolor=blue
}

\acsetup{
  make-links ,
  pages / display = first ,
  pages / fill    = {, },
  patch / floats  = false
}

\DeclareMathAlphabet{\mathcal}{OMS}{cmsy}{m}{n}
\SetMathAlphabet{\mathcal}{bold}{OMS}{cmsy}{b}{n}
\tcbset{highlight math style={boxsep=2mm,colback=white,colframe=black}}

\newlength{\figsize}
\title{\mytitle}
\author[1]{Michele Castellana}
\affil[1]{Institut Curie, PSL Research University, CNRS UMR168, France}

\begin{document}

\maketitle

\begin{abstract}
  We propose a novel \ac{RG} method for non mean-field models of spin glasses\revision{, which leads to the emergence of a novel order parameter}. Unlike previous approaches where the \ac{RG} procedure is based on  a priori notions on the system, our analysis follows a minimality principle, where no a priori assumption is made. We apply our approach to a spin-glass model built on a hierarchical lattice. In the \ac{RG} decimation procedure,  a novel order parameter spontaneously emerges from the  system symmetries,  and self-similarity features of the \ac{RG} transformation only. This order parameter is  the projection of the spin configurations on the ground state of the system. Kadanoff's majority rule for ferromagnetic systems is replaced by a more complex scheme, which involves such novel order parameter. The ground state thus acts as a pattern which translates spin configurations from one length scale to another. The rescaling \ac{RG} procedure is based on a minimal, information-theory approach and, combined with the decimation, it yields a complete \ac{RG} transformation.

  Below the upper critical dimension, the predictions for the critical exponent $\nu$, which describes the critical divergence of the correlation length,  are in excellent agreement with numerical simulations from both this and previous studies.
  Overall, this study opens new avenues in the understanding of the critical ordering of realistic spin glasses, and it can be applied to spin-glass models on a cubic lattice and nearest-neighbor couplings which directly model spin-glass materials, such as AuFe, CuMn and other  magnetic alloys.
\end{abstract}

\setlength{\parskip}{5pt plus 0pt minus 0pt}
\acresetall
\setcounter{table}{0}

\section{Introduction}\label{intro}

The nature of the glass and glass transition is believed to be  most interesting unsolved problem in solid-state theory \cite{anderson1995through}. \Ac{SG} models  have been introduced \cite{edwards1975theory} to describe dilute magnetic alloys with a small amount of magnetic impurity. A prototypical example is a  solution of Fe in Au, modeled by an array of spins of Fe randomly disposed  in the  Au lattice, and interacting with a potential which oscillates as a function of the spin-spin separation  \cite{cannella1972magnetic}.

The complex and rich behavior of models for  spin glasses has interested theoreticians for their challenging complexity and difficulty,  and opened new avenues in multiple  fields, such as physics \cite{parisi1983order,biroli2008thermodynamic}, mathematics \cite{talagrand2006parisi},  computational optimization \cite{mezard2002analytic} and neural networks \cite{schneidman2006weak}. The core of both the richness and complexity of \ac{SG} models stands in the presence of frustration: the oscillating inter-atom potential results in both ferromagnetic and antiferromagnetic spin-spin interactions, disposed according to a random pattern. This feature is responsible for the complex thermodynamical structure of mean-field \ac{SG} models \cite{sherrington1975solvable}, which involves a  hierarchical structure of mutually nested states  \cite{parisi1979infinite,parisi1983order,parisi2021nobel}.

Despite the notable theoretical advance in mean-field \ac{SG} models, the solution of realistic, non-mean-field \ac{SG} models has been proving to be extremely challenging for nearly fifty years now \cite{stein2004spin}. In this regard, the \ac{RG}, the mainstream method that allowed for a solution of a large variety non-mean-field models in statistical physics \cite{justin2002quantum,wilson1982the}, proves to be particularly difficult to use for such models. The core of this difficulty stands in frustration. In fact, for ferromagnetic systems, where frustration is absent, both the ground state and the low-energy excitations are known and physically intuitive:  They correspond to a state where spins are coherently aligned, and to droplets of reversed spins, respectively, and they naturally lead to the definition of a ferromagnetic order parameter---the local magnetization. Such order parameter is the only relevant degree of freedom in a spin block that needs to be retained in the infrared limit, and it allows for the notable reduction of degrees of freedom---and simplification---of the problem provided by the \ac{RG}.
\revision{It thus clear that the structure of the low-temperature phase and of the order parameter are intrinsically related. On top of the example of non-frustrated systems above, there is the case of mean-field spin glasses, where the multi-state, nested structure of the low-temperature phase results into a functional order parameter \cite{parisi1979infinite}. }

For frustrated systems such as \acp{SG}, neither the ground state nor the low-energy excitations are known, and the structure of the latter has been subject of debate for decades now \cite{newman2003finite}. As a result, the logic which leads to the \ac{RG} decimation and order parameter for ferromagnetic systems could not be applied to such frustrated systems.

Numerous \ac{RG} approaches have been proposed for \acp{SG}.
First, an \ac{RG} analysis for random models reminiscent of \ac{SG}s has been proposed \cite{theumann1980critical,theumann1980ferromagnetic} shortly after the original paper by Edwards and Anderson \cite{edwards1975theory}: However, such models are not a good representative of realistic, strongly-frustrated \ac{SG} systems, because spins on  the same hierarchical block interact with the same coupling, resulting in  weak frustration.
Second, an \ac{SG} version of models on hierarchical lattices built on diamond plaquettes \cite{berker1979renormalisation} has been studied with \ac{RG} methods, but it also yields a weakly frustrated system \cite{gardner1984spin}.

Third, a number of \ac{RG} approaches assume a priori some features on the model. For example, in a zero-temperature decimation scheme, odd spins in a one-dimensional \ac{SG} model are assumed to align with either of their neighbors, according to the strength and sign of the nearest-neighbor couplings, thus ignoring other spin-spin interactions \cite{monthus2014one}.
Other approaches assume, a priori, that some features of the mean-field theory hold in finite dimensions.   For example, in \cite{parisi2001renormalization} a block-spin decimation was defined in terms of a Kadanoff majority rule \cite{kadanoff1966scaling} for the overlap between spins in the block---the overlap being the quantity that defines the order parameter in the mean-field theory \cite{parisi1983order}. Along the same lines, multiple field-theoretical studies \cite{chen1977mean,bray1980renormalisation,kotliar1983one,temesvari2002generic,castellana2011renormalization,castellana2015hierarchical} are based on the replica field theory---the approach with which the mean-field solution has been originally formulated \cite{nishimori2001statistical,parisi1979infinite}. Finally, recently proposed \ac{RG} approaches in real space \cite{castellana2011real,angelini2013ensemble} are based on a decimation procedure which is also based on the overlap. The strong connection between these approaches and the mean-field description of the system may be ascribed to the substantial difficulty in  going beyond this picture, and finding the few physical degrees of freedom which stay relevant in the infrared limit for the system.

Rather than building our analysis on the mean-field picture, here we construct an \ac{RG} approach based only on the fundamental symmetries, thermodynamical features of the system, and on the self-similarity  properties  of the \ac{RG} transformation. The leitmotif of our analysis is a minimalistic approach, where no a priori physical picture of the model is used.
First, this reasoning yields to the emergence of a new order parameter, which has a transparent physical interpretation, and which entirely defines the \ac{RG} decimation procedure. Second, the rescaling procedure is realized by means of an information-theory approach, which allows us to determine the minimal rescaled spin-coupling distribution compatible with the thermodynamic features of the system. This idea is transformed into a well-posed constrained-optimization problem, which quantitatively determines the rescaled coupling distribution.
Combined, the rescaling and decimation procedure yield the \ac{RG} transformation \cite{wilson1974the}.

We apply this method to the \ac{HEA}, a \ac{SG} version of Dyson's hierarchical model for ferromagnetic systems, which proves to be particularly suited for \ac{RG} approaches \cite{dyson1969existence}. Like other one-dimensional \ac{SG} models with long-range interactions, the \ac{HEA} may elucidate the properties of short-range \ac{SG} models on a hypercubic lattice, which directly mimic physical \ac{SG} materials \cite{kotliar1983one}.

We show that, in the ferromagnetic limit, the order parameter reduces to the magnetization, and the decimation procedure reproduces the majority rule for decimation in ferromagnetic systems \cite{kadanoff1966scaling}. Finally, we show that the predictions of our method for the critical exponent $\nu$ are in excellent agreement with numerical simulations, from both this and previous studies, in the  non-mean-field region, where the equivalent of the system dimension lies below the upper critical dimension.

\section{Renormalization-group transformation}\label{rg_tr}

The \ac{HEA} is a \ac{SG} version of Dyson's hierarchical model---a ferromagnetic model of Ising spins built on a hierarchical lattice \cite{dyson1969existence,bleher1973investigation,collet1977a}. Dyson's  model is particularly suited for \ac{RG} studies, because its recursive structure  yields an exact \ac{RG} transformation. The Hamiltonian of the  \ac{HEA} is defined \cite{franz2009overlap}  by the recursion relation
\be\label{eq_H}
H_{k+1}[\bS] = H_k^\rmL[\bS_\rmL] + H_k^\rmR[\bS_\rmR] - \frac{1}{2^{\iexp k}} \sum_{i \in \rmL, j \in \rmR} J_{ij} S_i S_j.
\ee
In \cref{eq_H}, $\bS = \{ S_1, \cdots, S_{2^{k+1}} \}$ are Ising spins on a one-dimensional lattice with $2^{k+1}$ sites, and we use boldface for vectorial and matricial quantities. Also,  $\bS_\rmL = \{ S_1, \cdots, S_{2^{k}} \}$ and $H_k^\rmL[\bS_\rmL]$, denote spins and Hamiltonian in the left half of the lattice, and similarly for the right half. The third term in the right-hand side of \cref{eq_H} represent the interaction  between left and right half of the lattice,
and the spin couplings  $J_{ij}$ are independent, identically distributed random variables with zero mean and unit variance.
\Cref{eq_H} allows to build the Hamiltonian of an \ac{HEA} with $2^k$ spins recursively, starting with the initial condition $H_0^\rmL=H_0^\rmR=0$.
Finally, the exponent $\iexp$ sets the interaction range---the larger $\iexp$, the shorter the range. In what follows, we will consider the parameter range $\iexp> \iexp_{\infty} \equiv 1/2$ where the Hamiltonian \eqref{eq_H} is extensive, and the thermodynamic limit of the system exists \cite{franz2009overlap}.

According to the  analogy between one-dimensional models with long-range interactions and short-range models on a hypercubic lattice \cite{larson2010numerical,katzgraber2005probing,katzgraber2009study,leuzzi2009ising,banos2012correspondence,angelini2013ensemble},  the value of  $\iexp = \iexpl =1$ is analogous to the lower critical dimension of the \ac{HEA}.  In what follows we will say that for $\iexp < \iexpl$ and $\iexp > \iexpl$ the model lies above and below its lower critical dimension, respectively \cite{moore2010ordered,kotliar1983one}.

\subsection{Decimation}\label{sec_dec}

In this Section, we derive the block-spin decimation procedure, which reduces a four-spin \ac{HEA} model $\cM$ with a given sample of the couplings, to a two-spin \ac{HEA}   model $\cMp$. Non-primed and primed quantities, such as $\bS$ and $\bS'$, refer to models $\cM$ and $\cMp$, respectively. In particular, we will denote by $\bJ =\{ J_{ij} \}$ and $J'$ the couplings of $\cM$ and  $\cMp$, respectively.

The  decimation is achieved by means of a coarse-graining function, which we will denote by $\Phi_{\rmL}[\bS_{\rmL}]$, which maps a spin configuration $\bS_{\rmL}$ in the left half of model $\cM$ into a coarse-grained order parameter $\varphi_{\rmL}$, and similarly for the right half. An analog decimation occurs for model $\cMp$, with function $\Phi'_{\rmL}[\bS'_{\rmL}]$ and $\Phi'_{\rmR}[\bS'_{\rmR}]$.
The mapping between the two models results in the following relation between the thermal distributions of the order parameter $\varphi$:
\be
\label{eq_dec}
\langle \mathbb{I}(\Phi_\rmL[\bS_\rmL] = \varphi_\rmL) \; \mathbb{I}(\Phi_\rmR[\bS_\rmR] = \varphi_\rmR)\rangle  = \langle \mathbb{I}(\Phi'_\rmL[\bS'_\rmL] = \varphi_\rmL)\;  \mathbb{I}(\Phi'_\rmR[\bS'_\rmR] = \varphi_\rmR)\rangle',
\ee
where the indicator function $\mathbb{I}()$ is equal to unity if the condition in its argument is satisfied and zero otherwise. Here, $\langle  \cdot \rangle \equiv \frac{1}{Z} \sum_{\bS} e^{-\beta H[\bS]} \, \cdot$ is the Boltzmann average for model $\cM$,
its partition function and Hamiltonian are $Z \equiv \sum_{\bS} e^{-\beta H[\bS]}$ and $H$, respectively, $\beta = 1/T$ is the inverse temperature and, in what follows, we will set the Boltzmann constant $k_{\rm B}$ equal to unity.   Analogous definitions hold for  model $\cMp$.  \Cref{eq_H} implies that the Hamiltonians read
\begin{align}
  \label{eq_91}  H[\bS] =   & - J_{12} S_1 S_2  - J_{34} S_3 S_4 - \frac{1}{2^\iexp} ( J_{13} S_1 S_3+ J_{14} S_1 S_4+ J_{23} S_2 S_3+ J_{24} S_2 S_4), \\
  \label{eq_92}  H'[\bS'] = & - J' S'_1 S'_2.
\end{align}

The  expression of the order parameter in terms of the spin configurations is given by the coarse-graining function $\Phi_{\rmL(\rm R)}$, which we will \textit{ derive}
by leveraging the symmetry properties of the model.

As shown in \cref{sec_reduction}, \cref{eq_phi}, $\Phi'[S']$ is an odd function of $S'$. Also, given that the Ising spins  $\bS$ can take only two values, the most general form of $\Phi$ is  $\Phi_\rmL[\bS ] = A_\rmL + B_{\rmL \, 1} S_1 + B_{\rmL \, 2} S_2 + C_\rmL S_1 S_2,$ and similarly for the right half.
Given that $\Phi'[\bS']$ is an odd function of $\bS'$ and that the decimation must preserve the  order-parameter symmetry, $\Phi_\rmL$, $\Phi_\rmR$ must also be  odd functions of $\bS$. As a result, $A_\rmL = C_\rmL = 0$, and similarly for the right half.
As shown in \cref{sec_reduction}, the fundamental requirement that  the structure of the decimation relation and the parity of the functions $\Phi$s must be preserved across length scales, i.e., across models $\cM$ and $\cM'$, implies that the decimation relation \eqref{eq_dec} can be reduced to
\be\label{eq_dec2}
\blangle \Omega[\bS]  \brangle = \langle \Omega'[\bS']  \rangle',
\ee
where
\be
\label{eq_O}
\Omega[\bS] \equiv \Phi_\rmL[\bS_\rmL] \Phi_\rmR[\bS_\rmR],\,\,\,
\Omega'[\bS'] \equiv  \Phi_\rmL'[\bS'_\rmL] \Phi'_\rmR[\bS'_\rmR].
\ee

We will now determine the functions $\Phi$ and $\Phi'$, and write the decimation relation (\ref{eq_dec2}) explicitly. To achieve this, let us introduce the eight spin configurations  $\{ +, +, +, +\}$ , $\{ +, +, +, -\}$ , $\cdots$, $\{ +, -, -, -\} $ of model $\cM$,  where $\pm$ stands for $\pm1$. Such spin configurations are obtained by fixing the first spin $S_1$ to $+$ and by varying the other ones. We will denote this set of spin configurations by $\bS_1, \cdots \bS_8$, where the labels are assigned  in order of increasing energy, $H[\bS_1] < H[\bS_2] < \cdots < H[\bS_8]$. Here $\bS_1$ is the ground state, which we will denote by $\bsi$, and  $\bS_2,\cdots, \bS_8$   the excited states of $\cM$. The analogous construction is made for $\cMp$.

\begin{figure}
  \begin{center}
    \includegraphics[scale=0.075]{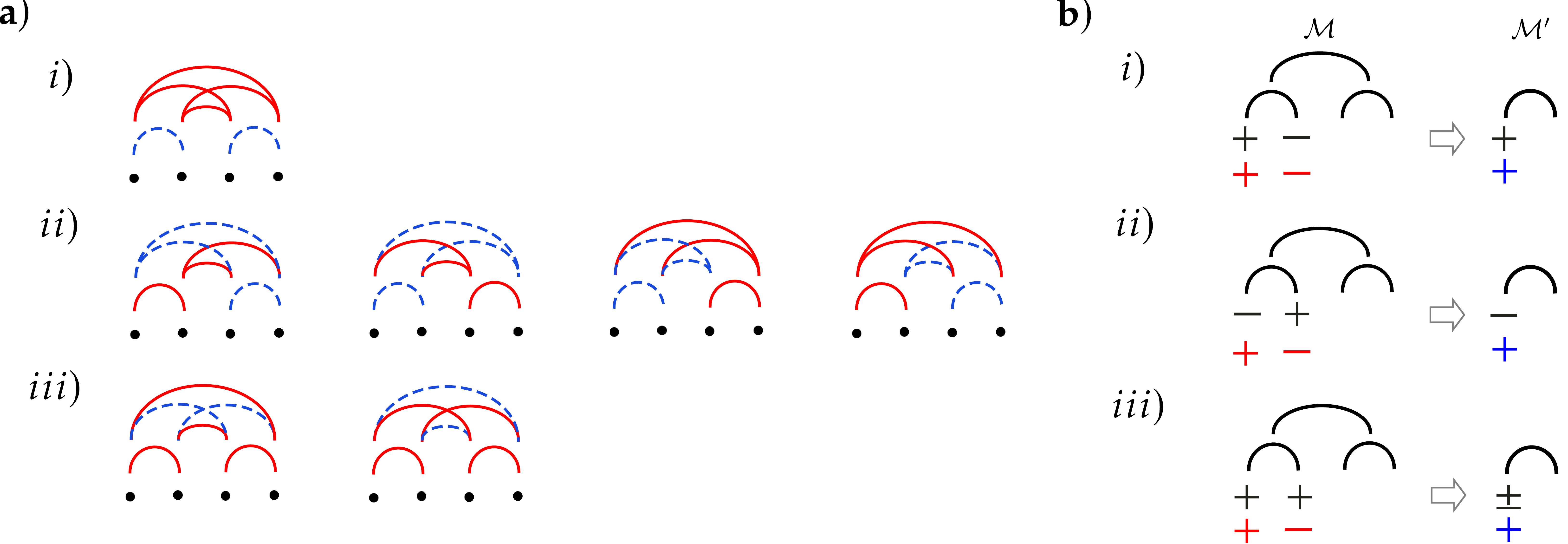}
    \mycaption{Energy excitations for the hierarchical Edwards-Anderson model and block-spin decimation}{\label{fig_ex}
      $\textbf A$) Energy excitations: The four spins of  model $\cM$ are represented by dots, and the coupling between a spin pair by  an arc connecting the dots. First- and second-level couplings are shown on bottom and top, and they correspond to the first two and last four terms in the right-hand side of \cref{eq_91}, respectively. Violated and non-violated couplings in the excited states are shown with red solid and blue dashed curves, respectively. Energy excitations are grouped according to the couplings that they violate:
      i) Excitation that violates all second-level couplings with respect to the ground state. ii) Excitations that violate one first-level  and two second-level couplings. iii) Excitations that violate two first-level and two second-level couplings.

      $\textbf B$) Block-spin decimation. For a given sample of the disordered couplings $\bJ$ and $J'$, models $\cM$ and $\cM'$, on the left and right,  are shown with their respective ground-state spin configurations, displayed in  red  and   blue, respectively. Three  spin configurations of $\cM$ and $\cM'$ are shown in black in panels $i$, $ii$ and $iii$, where only spins in the left half are shown for clarity.
      In each panel, the spin configuration of $\cM$ and $\cM'$ are related by the decimation relation \eqref{eq_dec}.
      $i)$ Left spins of $\cM$ are parallel to their ground state, and $\Phi_\rmL = +1$. As a result, the left spin of $\cM'$ is parallel to its own ground state, and $\Phi'_\rmL = +1$. $ii)$ Same as $i$, for spins antiparallel to the ground states, with $\Phi_\rmL = \Phi'_\rmL=-1$. $iii)$ Left spins of $\cM$ are neither parallel nor antiparallel to their ground state, i.e., $\Phi_\rmL = 0$. As a result, $\Phi'_\rmL = 0$, and the left spin in $\cM'$ is either parallel or antiparallel to the ground state of $\cM'$, with equal probability.

    }
  \end{center}
\end{figure}

To denote the excited states with respect to the ground state, we will use the notation $S_{\uarr \uarr \uarr \darr} \equiv \{\sigma_1,
  \sigma_2, \sigma_3, -\sigma_4\}$, $S_{\uarr \uarr \darr \uarr} \equiv \{\sigma_1,
  \sigma_2, -\sigma_3, \sigma_4\}$, etc. Given an excited state $\bS$ where, for example,  spin $S_i = - \sigma_i$ is flipped with respect to its ground-state value $\sigma_i$, we will say that this
state \textit{violates} the bonds $J_{ij}$, through which  spins $i$ and
$j$ interact.   We will characterize each excited state  by the set of couplings
$J_{ij}$ that it violates. For instance,  $S_{\uarr \uarr \darr \darr}$ violates all couplings on the second
hierarchical level, while it does not violate any coupling on the
first level, see panel $i$ of \cref{fig_ex}A. More generally, in \cref{fig_ex}A the spin configurations $\bS_2, \cdots, \bS_8$ are split into  three different groups, namely the excited states which violate:
\begin{enumerate*}[label=\roman*)]
  \item \label{g1} all second-level couplings, i.e.,  $S_{\uarr \uarr \darr \darr}$,
  \item \label{g2} one first-level coupling and two second-level couplings, i.e.,  $S_{\uarr \darr \uarr \uarr}$, $S_{\uarr \uarr \darr \uarr}$, $S_{\uarr \uarr \uarr \darr}$ and $S_{\uarr \darr \darr \darr}$,
  \item \label{g3} two first-level couplings and two second-level couplings, i.e.,  $S_{\uarr \darr \uarr \darr}$ and $S_{\uarr \darr \darr \uarr}$.
\end{enumerate*}

As shown in \cref{app_omega,sec_value_op,sec_proof}, the only order parameter which is consistent with the minimality principle and with scale invariance is such that that any pair of related states belongs to the same group, leading to
\begin{align}  \label{eq10}
  \Omega[S_{\uarr \darr \uarr \uarr}] =\, \Omega[S_{\uarr \uarr \darr \uarr}] =\ \Omega[S_{\uarr \uarr \uarr \darr}] =\, \Omega[S_{\uarr \darr \darr \darr}],\; \; \;\; \Omega[S_{\uarr \darr \uarr \darr}] =\, \Omega[S_{\uarr \darr \darr \uarr}].
\end{align}

Through the analysis of \cref{sol_eq_sys}, by leveraging the symmetry between left and right half of the model and solving \cref{eq3,eq10} for the coefficients $B_{\rmL
    }$, $B_{\rmR}$, we obtain the desired expression for the order parameter:
\be
\label{eq12}
\Phi_\rmL [\bS]= \frac{\sigma_1 S_1 + \sigma_2 S_2}{2}, \;
\Phi_\rmR [\bS] = \frac{\sigma_3 S_3 + \sigma_4 S_4}{2}.
\ee
Proceeding along the same lines for model $\cMp$, we obtain
\be\label{eq11}
\Phi'_\rmL [\bS']= \sigma'_1 S'_1,  \;
\Phi'_\rmR [\bS'] = \sigma'_2 S'_2,
\ee
see \cref{sol_eq_sys} for details. Taken together, \cref{eq12,eq11} yield the order parameter for the \ac{HEA}, and they constitute one of the main results of this work. \vspace{0.5cm}

The block-spin decimation  above yields a mapping between the probability distributions $p(\bJ)$ and $p'(J')$ of models $\cM$ and $\cMp$, respectively, which reads
\be\label{eq16}
p'(J') =  \frac{1}{2}\int \dJ p(\bJ) \left[ \delta \left(J' -  \mathcal{J}_\iexp(\bJ, T)\right) +  \delta \left(J' +  \mathcal{J}_\iexp(\bJ, T)\right)  \right],
\ee
where $\mathcal{J}_\iexp(\bJ, T)$ is given by \cref{eq_psi} and $\rmd \bJ \equiv \prod_{i<j}\rmd J_{ij}$, see \cref{app_omega,section_decimation_relation} for details.  Here, we require  $p'(J')$ to satisfy the \ac{SG} symmetry, i.e.,  to be  an even function of $J'$, throughout the \ac{RG} transformation. As a result, in \cref{eq16} we include both  solutions of \cref{eq26} with the $+$ and   $-$ sign, by assigning to them the same weight. This symmetry requirement will  be modified when we will consider the ferromagnetic limit of the \ac{RG} transformation, see \cref{sec_ferro}.\\

The order parameter in \cref{eq12,eq11}  depends on the spin-spin couplings $\bJ$ and $J'$   through the ground states only.
In particular, $\Phi[\bS]$ is a normalized scalar product between the spin vector $\bS$ and the ground state $\bsi$, and it thus reflects the alignment of spins with respect to the ground state \revision{$\bsi$}, and similarly for
$\Phi'$. \revision{Here, we recall that  $\bsi$ is the ground state of model $\cM$, which includes spin-spin couplings $\bJ$ with arbitrary sign. As a result, because of the  presence of frustration in $\bJ$, spins in $\bsi$ are not necessarily aligned, and similarly for $\bsi'$.  }

Such structure  of the order parameter allows us to interpret the
mapping between $\bS$ and $\bS'$ given by the decimation rule \eqref{eq_dec}, according to which, for a given $\bS$, spins of
$\cM'$ must be such that $\Phi'[\bS']$ matches $\Phi[\bS]$. For the
sake of simplicity, we will illustrate this point for the left half
of  models $\cM$ and $\cMp$ only; the same conclusions hold for the right half. As shown
in \cref{fig_ex}B, given a configuration of $\bS_\rmL$ in which $\bS_\rmL$  is either aligned or counter aligned with the ground state of model $\cM$,
$\bS'_\rmL$ aligns or counteraligns with the ground state of $\cM'$ in order to match the alignment or counteralignment above of $\bS_\rmL$. If $\bS_\rmL$ is neither parallel nor antiparallel to the  ground state of $\cM$, then $\bS'_\rmL$ is either parallel or antiparallel to the ground state of $\cM'$ with equal probability. \revision{As we pointed out above, in  the ground states $\bsi$ and $\bsi'$ spins are not necessarily aligned, as in the ground states of non-frustrated systems. The mapping between $\bS$ and $\bS'$ is thus built on the alignment with respect to these frustrated ground states, which depend on the coupling values $\bJ$ and $\bJ'$, respectively.}

\subsection{Rescaling}\label{sec_resc}

We will now show how to rescale $\cMp$ so as to re-obtain a four-spin model. The combination of the decimation procedure of \cref{sec_dec} and of such rescaling  will constitute one step of the \ac{RG} transformation.

We denote  models  $\cM$ and $\cMp$  at the $k$th \ac{RG} step  by  $\cM_k$ and $\cM'_k$, their couplings  by $\bJ_k
  \equiv \{ J_{ij \, k} \}$ and $J'_k$, and their coupling distributions by $p_k(\bJ_k)$ and $p'_k(J'_k)$,  respectively. In the decimation relation \eqref{eq16},  $p$ is replaced by $p_k$ and $p'$ by $p'_k$.
The rescaling procedure takes as an input a two-spin model $\cM'_k$, and produces as an output a four-spin model $\cM_{k+1}$, which will be decimated again at the $k+1$-th step to obtain a two-spin model $\cM'_{k+1}$, and so on, see \cref{fig_flow}.
In what follows, we will discuss how  to build the rescaled model $\cM_{k+1}$ from $\cM'_k$.
The coupling distribution of  $\cM_{k+1}$, in general, will not be the product of the single-coupling distributions of $\cM'_k$, because correlations are present, see \cref{sec_sol_corr}.

To determine the rescaled coupling distribution, we  leverage a fundamental, minimal physical feature of the \ac{HEA}.
Above the lower critical dimension,  i.e., for $\iexp < \iexpl$,  the variance of the  Hamiltonian (\ref{eq_H}) is extensive with respect to the system size. As shown in  \cref{fig_diag}, both analytical and numerical studies  show that in this region the  phase diagram of the \ac{HEA} is composed of \cite{moore2010ordered,kotliar1983one, castellana2015non,katzgraber2003monte,katzgraber2005probing}:
\begin{enumerate*}[label=(\alph*)]
  \item  \label{cond_b} a high-temperature region $T  > \Tc$,
  \item \label{cond_a} a critical point  $T = \Tc$,
  \item \label{cond_c} a low-temperature region $T  < \Tc$, where $\Tc \ra 0$ for $\iexp \ra \iexpl$.
\end{enumerate*}
In addition, the model displays no phase transition below the lower critical dimension, $\iexp > \iexpl$, where its Hamiltonian is subextensive \cite{franz2009overlap,moore2010ordered,kotliar1983one}.
Conditions \labelcref{cond_b,cond_a,cond_c}  are sufficient to  specify the  correlation structure of the rescaled couplings $\bJ_{k+1}$, and thus to set out the rescaling procedure.
In fact, such conditions  imply the two following relations at $\iexp = \iexpl$, see \cref{app-rescaling-equations} for details:
\begin{enumerate}
  \item \label{cond_aa} Because  $T=0$ is the critical temperature, the zero-temperature \ac{RG} transformation must leave unchanged the width of the coupling distribution.  In particular, when, at the $k$th \ac{RG} step  and $T=0$, model $\cM'_k$ is rescaled and then decimated into $\cM'_{k+1}$ , the expectation value of the low-energy excitations of $\cM'_{k+1}$ must match that of $\cM'_k$:
        \be\label{eq14}
        \bE[H'^{k+1}[\bS'_2]-H'^{k+1}[\bS'_1]]|_{\iexp = \iexpl, T=0} = \bE[H'^k[\bS'_2] - H'^k[\bS'_1]].
        \ee

  \item \label{cond_bb} For  $T\gtrsim 0$,  the \ac{RG} transformation must shrink the width of the coupling distribution. As a result, at any step $k$,  the  low-energy excitations of the rescaled model $\cM_{k+1}$, i.e., the left-hand side  of \cref{eq14}, must be a non-increasing function of temperature:
        \be\label{eq17}
        \left. \frac{\pa}{\pa T}  \bE[H'^{k+1}[\bS'_2]-H'^{k+1}[\bS'_1]]\right |_{\iexp = \iexpl,T=0} \leq 0.
        \ee
\end{enumerate}

Following the minimality principle of our analysis, we seek  $p_{k+1}$ as the simplest coupling distribution which is consistent with  \cref{eq14,eq17} \cite{jaynes1957information}.
We will consider the distribution  $p_{k+1}^\ast(\bJ) \equiv \prod_{i<j}p'_k(J_{ij})$ of independent couplings  as the simplest choice for the coupling distribution.
In fact, $p_{k+1}^\ast$ involves no inter-coupling correlations, i.e., it contains a minimal amount of information.
On the other hand, $p_{k+1}(\bJ)$  incorporates a larger amount of information because it involves inter-coupling correlations: Knowing the actual, correlated distribution $p_{k+1}$  rather than $p_{k+1}^\ast$  thus yields an `information gain.'
This reasoning naturally leads to  the Kullback-Leibler divergence  between  $p_{k+1} $ and $p_{k+1}^\ast$, $D[p_{k+1} || p_{k+1}^\ast] \equiv \int \dJ p_{k+1}(\bJ) \log [p_{k+1}(\bJ)  / p_{k+1}^\ast(\bJ) ]$ \cite{kullback1951information}.
Besides its interpretation as a statistical distance between  distributions \cite{amari2016information}, in information theory and machine learning $D[p_{k+1} || p_{k+1}^\ast]$ represents the information that is lost when  $p_{k+1}^\ast$ is used to approximate $p_{k+1}$, or the  information gain obtained  if $p_{k+1}$ is used instead of $p_{k+1}^\ast$ \cite{burnham2003model}.
Following a minimality principle, we  seek $p_{k+1}$ as the distribution which yields as a little information gain with respect to $p_{k+1}^\ast$ as possible,  which satisfies the physical conditions \eqref{eq14}  and \eqref{eq17}, and  is properly normalized. The distribution $p_{k+1}$ is thus the solution of the following optimization problem:

\be \label{resc1}
\begin{aligned}
  \min_{p_{k+1}}D[p_{k+1} || p_{k+1}^\ast]                                           &   \\
  \textrm{subject to Eqs. \eqref{eq14}, \eqref{eq17} and} \int \dJ p_{k+1}(\bJ)  = 1 & ,
\end{aligned}
\ee
where the left-hand sides of \cref{eq14,eq17} depend on  $p_{k+1}$ through  \cref{eq16}, which relates $p'_{k+1}$ to $p_{k+1}$.

As shown in \cref{app0,app1,sol_opt_pr,app_unique}, the solution of the optimization problem can be worked out explicitly, and it is given by \cref{eq31}.

\section{Renormalization-group flow and fixed distributions}\label{sec_flow}

The combination of the decimation and rescaling procedures define a temperature-dependent \ac{RG} transformation  for the spin-coupling probability distribution $p'_k$: $p'_k(J') \ra p'_{k+1}(J')$, see \cref{fig_flow}.

We study the flow  of the \ac{RG} transformation for $p'_k$, by approximating $p'_k$ as a finite number of parameters, study the  flow of such parameters, and recover the exact results as the number of parameters goes to infinity.
Given that $p'(J')$ is even throughout the \ac{RG} transformation, we consider only half of the image of  its \ac{CDF}, i.e., the interval $[0,1/2]$, partition it into $N+1$ intervals, and thus parametrize $p'$  in terms of its quantiles  \cite{blitzstein2015introduction} ${\bm \qJ} \equiv \{ \qJ_1, \cdots, \qJ_N\}$, see \cref{sec_disc} for details. Given the quantiles $\bm \qJ^k$ of $p'_k$, we obtain the distribution  $p_{k+1}$ by applying to $p'_k$  the rescaling  procedure set out in \cref{sec_resc}. We then apply to $p_{k+1}$ the decimation procedure of \cref{sec_dec}, and obtain the quantiles ${\bm \qJ}^{k+1}$ of $p'_{k+1}$ by solving \cref{eq37}, see \cref{sec_quant_dec} for details. This procedure results in  a temperature-dependent mapping between  quantiles
${\bm \qJ}^k \ra {\bm \qJ}^{k+1}$ for each \ac{RG} step.

The  resulting fixed-distribution structure of the \ac{RG} transformation  presents the same topology as in ferromagnetic systems \cite{wilson1974the}: The high- and low-temperature fixed distributions are attractors which are reached at high and low temperatures, respectively, see \cref{fixed-distribution-structure,sec_lh}.

In addition to the fixed distributions of \cref{sec_lh}, we seek a critical fixed distribution with a finite width \cite{wilson1975renormalization}. To achieve this, we iterate the \ac{RG} transformation at a given value of $\iexp$ and, at each step $k$, we set  the temperature by imposing that the width of $p'$ at the $k+1$-th step equals that at the  $k$th step:
$\int \rmd J' p'_{k+1}(J') |J'| =  \int \rmd J' \, p'_{k}(J') |J'| =  \int  \rmd \bJ \, p_{k+1}(\bJ) \mathcal{J}_\iexp(\bJ, T)$.
In the last equality we rewrote the expectation value of $|J'|$ with  $p'_{k+1}$ as per  \cref{eq16}, so as to bring out its temperature dependence.
We solved the relation above for $\beta$ with stochastic-approximation methods, see \cref{sec_num} for details. By then iterating the \ac{RG} transformation, $p_k$ converges,  for large $k$, to a finite fixed distribution, and $\beta$ converges to a finite value $\betac$. \Cref{fig_cr} shows the critical fixed distribution for multiple values of $\iexp$, whose stability will be discussed below.

\section{Predictions}\label{section_predictions}

We will first study the  predictions of the \ac{RG} procedure in some specific limits, see \cref{part_limits} for details.  In \cref{low_crit_dim} we demonstrate that the predicted critical temperature satisfies the lower-critical-dimension limit  for the critical temperature, $\Tc \ra 0$ for $\iexp \ra \iexpl$ \cite{moore2010ordered}. In addition, \cref{sec_ferro} shows that, in the ferromagnetic limit, we recover  the \ac{RG} transformation of the ferromagnetic version of the \ac{HEA} \cite{dyson1969existence}. Finally, in \cref{zero_temp_first} we show that at zero temperature and in the approximation where only first-level couplings are considered, our \ac{RG} transformation reduces to an \ac{RG} decimation procedure proposed recently \cite{monthus2014one}.

\begin{figure}
  \begin{center}
    \includegraphics[width=0.95\textwidth]{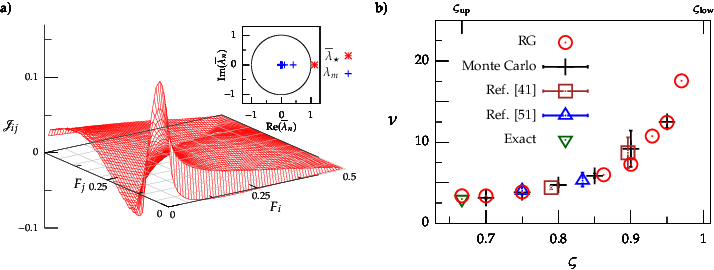}
    \mycaption{Linearization of the \acf{RG} transformation and critical exponent}{\label{fig_li}
      $\textbf A$) Jacobian $\bm \jacs$ of the scaled \ac{RG} transformation evaluated at the scaled critical fixed distribution ${\bm \qJs}^k = \betac \bm{\qJ}_{\rm c}$, where  $\bm \qJ$ are  the  quantiles of the spin-coupling distribution, as a function of the values $F_i$, $F_j$ of the relative cumulative distribution function, for $\iexp = 0.8$. Here the cumulative-distribution-function values $F_i$, $F_j$  serve as labels for the Jacobian rows and columns, respectively, and $i,j = 1,\cdots, N$, where $N$ is the number of bins of the discretization. Inset: eigenvalues $\lambdas_m$ of $\bm \jacs$ with norm smaller than unity (blue), the eigenvalue $\olambdaast$ with norm larger than unity (red), and the unit disk (black).
      $\textbf B$) Critical exponent $\nu$, which describes the divergence of the correlation length, as a function of the coupling-range exponent $\iexp$. The values of $\nu$ obtained from the renormalization-group method are shown as red circles. Monte Carlo simulation predictions for $\nu$ from this work (black crosses) have been generated from the Hierarchical Edwards-Anderson model with power-law interaction decay  for $\iexp \leq 0.8$, and  with fixed coordination number for $\iexp > 0.8$---see \cref{sec_mc}. Monte Carlo predictions for $\nu$ from previous studies \cite{leuzzi2008dilute,banos2012correspondence}, obtained from  diluted one-dimensional Ising spin glasses with power-law interactions, are shown as brown squares and blue triangles.  \revision{We also show the upper and   lower critical dimensions, $\iexpu$ and $\iexpl$, respectively, and the exact value of $\nu$ at $\iexpu$ (green triangle).}
    }
  \end{center}
\end{figure}

We  will now discuss the predictions of our \ac{RG} method  for the critical exponent $\nu$  related to the  divergence of the correlation length \cite{wilson1974the}.
We linearize the \ac{RG} transformation at the critical fixed distribution, see \cref{scaled_rg_tr,section_jacobian} for details.
Introducing the scaled quantiles
$\bm{\qJs}^k \equiv \beta \bm{\qJ}^k$ if the Jacobian $\bm \jacs$ of the scaled \ac{RG} transformation ${\bm \qJs}^k \ra {\bm \qJs}^{k+1}$  has at least one eigenvalue with norm larger than unity,  then $\nu$  is obtained from the eigenvalue $ \olambdaast$ with the largest norm by means of the relation $\olambdaast = 2^{1/\nu}$ \cite{wilson1974the}.
In \cref{fig_li}A we show the linearization of the \ac{RG} transformation at the critical fixed distribution for $\iexp = 0.8$, obtained with $N=2^6$ bins, and $S = 2^{12}$ and $\rmiterations =  2^{22}$ as parameters for the numerical solution of the \ac{RG} equations with stochastic-approximation methods, see \cref{sec_num}.  There is one relevant eigenvalue $ \olambdaast$, implying that the critical fixed distribution is unstable. The  eigenvectors relative to $ \olambdaast$ are shown in \cref{fig_li_bis}.

The numerical value of $\nu$  as a function of  $\iexp$ is shown in \cref{fig_li}B, which constitutes one of the main results of this work.
In order to test the prediction of the \ac{RG} approach, we performed extensive \ac{MC} simulations for two diluted versions of the \ac{HEA}, see \cref{sec_mc} for details, whose predictions for $\nu$ are shown in \cref{fig_li}B.
In addition, \cref{fig_li}B shows values of $\nu$ from previous \ac{MC} simulations for two one-dimensional \ac{SG} models where the interaction strength decays with distance $r$ as $\sim r^{-\iexp}$  \cite{leuzzi2008dilute,banos2012correspondence}.  The three simulated models above are supposed to belong to the same universality class as the \ac{HEA} studied with the \ac{RG} method \cite{katzgraber2006universality,castellana2011renormalization}.

\acresetall
\section{Discussion}\label{the_discussion}

We propose a \ac{RG} method for non-mean-field models of spin glasses\revision{, which leads to the emergence of a novel order parameter.} We focus on a spin-glass model built on a hierarchical lattice, the \ac{HEA}.
Unlike previous methods \cite{monthus2014one,parisi2001renormalization,chen1977mean,bray1980renormalisation,kotliar1983one,temesvari2002generic,castellana2011renormalization,castellana2015hierarchical,castellana2011real,angelini2013ensemble}, ours follows a minimality principle, where no a priori assumption is made on the physics of the system. In the decimation procedure,  the order parameter spontaneously emerges solely from the fundamental symmetries of the system and from the requirement that the \ac{RG} transformation preserves the order parameter and model structure across  length scales. Such order parameter is given by the projection of the spin configuration on the ground state of the system. As a result, Kadanoff's majority rule \cite{kadanoff1966scaling} is replaced by a more complex scheme: Rather than letting a block of spins point up (down) if the majority of the spins in the block are up (down), here a spin block  is set parallel (antiparallel) to its ground state, if the majority of the spins in the block are parallel (antiparallel) to their ground state, see \cref{fig_ex}B. The ground state thus acts as an underlying pattern which translates spin configurations from one length scale to another. \revision{Given the random signs in the spin-spin couplings, in such ground state spins need not be aligned as in non-frustrated, e.g., ferromagnetic systems.}
The decimation is then followed by a  rescaling procedure, based on an information-theory approach.

Combined, the decimation and rescaling procedure yield an \ac{RG} transformation for the probability distribution of spin couplings. In the limit where spin-spin couplings are ferromagnetic, such transformation reproduces the magnetization order parameter and Kadanoff's majority rule. The \ac{RG} space of the transformation presents a stable, high-temperature fixed distribution, a low-temperature fixed distribution, and an unstable critical fixed distribution.
\revision{
  The \ac{RG} predictions for the critical exponent $\nu$, which describes the critical divergence of the correlation length, are shown in \cref{fig_li}B. First, the \ac{RG} method is capable of  predicting the value of $\nu$  close to the lower  critical dimension---a region practically unaccessible to numerical simulations because of the long equilibration times  \cite{young2006numerical,leuzzi2008dilute,banos2012correspondence}. In this limit, the \ac{RG} predictions are compatible with  $\nu \ra 0$ for $\iexp \ra \iexpl$: This is the expected limit for the critical exponent at the lower critical dimension, where semi-analytical treatments are possible \cite{moore2010ordered}. Second, in the region between the upper and lower critical dimension,
}
the \ac{RG} predictions are in excellent agreement with numerical simulations from both this and previous studies \cite{leuzzi2008dilute,banos2012correspondence}, see \cref{fig_li}B.

\revision{Given the strong connection between order parameter and   low-temperature phase, our approach provides a new framework to identify the  low-temperature structure of spin-glass models  beyond the mean-field approximation, as well as their critical ordering}, and it opens multiple future directions.
In addition to the  developments discussed in  \cref{sec_supp_discussion}, our analysis may shed light on a long-standing debate---the nature of the low-temperature phase \cite{stein2004spin} of non-mean-field spin-glass models. In this regard, two mainstream  theories, the replica-symmetry breaking \cite{parisi1983order} and the droplet \cite{fisher1986ordered,fisher1988equilibrium} picture, predict the existence of two versus infinitely many pure states at low temperatures, respectively \cite{fisher1987absence}. Such theories markedly differ  in the presence of a magnetic field. Below the upper critical dimension, the replica-symmetry-breaking picture predicts a phase transition in a magnetic field \cite{mezard1987spin}, while in the droplet picture this transition is wiped out by the field \cite{fisher1988equilibrium}. By extending our analysis to this scenario, one may assess the existence of an unstable, critical fixed distribution associated with a phase transition, and characterize the nature of this instability through the spectral analysis of \cref{fig_li,fig_li_bis}. This future direction  may also yield a novel order-parameter structure in the presence of a field. Indeed,  while the order parameter \eqref{eq12} depends on the Hamiltonian parameters---the spin-spin couplings $\bJ$---through the ground state, the order parameter in a field may also display a nontrivial dependence on the field itself.

\revision{
The predictions of our analysis on the nature of the low-temperature phase can be also investigated through the stiffness exponent $\theta$,  which describes the low-energy excitations of the system \cite{fisher1988equilibrium}.
In order to compute $\theta$, one may consider the energy gap above the ground state $\bE[H_k'[\bS'_2] - H_k'[\bS'_1]]  \sim 2^{k \theta}$, and obtain $\theta$ by fitting the energy gap  with respect to the \ac{RG} step $k$.  The \ac{RG} results for the stiffness exponent may then be compared with the prediction $\theta = 1 - \iexp$  for one-dimensional models with long-range interactions \cite{monthus2014one}.
}

Finally, our method may be extended to spin-glass models on a hypercubic lattice with nearest-neighbor interactions, such as the Edwards-Anderson model \cite{edwards1975theory}, which mimics the disordered, short-range interaction potential in spin-glass alloys such as AuFe or CuMn  \cite{katori1994experimental,nair2007critical}, so as to directly relate the \ac{RG} predictions to experimental data.

\section*{Acknowledgments}
We would like to thank A. Barra, I. A. Campbell, J.-F. Joanny, V. Martin-Mayor,  M. A. Moore, D. J. Papoular, J. Prost and F. Zamponi for useful discussions.
This work was granted access to the computational resources supported by Institut Curie, and to  the HPC resources of MesoPSL financed by the Region \^{I}le de France and the project Equip@Meso (reference ANR-10-EQPX-29-01) of the programme Investissements d’Avenir supervised by the Agence Nationale pour la Recherche.

\bibliographystyle{unsrt}
\bibliography{bibliography}



\section*{Supplementary material}

# Supplementary Material for 'Order parameter for non-mean-field spin glasses'

Michele Castellana[1]

[1]Institut Curie, PSL Research University, CNRS UMR168, France

December 9, 2025

# Contents

# List of Figures

## S1  Reduction of the decimation relation

Given that $S'$ can take only two values, the most geenral form of $\Phi'$ is

$$\Phi'[S'] = a + bS', \tag{S1}$$

where, for the sake of clarity, in what follows we will drop the left and right subscripts unless necessary. The indicator functions in the right-hand side of Eq. (2) can thus be written as

$$\begin{aligned}\mathbb{I}(\Phi'[S'] = \varphi) &= \mathbb{I}(a + bS' = \varphi) \\ &= \mathbb{I}\left(\frac{b}{|b|}S' = \frac{\varphi - a}{|b|}\right) \\ &\to \mathbb{I}(\Phi'[S'] = \varphi),\end{aligned} \tag{S2}$$

where in the last line of Eq. (S2) we re-defined $\Phi'$ and $\varphi$:

$$\Phi'[S'] \to \frac{b}{|b|}S', \tag{S3}$$

$$\frac{\varphi - a}{|b|} \to \varphi. \tag{S4}$$

Setting $b/|b| \equiv \zeta = \pm 1$, the quantity

$$\Phi'[S'] = \zeta S' \tag{S5}$$

in the last line of Eq. (S2) can be equal to $\pm 1$ only, implying that $\varphi$ may take the values $+1$ and $-1$ only. As a result, the identity function $\mathbb{I}$ can be written in the form

$$\mathbb{I}(\Phi'[S'] = \varphi) = \frac{1 + \Phi'[S']\varphi}{2}. \tag{S6}$$

Given that the renormalization group (RG) transformation must preserve its structure across multiple length scales, i.e., no element in its formal definition may tell apart one scale from another, the identity function $\mathbb{I}$ of model $\mathcal{M}$ that enters the left-hand side (LHS) of Eq. (2) must have the same form (S6) as the identity function of model $\mathcal{M}'$:

$$\mathbb{I}(\Phi[S] = \varphi) = \frac{1 + \Phi[S]\varphi}{2}. \tag{S7}$$

By using Eq. (S7) we can thus rewrite the LHS of Eq. (2) as

$$\langle \mathbb{I}(\Phi_L[S_L] = \varphi_L)\, \mathbb{I}(\Phi_R[S_R] = \varphi_R)\rangle = \frac{1}{4}(1 + \langle\Phi_L[S_L]\rangle\varphi_L + \langle\Phi_R[S_R]\rangle\varphi_R + \langle\Phi_L[S_L]\Phi_R[S_R]\rangle\varphi_L\varphi_R). \tag{S8}$$

Given that $\Phi'[S']$ is an odd function of $S'$ and that the decimation must preserve such symmetry, $\Phi[S]$ must be an odd function of $S$, see Section 2.1. As a result, the second and third terms in the right-hand side of Eq. (S8) vanish. By using Eq. (6) and Eqs. (S6) and (S8), the spin-decimation relation (2) is reduced to Eq. (5).

Finally, we observe that, by using Eq. (S5), Eq. (5) can be written as

$$\begin{aligned}\langle\Omega[S]\rangle &= \zeta_L\zeta_R\langle S'_1 S'_2\rangle' \\ &= \zeta_L\zeta_R\,\mathrm{sgn}(J')\frac{1 - e^{-\beta(H'[S'_2] - H'[S'_1])}}{1 + e^{-\beta(H'[S'_2] - H'[S'_1])}}.\end{aligned} \tag{S9}$$

## S2  Value of the order parameter on the ground state

We will now derive an important property of the function $\Omega$, defined by Eq. (6), when evaluated on the ground state (GS):

$$\Omega[S_1] = \pm 1, \tag{S10}$$

where the $\pm 1$ sign will denote two distinct cases in what follows, and the analog of Eq. (S10) holds for model $\mathcal{M}'$.

We will now prove Eq. (S10), by using the *reductio ad absurdum*. Solving Eq. (S9) for the energy gap $H'[S'_2] - H'[S'_1]$, we obtain

$$H'[S'_2] - H'[S'_1] = \frac{\zeta_L \zeta_R \, \text{sgn}(J')}{\beta} \log \frac{1 + \langle \Omega[S] \rangle}{1 - \langle \Omega[S] \rangle}$$
$$= 2\zeta_L \zeta_R \, \text{sgn}(J') \mathcal{J}_\varsigma(J, T), \tag{S11}$$

where $\mathcal{J}_\varsigma$ is given by

$$\mathcal{J}_\varsigma(J, T) \equiv \frac{1}{2\beta} \log \frac{\sum_p e^{-\beta(H[S_p] - H[S_1])}(1 + \Omega[S_p])}{\sum_p e^{-\beta(H[S_p] - H[S_1])}(1 - \Omega[S_p])}. \tag{S12}$$

If Eq. (S10) did not hold, Eq. (S11) for large $\beta$ would imply that $H'[S'_2] - H'[S'_1]$ vanishes identically:

$$H'[S'_2] - H'[S'_1] = \frac{\zeta_L \zeta_R \, \text{sgn}(J')}{\beta} \log \frac{1 + \Omega[S_1] + \mathcal{O}(e^{-\beta(\epsilon_2 - \epsilon_1)})}{1 - \Omega[S_1] + \mathcal{O}(e^{-\beta(\epsilon_2 - \epsilon_1)})}$$
$$\xrightarrow{\beta \to \infty} 0, \tag{S13}$$

where we have set

$$\epsilon_p \equiv H[S_p], \tag{S14}$$

and we denote by $\mathcal{O}$ the orders of magnitude in the small-$T$ limit. Given that Eq. (S13) is physically absurd, Eq. (S10) must hold.

We will now show that the analog of Eq. (S10) holds for $\Omega'$. To achieve this, we observe that, in the zero-temperature limit $T \to 0$, Eqs. (S10) and (S11) yield

$$H'[S'_2] - H'[S'_1] = \pm \zeta_L \zeta_R \, \text{sgn}(J')(H[S_2] - H[S_1]). \tag{S15}$$

Given that the energy gaps $H[S_2] - H[S_1]$ and $H'[S'_2] - H'[S'_1]$ are both positive, Eq. (S15) implies

$$\zeta_L \zeta_R \, \text{sgn}(J') = \pm 1. \tag{S16}$$

We substitute Eqs. (S5) and (S16) into the second relation in Eq. (6) and obtain:

$$\Omega'[S'_1] = \zeta_L \zeta_R S'_{1L} S'_{1R}$$
$$= \pm \text{sgn}(J') S'_{1L} S'_{1R} \tag{S17}$$
$$= \pm 1,$$

where in the last line we used the fact that, in the GS, spins $S'_{1L}$ and $S'_{1R}$ are parallel and antiparallel if $J'$ is positive or negative, respectively.

For future reference, we observe that Eqs. (S15) and (S16) imply that the lowest energy gaps of $\mathcal{M}$ and $\mathcal{M}'$ match:

$$H[S_2] - H[S_1] = H'[S'_2] - H'[S'_1], \tag{S18}$$

meaning that the energy gap is invariant across length scales, consistently with the general framework of our analysis.

## S3 Value of the order parameter on the excited states

The values that $\Omega$ takes when evaluated on the excited states can be determined as follows.

We set $\sigma \equiv S_1$, and similarly for $\mathcal{M}'$. First, if $\Omega$ takes the same value on a pair of excited states, i.e., $\Omega[S_p] = \Omega[S_q]$, we will say that $S_p$ and $S_q$ are *related*. We may then tell apart the following possibilities:

1. All excited states are related:

$$\Omega[S_2] = \Omega[S_3] = \cdots = \Omega[S_8]. \tag{S19}$$

The symmetry between left and right half of the model combined with Eq. (S19) imply that

$$B_{L\,i} = B_{R\,i} = 0, \; i = 1, 2, \tag{S20}$$

where we recall that $B_L$ and $B_R$ are defined by

$$\Phi_L[S] = A_L + B_{L1}S_1 + B_{L2}S_2 + C_L S_1 S_2, \tag{S21}$$

see Sections 2.1 and S4 for details. Combining Eqs. (S20) and (S21) and the relation $A_L = C_L = 0$ of Section 2.1, we obtain that $\Phi_L = \Phi_R = 0$, which implies that this case is not physically meaningful, and it will be ruled out.

2. Only some excited states are related:

   (a) Given any pair of related states, its states belong to the same group.

   Some of the excitation groups shown in Fig. 1A may thus contain at least one pair of related states, while others may not. Given that there is no a priori rationale to establish which excitations groups contain at least one pair of related states, all such groups must contain at least one pair of related states. Proceeding along the same lines, given that there is no a priori rationale to select specific pairs of related states in a group, all states within a group must be related. As a result, we obtain Eq. (7).

   (b) Given any pair of related states, its states belong to different groups.

   Proceeding along the lines of 2a, there is no a priori rationale to establish which pairs of excitations groups contain at least one pair of related states, nor to select specific pairs of related states in a pair of groups. As a result, given any pair of excitation groups, all states in the first member of the pair must be related to all states in the second member of the pair. As a result, we are led back to case 1, which is ruled out.

   (c) Given any pair of related states, its states may belong to to either the same group, or to different groups. Given that there is no a priori rationale to chose whether such related states belong to the same group or to different groups, we are led to either case 2a or 2b and, given that case 2b is ruled out, we are led to case 2a.

3. No excited state are related.

   In this case
   $$\Omega[S_i] = O_i, \; i = 2, \cdots, 8, \tag{S22}$$

   where $O_i$ are different, independent values. Given that Eq. (S22) is a set of seven independent equations for four variables $B_{L\,i}$, $B_{R\,i}$, the system (S22) is overdetermined, and this case is ruled out.

As a result of the analysis above, 2a is the only viable case.

## S4  Order parameter and the energy-excitation groups

Let us introduce the reduced spin variables $\tau$, which will allow us to rewrite the equations in a form which is independent on the GS $\sigma$. We set

$$\tau_i \equiv S_i \sigma_i, \tag{S23}$$

which represent the alignment of $S$ with respect to the GS $\sigma$. We also set

$$\begin{aligned}\phi_L[\tau] &\equiv \Phi_L[S] \\ &= b_{L1}\tau_1 + b_{L2}\tau_2,\end{aligned} \tag{S24}$$

where

$$b_{L\,i} \equiv \sigma_{L\,i} B_{L\,i}, \tag{S25}$$

and similarly for the right half, and $\sigma_{L\,i} \equiv \sigma_i$, $\sigma_{R\,i} \equiv \sigma_{2+i}$. Also, we set

$$\omega[\tau] \equiv \Omega[S]. \tag{S26}$$

If case 1 of Section S3 holds, Eqs. (S19) and (S26) thus imply that $\omega[\tau_2] = \omega[\tau_3] = \cdots = \omega[\tau_8]$, which implies the following system of equations for $b_L$, $b_R$

$$\begin{cases} b_{L1}(b_{R1} + b_{R2}) = 0, \\ b_{R2}(b_{L1} - b_{L2}) = 0, \\ b_{R2}(b_{L1} + b_{L2}) = 0, \\ (b_{L1} - b_{L2})(b_{R1} - b_{R2}) = 0, \\ (b_{L1} + b_{L2})(b_{R1} - b_{R2}) = 0. \end{cases} \quad (S27)$$

We will solve the system of equations (S27) as follows, by considering the following cases:

(i) If
$$b_{L1} = 0. \quad (S28)$$

Then Eq. (S27) reduces to
$$\begin{cases} b_{L2} b_{R2} = 0 \\ b_{L2} b_{R1} = 0 \end{cases}. \quad (S29)$$

(a) If
$$b_{L2} \neq 0, \quad (S30)$$
then Eq. (S29) yields
$$b_{R1} = b_{R2} = 0. \quad (S31)$$
This solution is not valid, because Eqs. (S28), (S30) and (S31) would imply that $\Phi_L$ and $\Phi_R$ have different forms, and this would violate the symmetry between the left and right half of the system.

(b) If
$$b_{L2} = 0, \quad (S32)$$
by symmetry, Eqs. (S28) and (S32) imply $b_{R1} = b_{R2} = 0$.

(ii) If
$$b_{L1} \neq 0, \quad (S33)$$
Eq. (S27) implies
$$\begin{cases} b_{R1}(b_{L1} + b_{L2}) = 0 \\ b_{R1}(b_{L1} - b_{L2}) = 0 \end{cases}. \quad (S34)$$

We then have the following cases:

(a) If $b_{R1} \neq 0$, Eq. (S34) implies that $b_{L1} = b_{L2} = 0$, which contradicts Eq. (S33), and thus rules out this case.

(b) The case
$$b_{R1} = 0 \quad (S35)$$
is not valid either: in fact, Eqs. (S33) and (S35) would explicitly break the symmetry between the left and right half of the system.

It follows that (i)b is the only viable case and all $b$s vanish, which, combined with Eq. (S25), implies Eq. (S20).

## S5 Solution for the order-parameter coefficients

In this Section, we will work out the expression for the order parameter in case 2a of Section S3.

By substituting Eq. (S26) in Eq. (7), we obtain

$$\begin{cases} (b_{L1} + b_{L2})(b_{R1} + b_{R2}) = \pm 1, \\ (b_{L1} - b_{L2})(b_{R1} + b_{R2}) = 0, \\ b_{L2} b_{R1} - b_{L1} b_{R2} = 0, \\ b_{L1} b_{R1} - b_{L2} b_{R2} = 0, \\ (b_{L1} - b_{L2})(b_{R1} - b_{R2}) = 0. \end{cases} \quad (S36)$$

To solve Eq. (S36), let us consider its last line and the following cases

(i)
$$b_{L1} = b_{L2}. \tag{S37}$$

Equation (S36) yields
$$\begin{cases} 2b_{L2}(b_{R1} + b_{R2}) = \pm 1, \\ b_{L2}(b_{R1} - b_{R2}) = 0, \end{cases} \tag{S38}$$

which implies
$$b_{R1} = b_{R2}, \tag{S39}$$
$$4b_{L2}b_{R1} = \pm 1. \tag{S40}$$

By substituting Eqs. (S37), (S39) and (S40) into Eqs. (6) and (S24), we obtain
$$\begin{aligned} \Omega[S] &= \pm \frac{1}{4}(\tau_1 + \tau_2)(\tau_1 + \tau_2) \\ &= \phi_L[\tau_L]\phi_R[\tau_R] \end{aligned} \tag{S41}$$

The symmetry between the left left and right half of $\mathcal{M}$, combined with Eq. (S41), implies that the only viable possibility which leads to a symmetric, real-valued order parameter is
$$\begin{aligned} \phi_L[\tau] &= \frac{\tau_1 + \tau_2}{2}, \\ \phi_R[\tau] &= \frac{\tau_3 + \tau_4}{2} \end{aligned} \tag{S42}$$

combined with the choice of the + sign. By substituting Eq. (S23) in Eq. (S42) and Eq. (S42) in Eq. (S24), we obtain Eq. (8).

(ii)
$$b_{R1} = b_{R2}. \tag{S43}$$

By substituting Eq. (S43) into Eq. (S36), we obtain Eq. (S38) with left ↔ right. Proceeding along the lines of (i), we obtain that the + sign must hold in Eq. (S41), and we obtain Eq. (8).

We will now proceed along the same lines for model $\mathcal{M}'$, for which we will obtain the equivalent of Eq. (8). Given that $\mathcal{M}'$ is a two-spin model, at the GS its spins are parallel or anti-parallel if their coupling $J'$ is positive or negative, respectively:
$$\sigma'_1 \sigma'_2 = \text{sgn}(J'). \tag{S44}$$

The second relation in Eq. (6) thus yields
$$\begin{aligned} \Omega'[S'] &= \zeta_L \zeta_R S'_L S'_R \\ &= \text{sgn}(J') S'_L S'_R \\ &= (\sigma'_1 S'_1)(\sigma'_2 S'_2), \end{aligned} \tag{S45}$$

where in the first line we used Eq. (S5), in the second line we substituted Eq. (S16) with the + sign as discussed above, and in the last line we used Eq. (S44). By comparing the last line of Eq. (S45) with the second relation in Eq. (6), we obtain Eq. (9).

## S6 Decimation relation

By substituting in Eq. (S16) the + sign, see Section S5, and using Eq. (S16) into Eq. (S11), we obtain
$$\mathcal{J}_\varsigma(J, T) \geq 0. \tag{S46}$$

Also, given that, according to Eqs. (4) and (S44),
$$H'[S'_2] - H'[S'_1] = 2|J'|, \tag{S47}$$

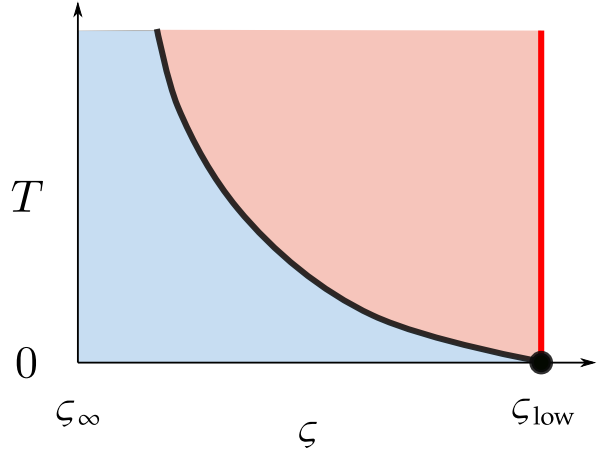

**Figure S1: Cartoon picture of the phase diagram of the hierarchical Edwards-Anderson model.** Critical temperature (black curve) as a function of $\varsigma_\infty \leq \varsigma \leq \varsigma_{\text{low}}$, where $\varsigma_\infty$ and $\varsigma_{\text{low}}$ are the value of the interaction parameter below which the thermodynamic limit is defined, and the lower critical dimensions of the model, respectively. The critical line separates the high- and low-temperatures phases (dim red and blue areas, respectively). The phase diagram of the system for $\varsigma = \varsigma_{\text{low}}$, is composed of a critical point $T = 0$ (black dot), and a high-temperature phase (red line).

Eq. (S11) can be rewritten as
$$|J'| = \mathcal{J}_\varsigma(\mathbf{J}, T). \tag{S48}$$
which implies
$$J' = \pm \mathcal{J}_\varsigma(\mathbf{J}, T), \tag{S49}$$
where $\mathcal{J}_\varsigma$ is given by Eq. (S12) and, for the sake of clarity, we observe that the $\pm$ sign no longer denotes the $\pm$ sign in Eq. (S10), which has been fixed to $+$ in Section S5. As we pointed out in Section 2.1, the $+$ and $-$ solution in Eq. (S49) will be given the same weight, so as to preserve the symmetry of the spin-coupling distribution.

Finally, the relation (S49) between the couplings of $\mathcal{M}$ and $\mathcal{M}'$ implies Eq. (10) for the probability distributions $p(\mathbf{J})$ and $p'(\mathbf{J}')$.

## S7 Independent couplings

A tentative way to build the coupling distribution of model $\mathcal{M}_{k+1}$ would be to draw independently each coupling $J_{k+1\,ij}$ with the distribution $p'_k(J'_k)$ of $\mathcal{M}'_k$, see Fig. S2, thus writing
$$p_{k+1}(\mathbf{J}) \to p^*_{k+1}(\mathbf{J}) \equiv \prod_{i<j} p'_k(J_{ij}). \tag{S50}$$

If the couplings of model $\mathcal{M}_{k+1}$ were drawn according to Eq. (S50), i.e., by neglecting inter-coupling correlations, when $\mathcal{M}_{k+1}$ is further decimated, the width of the resulting coupling $J'_{k+1}$ would shrink to zero as the RG transformation is iterated for all $T \geq 0$. This would violate the condition that, above lower critical dimension, i.e., for $\varsigma < \varsigma_{\text{low}}$ [2–4], the RG transformation at zero temperature must increase the width of the coupling distribution:
$$\mathbb{E}[J'^2_k] \stackrel{k \to \infty}{\to} \infty, \tag{S51}$$
where in what follows we denote by $\mathbb{E}[]$ the expectation value taken with the probability distributions of the couplings which appear in its argument, e.g., $p'_k(J'_k)$ in Eq. (S51).

In order to understand the feature above, we observe that the coupling $J'_k$ in Eq. (S51) is related to the energy gap of the rescaled model by
$$2|J'_k| = H_k[\mathbf{S}_2] - H_k[\mathbf{S}_1], \tag{S52}$$
see Eqs. (S18) and (S47). We thus conclude that inter-coupling correlations play an important role in constructing a correct rescaling procedure, and that drawing the couplings independently results in an underestimate of

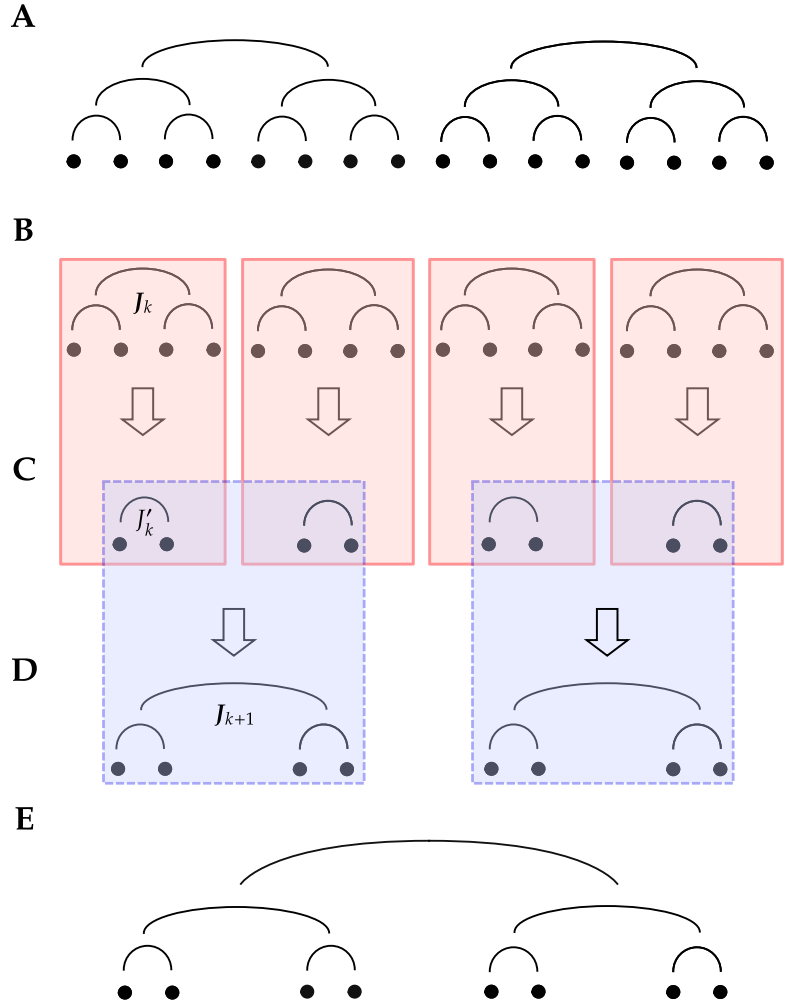

**Figure S2: Renormalization-group (RG) flow for the hierarchical Edwards-Anderson model (hierarchical Edwards-Anderson model (HEA)).** Diagram of the RG procedure in the approximation where a four-spin model is reduced to a two-spin model. **A)** An infinite HEA is considered, of which only two eight-spin blocks are shown for clarity. **B)** Four-spin models with couplings $J_k$ at the first and second hierarchical levels are considered, while couplings at higher levels are neglected. **C)** Each four-spin model is reduced to a two-spin model with coupling $J'_k$ through the decimation procedure described in Section 2.1 (red area with solid boundaries). **D)** Pairs of two-spin systems are coupled so as to obtain a four-spin system with couplings $J_{k+1}$, by means of the rescaling procedure described in Section 2.2 (blue area with dashed boundaries). **E)** Couplings at the third hierarchical level and higher, which have been neglected in **B**, are reintroduced, and they couple the four-spin systems of **D**. As a result, we obtain a new infinite HEA with the same form as **A**, in which the short-wavelength degrees of freedom in **A** have been integrated out [1].

the energy gap $H[S_2] - H[S_1]$ of the rescaled model. In the presence of these correlations, the probability distribution $p_{k+1}(J)$ is not given by a product of single-coupling distributions as per Eq. (S50)—see Eq. (S67). While these correlations are expected to be irrelevant in the mean-field region of the model, they may become important [5, 6] close to the lower critical dimension.

## S8  Interpretation of the rescaling equations

In Eqs. (11) and (12) and in what follows, it is important to clarify the meaning of the suffix

$$\varsigma = \varsigma_{\text{low}}, \ T = 0. \tag{S53}$$

In this regard, we recall that the coupling distribution $p'_k$ is the result of multiple RG steps, each of which is performed with a given value of the parameters $\varsigma$ and $T$; $p'_k$ thus depends on these parameters. However, as we discussed in points 1 and 2 of Section 2.2, the parameter setting (S53) concerns the decimation procedure $\mathcal{M}_{k+1} \to \mathcal{M}'_{k+1}$ only, not the proceeding RG steps. As a result, the subscript (S53) is meant to be applied, for instance, to $H^{k+1}$ and $H'^{k+1}$, but not to $H'_k$ nor $p'_k$.

## S9  Zero-temperature limit

In what follows, we will work out Eq. (S55), i.e., the zero-temperature limit of Eq. (S12). Given that here we consider the decimation of Eq. (S12) as part of the $k+1$th RG step, we set $H[S] \to H^{k+1}[S]$.

By expanding the right-hand side of Eq. (S12) for large $\beta$, we obtain

$$\begin{aligned}
\mathcal{J}_\varsigma(J, T) &= \frac{1}{2\beta} \log \frac{2 + \mathcal{O}(e^{-\beta(\epsilon_2 - \epsilon_1)})}{e^{-\beta(\epsilon_2 - \epsilon_1)}(1 - \Omega[S_2]) + \mathcal{O}(e^{-\beta(\epsilon_3 - \epsilon_1)})} \\
&= \frac{1}{2\beta} \left[ \log 2 + \mathcal{O}(e^{-\beta(\epsilon_2 - \epsilon_1)}) + \beta(\epsilon_2 - \epsilon_1) - \log(1 - \Omega[S_2]) + \mathcal{O}(e^{-\beta(\epsilon_3 - \epsilon_1)}) \right] \\
&\xrightarrow{T \to 0} \frac{\epsilon_2 - \epsilon_1}{2},
\end{aligned} \tag{S54}$$

In the first line of Eq. (S54), we used Eq. (8) with the + sign in the numerator. In the denominator, we considered $\Omega[S_2]$. In general, it is easy to prove that $\Omega[S_p] \neq 1$ for $p > 1$. In fact, if $\Omega[S_p] = 1$ for $p > 1$, according to Eq. (6), we would have the following possibilities:

- $\Phi_L[S_{pL}] = \Phi_R[S_{pR}] = 1$. According to Eq. (8), this case would imply that $S_p = \sigma$, and it is thus ruled out because $p > 1$.

- $\Phi_L[S_{pL}] = \Phi_R[S_{pR}] = -1$. According to Eq. (8), this case would imply that $S_p = -\sigma$, which does not correspond to any of the possible values of $S_p$, see Section 2.1: as a result, this case is ruled out.

In the second line of Eq. (S54), we thus expanded the arguments of the logarithms by taking into account the relation $\Omega[S_2] \neq 1$, and in the last time we took the limit $T \to 0$ and, by using Eq. (S14), we obtain Eq. (S55).

## S10  High-temperature-phase constraint

Here, we will show that the constraint (12) for the high-temperature phase at the lower critical dimension is equivalent to the condition (S64).

We first write the the zero-temperature limit of Eq. (S12), which reads

$$\mathcal{J}_\varsigma(J, 0) = \frac{H^{k+1}[S_2] - H^{k+1}[S_1]}{2}, \tag{S55}$$

see Section S9. We then rewrite the average energy gap of $\mathcal{M}'_{k+1}$ as follows:

$$\mathbb{E}[H'^{k+1}[S'_2] - H'^{k+1}[S'_1]]\big|_{\varsigma=\varsigma_{\text{low}}, T=0} = 2\int dJ' \, [p'_{k+1}(J')]\big|_{\varsigma=\varsigma_{\text{low}}, T=0} |J'|$$

$$= 2\int dJ \, p_{k+1}(J) \mathcal{J}_{\varsigma_{\text{low}}}(J, 0) \tag{S56}$$

$$= \int dJ \, p_{k+1}(J) \, [H^{k+1}[S_2] - H^{k+1}[S_1]]\big|_{\varsigma=\varsigma_{\text{low}}},$$

where in the second line we substituted Eq. (10) and integrated over $J'$, and in the last line we used Eq. (S55).

Let us now proceed and prove the equivalence between Eq. (12) and Eq. (S64). Given that $\Omega[\sigma] = 1$, $\Omega[S_{\uparrow\uparrow\downarrow\downarrow}] = -1$, and that $\Omega$ vanishes on the other excited states, we rewrite Eq. (S12) as

$$\mathcal{J}_{\varsigma}(J, T) = \frac{1}{2\beta} \log \frac{2 + \sum_{p>1} e^{-\beta(\epsilon_p - \epsilon_1)} - e^{-\beta(\epsilon_{\uparrow\downarrow} - \epsilon_1)}}{\sum_{p>1} e^{-\beta(\epsilon_p - \epsilon_1)} + e^{-\beta(\epsilon_{\uparrow\downarrow} - \epsilon_1)}}, \tag{S57}$$

where we have set

$$\epsilon_{\uparrow\downarrow} \equiv H^{k+1}[S_{\uparrow\uparrow\downarrow\downarrow}], \tag{S58}$$

and in the numerator we have added and subtracted $e^{-\beta(\epsilon_{\uparrow\downarrow} - \epsilon_1)}$.

Setting

$$\mathcal{E} \equiv \{J \mid S_{\uparrow\uparrow\downarrow\downarrow} = S_2\}, \tag{S59}$$

the two following cases may occur depending on the sample $J$:

- $J \in \mathcal{E}$: By deriving Eq. (S57), we obtain

$$\frac{\partial \mathcal{J}_{\varsigma}}{\partial T} = \frac{1}{2} \log[1 + \mathcal{O}(e^{-\beta(\epsilon_3 - \epsilon_2)})] - \frac{\beta}{4}[(\epsilon_3 - \epsilon_2)e^{-\beta(\epsilon_3 - \epsilon_2)} + \mathcal{O}(e^{-2\beta(\epsilon_3 - \epsilon_2)}) + \mathcal{O}(e^{-\beta(\epsilon_3 - \epsilon_1)})$$
$$+ \mathcal{O}(e^{-\beta[(\epsilon_3 - \epsilon_2) + (\epsilon_3 - \epsilon_1)]}) + \mathcal{O}(e^{-\beta(\epsilon_4 - \epsilon_2)})] \tag{S60}$$
$$= -\frac{\beta}{4}(\epsilon_3 - \epsilon_2)e^{-\beta(\epsilon_3 - \epsilon_2)} + \cdots,$$

where, in the first line, the first and second term are obtained by deriving $1/\beta$ and the logarithm in Eq. (S57), respectively. Finally, in the second line and in what follows, we denote by $\cdots$ terms that are subleading with respect to the preceding term for small $T$.

- $J \notin \mathcal{E}$: Proceeding along the same lines as in Eq. (S60), we have

$$\frac{\partial \mathcal{J}_{\varsigma}}{\partial T} = \frac{1}{2}[\log 2 + \mathcal{O}(e^{-\beta(\epsilon_2 - \epsilon_1)}) + \mathcal{O}(e^{-\beta(\epsilon_3 - \epsilon_2)})] - \frac{\beta}{4}[\mathcal{O}(e^{-\beta(\epsilon_2 - \epsilon_1)}) + \mathcal{O}(e^{-\beta(\epsilon_3 - \epsilon_2)})]$$
$$= \frac{\log 2}{2} + \cdots. \tag{S61}$$

Combining Eqs. (S56), (S60) and (S61), we obtain

$$\frac{\partial}{\partial T}\mathbb{E}[H'^{k+1}[S'_2] - H'^{k+1}[S'_1]] =$$
$$2\int dJ \, p_{k+1}(J) \frac{\partial \mathcal{J}_{\varsigma}(J, T)}{\partial T} =$$
$$2\int dJ \, p_{k+1}(J) \left\{ \mathbb{I}(S_2 = S_{\uparrow\uparrow\downarrow\downarrow})\left[-\frac{\beta}{4}(\epsilon_3 - \epsilon_2)e^{-\beta(\epsilon_3 - \epsilon_2)} + \cdots\right] + \mathbb{I}(S_2 \neq S_{\uparrow\uparrow\downarrow\downarrow})\left[\frac{\log 2}{2} + \cdots\right]\right\} = \tag{S62}$$
$$2\left\{-\mathbb{E}\left[\mathbb{I}(S_2 = S_{\uparrow\uparrow\downarrow\downarrow})\frac{\beta}{4}(\epsilon_3 - \epsilon_2)e^{-\beta(\epsilon_3 - \epsilon_2)}\right] + \cdots + \mathbb{E}[\mathbb{I}(S_2 \neq S_{\uparrow\uparrow\downarrow\downarrow})]\frac{\log 2}{2} + \cdots\right\}.$$

Given Eq. (S14), and given that the excited states are in order of increasing energy $H[S_1] < H[S_2] < \cdots < H[S_8]$, we have $\epsilon_3 > \epsilon_2$, implying that the first term in the last line of Eq. (S62) vanishes for $T \to 0$. On the other hand, the third term is non-negative, and independent of $T$. As a result, the last line in Eq. (S62) shows that the condition $\frac{\partial}{\partial T}\mathbb{E}[H'^{k+1}[S'_2] - H'^{k+1}[S'_1]]\big|_{\varsigma=\varsigma_{\text{low}}, T=0} \leq 0$ is equivalent to Eq. (S64).

## S11 Solution of the minimization problem

In order to solve the optimization problem above, in what follows we rewrite the constraints (11) and (12) as functions of the distribution $p_{k+1}$.

For the sake of clarity, in what follows we will omit the subscript 'low' in $\varsigma$, $H^{k+1}[\boldsymbol{S}_2] - H^{k+1}[\boldsymbol{S}_1]$, $\boldsymbol{S}_2$ and $\boldsymbol{S}_{\uparrow\uparrow\downarrow\downarrow}$, implying that these quantities are evaluated at the lower critical dimension. As for Eq. (11), we rewrite it as

$$\int d\boldsymbol{J}\, p_{k+1}(\boldsymbol{J}) \left( \mathcal{J}_\varsigma(\boldsymbol{J}, 0) - \mathbb{E}[|J'|] \right) = 0, \tag{S63}$$

where we rewrote the LHS as per Eq. (S56), used Eq. (S47) in the right-hand side, and moved $\mathbb{E}[|J'|]$ in the integral by using the normalization condition in Eq. (13). As for Eq. (12), in Section S10 we have shown that it can be rewritten as

$$\left[ \int d\boldsymbol{J}\, p_{k+1}(\boldsymbol{J}) \mathbb{I}(\boldsymbol{S}_2 \neq \boldsymbol{S}_{\uparrow\uparrow\downarrow\downarrow}) \right]\bigg|_{\varsigma = \varsigma_{\text{low}}} = 0, \tag{S64}$$

where $\boldsymbol{S}_{\uparrow\uparrow\downarrow\downarrow}$ is the excited state of type i) of $H^{k+1}$ obtained from the GS by flipping the right half of spins, see Section 2.1 and panel $i$ of Fig. 1A.

From the mathematical standpoint, the equality condition (S64), with which we will replace the inequality (12), substantially simplifies the minimization problem (13), whose solution would be particularly involved if inequality constraints were present [7, 8]. From the physical standpoint, the condition (S64) means that $p_{k+1}(\boldsymbol{J})$ is nonzero only for the samples $\boldsymbol{J}$ for which the first excited state of $\mathcal{M}_{k+1}$ is equal to $\boldsymbol{S}_{\uparrow\uparrow\downarrow\downarrow}$, and the set of these couplings will be denoted by $\mathcal{E}$, defined by Eq. (S59). Given that, in the decimated model $\mathcal{M}'_k$, the first excited state is also obtained by flipping the right spin in the GS, the constraint (S64) imposes that the structure of the first excited state is preserved through the rescaling procedure.

We solve the constrained optimization problem above given by Eqs. (13), (S63) and (S64) and by the normalization condition in Eq. (13), by introducing the Lagrange function

$$\begin{aligned}L[p_{k+1}, \eta, \mu, \psi] \equiv & D[p_{k+1} || p^*_{k+1}] - \eta \int d\boldsymbol{J}\, p_{k+1}(\boldsymbol{J}) \left( \mathcal{J}_\varsigma(\boldsymbol{J}, 0) - \mathbb{E}[|J'|] \right) \\ & - \mu \int d\boldsymbol{J}\, p_{k+1}(\boldsymbol{J}) \mathbb{I}(\boldsymbol{S}_2 \neq \boldsymbol{S}_{\uparrow\uparrow\downarrow\downarrow}) - \psi \left( \int d\boldsymbol{J}\, p_{k+1}(\boldsymbol{J}) - 1 \right),\end{aligned} \tag{S65}$$

where $\eta$, $\mu$ and $\psi$ are the Lagrange multipliers corresponding to constraints (S63), (S64), and to the normalization condition in Eq. (13), respectively. By differentiating Eq. (S65) with respect to $p_{k+1}$ and $\eta$ we obtain, respectively, the stationarity conditions

$$p_{k+1}(\boldsymbol{J}) = \frac{1}{\mathcal{Z}} p^*_{k+1}(\boldsymbol{J}) \exp\left[ \eta \mathcal{J}_\varsigma(\boldsymbol{J}, 0) + \mu \mathbb{I}(\boldsymbol{S}_2 \neq \boldsymbol{S}_{\uparrow\uparrow\downarrow\downarrow}) \right], \tag{S66}$$

and Eq. (S63). Finally, by differentiating Eq. (S65) with respect to $\mu$ and $\psi$ we find Eq. (S64) and the normalization condition in Eq. (13), respectively. Combining Eqs. (S64) and (S66), we obtain the solution

$$p_{k+1}(\boldsymbol{J}) = \begin{cases} \frac{1}{\mathcal{Z}} p^*_{k+1}(\boldsymbol{J}) \exp\left[ \eta \mathcal{J}_\varsigma(\boldsymbol{J}, 0) \right] & \text{if } \boldsymbol{J} \in \mathcal{E}, \\ 0 & \text{otherwise,} \end{cases} \tag{S67}$$

in which the multiplier $\mu$ is incorporated into the integration constraint, and and the Lagrange multiplier $\eta$ is determined by Eq. (S63) whose solution, if any, is unique—see Section S12. Finally, the normalization condition in (13) yields the normalization constant $\mathcal{Z}$, see Eq. (S84).

## S12 Uniqueness of the solution for the Lagrange multiplier

By differentiating the LHS of Eq. (S63) with respect to $\eta$, we obtain

$$\begin{aligned}\frac{\partial}{\partial \eta} \int d\boldsymbol{J}\, p_{k+1}(\boldsymbol{J}) \left( \mathcal{J}_\varsigma(\boldsymbol{J}, 0) - \mathbb{E}[|J'|] \right) & = \frac{\partial}{\partial \eta} \left[ \frac{1}{\mathcal{Z}} \int_{\boldsymbol{S}_2 = \boldsymbol{S}_{\updownarrow}} d\boldsymbol{J}\, p^*_{k+1}(\boldsymbol{J}) e^{\eta \mathcal{J}_\varsigma(\boldsymbol{J}, 0)} \mathcal{J}_\varsigma(\boldsymbol{J}, 0) \right] \\ & = \mathbb{E}[[\mathcal{J}_\varsigma(\boldsymbol{J}, 0)]^2] - [\mathbb{E}[\mathcal{J}_\varsigma(\boldsymbol{J}, 0)]]^2 \\ & \geq 0.\end{aligned} \tag{S68}$$

Given that Eq. (S68) is a non-decreasing function of $\eta$, then if Eq. (S63) has a solution for $\eta$, this solution is unique.

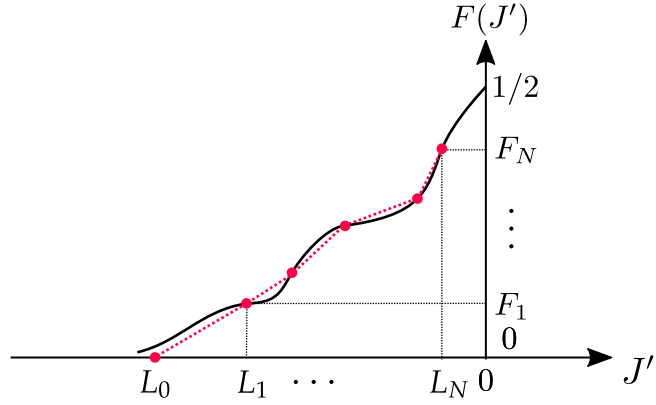

**Figure S3: Quantiles of the coupling distribution.** Cumulative distribution function (CDF) $F(J')$ as a function of the spin-spin coupling $J'$ (black curve). Because the probability density function (PDF) associated with $F$ is even, only the interval $0 < F < 1/2$ is shown. The quantiles $L_i$ are obtained as the inverse of the CDF at $F = F_i$ (red dots), with $i = 1, \cdots, N$ and, for large $N$, the full set of quantiles specifies the shape of the CDF. The CDF is approximated by a piecewise-linear function (red dashed lines), where the quantile $L_0$ denotes the boundary of the support of the PDF.

## S13 Discretization

We denote by $F(J')$ the cumulative distribution function (CDF) of $p'$

$$F(J') \equiv \int dL\, p'(L)\, \mathbb{I}(L < J'), \tag{S69}$$

where in what follows we omit the subscript $k$ for the sake of clarity. Because $p'(J')$ is even throughout the RG transformation, we consider only half of the image of the CDF, i.e., the interval $[0, 1/2]$, and partition it into $N+1$ intervals

$$0 < F_1 < F_2 < \cdots < F_N < 1/2. \tag{S70}$$

As shown in Fig. S3, we then parametrize $p'$ in terms of its quantiles [9]

$$L_1 < L_2 < \cdots < L_N < 0, \tag{S71}$$

which are defined as the solutions of

$$F(L_i) = F_i, \quad 1 \leq i \leq N. \tag{S72}$$

We then write the density $p'$ as a piecewise-constant function, by expressing it in terms of its quantiles $L$ as follows:

$$p'(J') = \frac{F_2}{L_2 - L_0}\mathbb{I}(L_0 < J' < L_2) + \sum_{i=2}^{2N-2} \frac{F_{i+1} - F_i}{L_{i+1} - L_i}\mathbb{I}(L_i < J' < L_{i+1}) \\ + \frac{1 - F_{2N-1}}{L_{2N+1} - L_{2N-1}}\mathbb{I}(L_{2N-1} < J' < L_{2N+1}). \tag{S73}$$

Given that $p'$ is an even function, in Eq. (S73) we have set, for $1 \leq i \leq N$,

$$F_{2N+1-i} \equiv 1 - F_i, \quad L_{2N+1-i} \equiv -L_i \tag{S74}$$

and we have made an approximation, which is exact for large $N$, by introducing the boundary quantiles

$$L_0 \equiv \frac{F_1 L_2 - F_2 L_1}{F_1 - F_2}, \tag{S75}$$

$$L_{2N+1} \equiv \frac{(1 - F_{2N-1})L_{2N} - (1 - F_{2N})L_{2N-1}}{F_{2N} - F_{2N-1}} = -L_0. \tag{S76}$$

which are defined by the condition that $p'$ is identically zero outside the interval $[L_0, L_{2N+1}]$, see Fig. S3 and Section S13.1 for details.

### S13.1 Boundary quantiles

In what follows, we will determine the value of the boundary quantiles $L_0$, $L_{2N+1}$. As for $L_0$, we approximate $F(L)$ for $L < L_1$ as a linear function with slope equal to the slope of $F(L)$ at $L = L_1$, i.e., $(F_2 - F_1)/(L_2 - L_1)$. As a result, for $L < L_1$

$$F(L) = F_1 + \frac{F_2 - F_1}{L_2 - L_1}(L - L_1). \tag{S77}$$

The boundary quantile $L_0$ is defined as the value of $L$ at which the right-hand side of Eq. (S77) vanishes, and it is given by Eq. (S75). Proceeding along the same lines, we obtain Eq. (S76) for $L_{2N+1}$, which is defined as the value of $L$ at which $F(L)$ equals unity.

### S13.2 Quantiles of the decimated distribution

Here, we will derive the system of equations that determines the quantiles of the decimated distribution $p'_{k+1}$. For $1 \leq i \leq N$, Eq. (10) implies

$$\begin{aligned}
\int dJ' \, p'_{k+1}(J') \mathbb{I}(J' < L_i^{k+1}) &= \int dJ' dJ \, p_{k+1}(J) \frac{1}{2} \left[ \delta(J' - \mathcal{J}_\varsigma(J,T)) + \delta(J' + \mathcal{J}_\varsigma(J,T)) \right] \mathbb{I}(J' < L_i^{k+1}) \\
&= \frac{1}{2} \int dJ \, p_{k+1}(J) \mathbb{I}(-\mathcal{J}_\varsigma(J,T) < L_i^{k+1}) \\
&= F_i.
\end{aligned} \tag{S78}$$

In order to obtain the first line of Eq. (S78), we multiplied both sides of Eq. (10) by the indicator function $\mathbb{I}(J' < L_i^{k+1})$. In the second line we integrated with respect to $J'$ and observed that, given Eq. (S46) and the fact that the indicator function imposes that $J' < L_i^{k+1} \leq 0$, only the second delta function in the first line contributes to the integral. Finally, in the third line we used Eqs. (S69) and (S72). The second and third line of Eq. (S78) yield the desired system of equations.

## S14 Fixed-distribution structure

In this Section we will analyze the fixed distributions of the RG transformation. By iterating it at high and low temperatures, the transformation flows to a high- and a low-temperature fixed distribution in which the width of $p'_k$ goes to zero and infinity, respectively [10], see Section S17 for details.

To characterize the stability of any fixed distribution $p'_*$, we consider the Jacobian $\mathcal{K}_{ij} \equiv \partial L_i^{k+1}/\partial L_j^k$ evaluated at $p'_*$, where the explicit expression is given in Eq. (S101), see Sections S18 and S19 for details. We write $\mathcal{K}$ in terms of its eigenvalues $\lambda_n$ and its respective left and right eigenvectors $v_L^n$ and $v_R^n$, respectively, as $\mathcal{K}_{ij} = \sum_n \lambda_n v_{R\,i}^n v_{L\,j}^n$. If there is at least one $\lambda_n$ with $|\lambda_n| > 1$, then the distribution under consideration is unstable, otherwise it is stable.

The Jacobian at the high-temperature fixed distribution is shown in Fig. S4, for $\varsigma = 0.8$. Such graph indicates that, for large enough $N$, the Jacobian tends to a smooth function of its arguments, thus validating the overall discretization procedure of Section 3. In Fig. S4B we show the eigenvalues of $\mathcal{K}$: All eigenvalues have norm smaller than unity, implying that the high-temperature fixed distribution is stable.

In the low-temperature regime there is no finite fixed distribution as in the high-temperature case: The RG transformation tends to a distribution with infinite width for $k \to \infty$, along the lines of the RG flow at zero temperature in ferromagnetic systems [1]. In order to show this, we iterated the RG transformation at $T = 0$, and studied the stability of the coupling distribution obtained after a few iterations. The Jacobian evaluated at such distribution is shown in Fig. S5 for $\varsigma = 0.8$, and it presents one eigenvalue with norm larger than unity. Distributions along the zero-temperature RG flow are thus unstable.

## S15 Renormalization-group transformation for scaled distributions

In this Section we will show how to perform the RG transformation for the rescaled coupling distribution $\bar{p}'_k$.

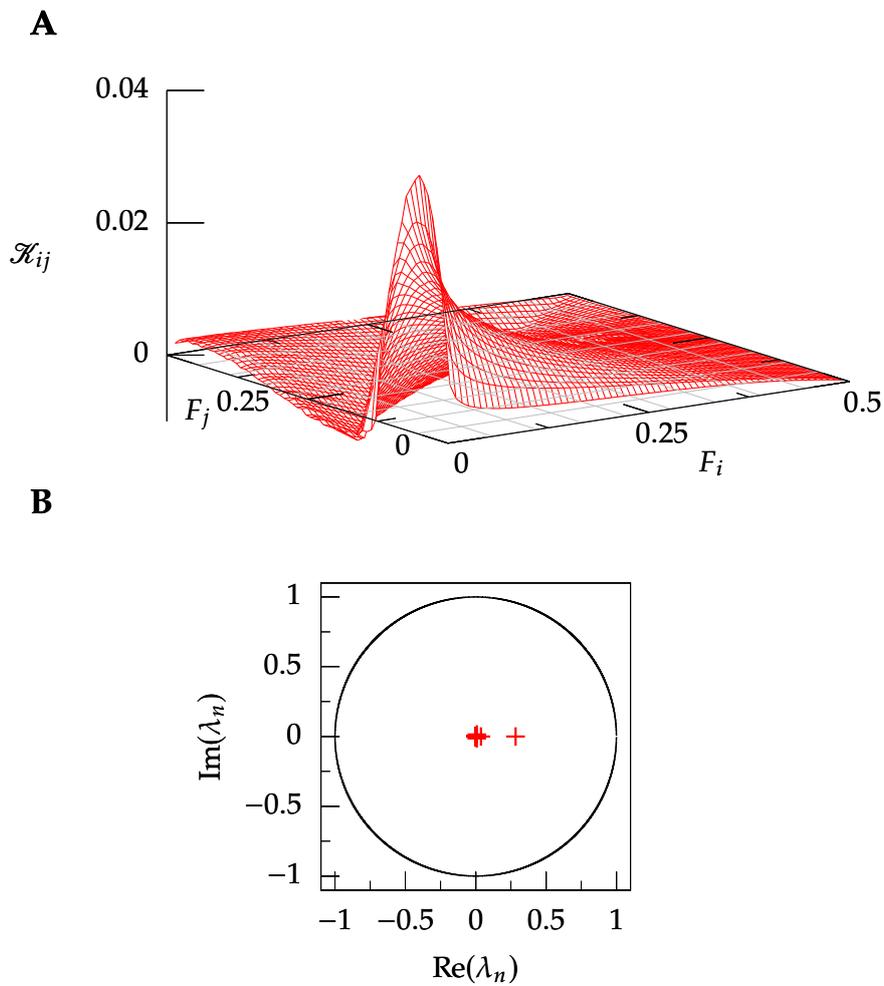

**Figure S4: Linearization of the renormalization-group transformation at the high-temperature fixed distribution.** **A)** Jacobian $\mathcal{K}_{ij} = \partial L_i^{k+1}/\partial L_j^k$ evaluated ad the high-temperature fixed distribution, cf. Fig. S7, where $\boldsymbol{L}^k$ are the quantiles of the spin-coupling distribution, as a function of the values $F_i$, $F_j$ of the relative cumulative distribution function, for $\varsigma = 0.8$, $N = 2^6$, $S = 2^8$, $M = 2^{22}$ and $i, j = 1, \cdots, N$, see Section S16. Here the cumulative distribution function values $F_i$, $F_j$, serve as labels for the Jacobian rows and columns, respectively. **B)** Eigenvalues of the Jacobian in **A** in the complex plane (red), and unit disk centered at the origin (black curve).

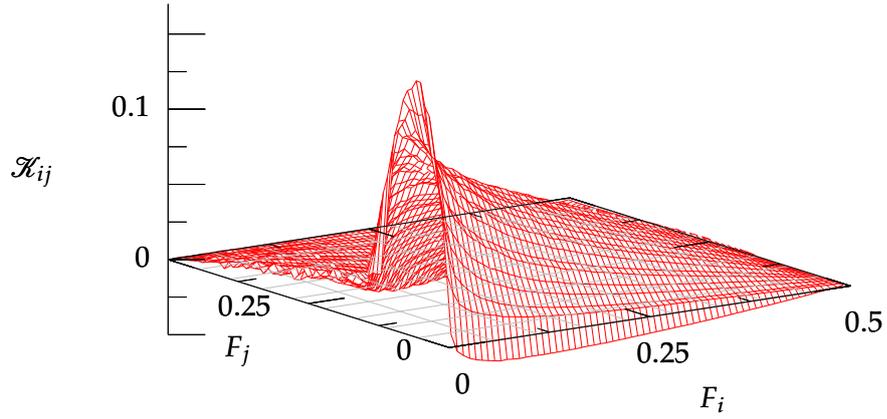
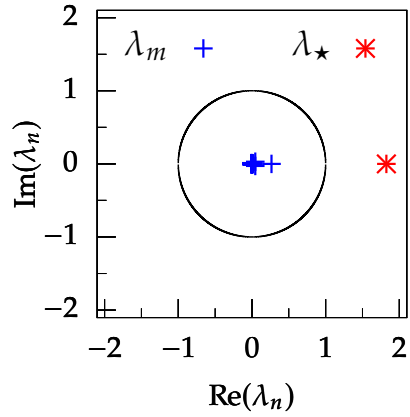
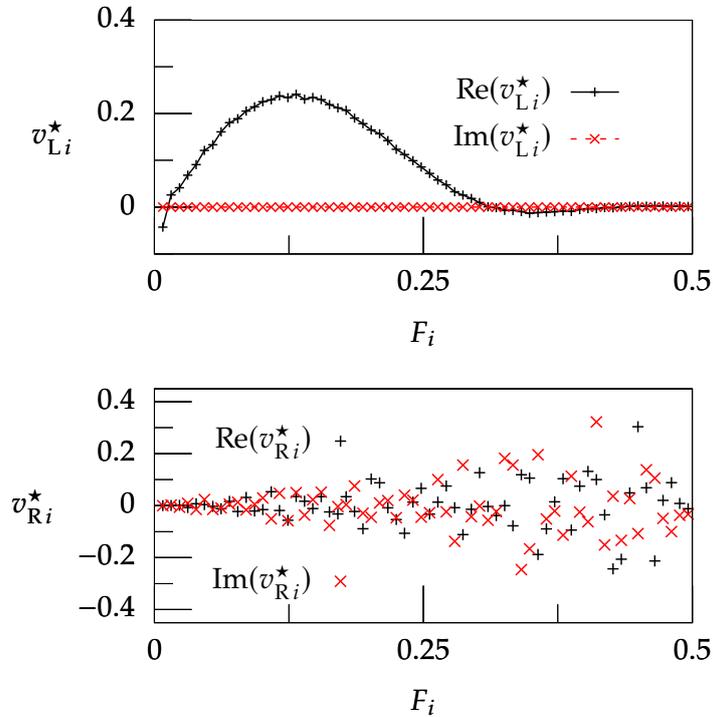

**Figure S5: Linearization of the renormalization-group transformation at zero temperature**. The Jacobian is evaluated at the coupling distribution obtained by iterating a finite number of times the renormalization-group transformation at zero temperature. **A**) and **B**): Same as Fig. S4, for $\varsigma = 0.8$, $N = 2^6$, $S = 2^8$ and $Q = 2^{18}$. In **B**, the eigenvalues $\lambda_n$ with norm smaller than unity are shown in blue, and the eigenvalue $\lambda_\star$ with norm larger than unity in red. **C**) Eigenvectors corresponding to $\lambda_\star$: In the top and bottom panel we show the components $v^\star_{L\,i}$, $v^\star_{R\,i}$ of the left and right eigenvector, respectively, as functions of the value $F_i$ of the cumulative distribution function of the spin-coupling distribution. The real and imaginary part of the components are shown in black and red, respectively. Such eigenvectors characterize the instability related to $\lambda_\star$.

## S15.1 Rescaling

Let us consider the procedure set out in Section 2.2, where a model with coupling distribution $p'_k$ is rescaled so as to obtain a model with coupling distribution $p_{k+1}$. For the sake of clarity, we will drop the subscripts $k$ and $k+1$.

We introduce a coupling distribution obtained by scaling $p'$ by a factor $s > 0$:

$$\overline{p}'(J') \equiv sp'(sJ'). \tag{S79}$$

In what follows, we will consider two rescaling procedures: First, the rescaling procedure of Section 2.2 applied to the original distribution $p'$, which yields the Lagrange multiplier $\eta$ and $p(J)$. Second, the rescaling procedure of Section 2.2 applied to $\overline{p}'$, whose quantities will be denoted by an overscore, e.g., $\overline{p}(J)$, $\overline{\eta}$. We will denote the two rescaling procedures above by $\mathcal{R}$ and $\overline{\mathcal{R}}$, respectively:

$$p' \xrightarrow{\mathcal{R}} p, \quad \overline{p}' \xrightarrow{\overline{\mathcal{R}}} \overline{p}. \tag{S80}$$

We will show that $\eta$ is related to $\overline{\eta}$ by

$$\eta = \overline{\eta}/s. \tag{S81}$$

To achieve this, we consider $\overline{\mathcal{R}}$ and $\overline{\eta}$. By construction, $\overline{\eta}$ satisfies Eq. (S63) for $\overline{\mathcal{R}}$:

$$\int dJ\, \overline{p}(J)\left(\mathcal{J}_{\varsigma_{\text{low}}}(J, 0) - \overline{\mathbb{E}}[|J'|]\right) = 0, \tag{S82}$$

where we rewrote the energy gap as per Eq. (S55).

Given the solution $\overline{\eta}$ above, in what follows we will show that, if $\eta$ is related to $\overline{\eta}$ as per Eq. (S81), then $\eta$ satisfies the stationarity condition (S63) for $\mathcal{R}$. We rewrite the normalization factor $\overline{\mathcal{Z}}$ as

$$\begin{aligned}
\overline{\mathcal{Z}} &= \int_{\overline{\mathcal{E}}} dJ \left[\prod_{i<j}(sp'(sJ_{ij}))\right] e^{s\eta \mathcal{J}_{\varsigma_{\text{low}}}(J,0)} \\
&= \int_{\mathcal{E}} dJ\, p^*(J)\, e^{\eta \mathcal{J}_{\varsigma_{\text{low}}}(J,0)} \\
&= \mathcal{Z},
\end{aligned} \tag{S83}$$

where in the first line we used the relations

$$\mathcal{Z} = \int_{\mathcal{E}} dJ\, p^*_{k+1}(J)\, \exp\left[\eta \mathcal{J}_\varsigma(J,0)\right]. \tag{S84}$$

and the definition in (S50), in which we substituted Eqs. (S79) and (S81), and we rewrote the energy gap in terms of $\mathcal{J}_{\varsigma_{\text{low}}}$ by using Eq. (S55). In the second line we changed integration variables setting $sJ_{ij} \to J_{ij}$ and we observed that this change of variables does not alter the condition $J \in \mathcal{E}$ on the integration domain: In fact, if all couplings $J$ are scaled by the same factor $s$, the energy levels are scaled accordingly, and the corresponding excited states stay unchanged. Also, in the second line we used the fact that, for $T = 0$, $\mathcal{J}_\varsigma$ is a homogeneous function of the couplings:

$$s\mathcal{J}_\varsigma(J, 0) = \mathcal{J}_\varsigma(sJ, 0), \tag{S85}$$

which follows from the definition (S55), the linearity of the Hamiltonian (1) with respect to $J$, and the fact that the states $S_1, \cdots, S_8$ are invariant under a rescaling of all $J_{ij}$s by $s$. Finally, in the last line we used the definition (S84) of $\mathcal{Z}$.

By rewriting explicitly Eq. (S67) as

$$\overline{p}(J) = \begin{cases} \frac{1}{\overline{\mathcal{Z}}} \prod_{i<j} \overline{p}'(J_{ij}) \exp\left(\overline{\eta} \mathcal{J}_{\varsigma_{\text{low}}}(J, 0)\right) & \text{if } J \in \mathcal{E}, \\ 0 & \text{otherwise,} \end{cases} \tag{S86}$$

combining Eqs. (S79), (S81), (S83), (S85) and (S86) and proceeding along the same lines, we recover the definition (S67) for $p(J)$, and obtain

$$\overline{p}(J) = s^6 p(sJ), \tag{S87}$$

where the sixth power of $s$ comes from the fact that the four-spin model has six couplings $J_{ij}$. As a result, Eq. (S63), which defines the solution $\eta$, is satisfied:

$$\int dJ\, p(J) \left(\mathcal{J}_{\varsigma_{\text{low}}}(J,0) - \mathbb{E}[|J'|]\right) = s^6 \int dJ\, p(sJ)(\mathcal{J}_{\varsigma_{\text{low}}}(sJ,0) - \mathbb{E}[|J'|]) \\ = \int dJ\, \overline{p}(J)(s\mathcal{J}_{\varsigma_{\text{low}}}(J,0) - s\overline{\mathbb{E}}[|J'|]) \\ = 0, \tag{S88}$$

where in the first line we changed integration variables setting $J_{ij}/s \to J_{ij}$, in the second line we used Eq. (S87) and the homogeneity relation (S85), and observed that Eq. (S79) implies

$$\mathbb{E}[|J'|] = s\,\overline{\mathbb{E}}[|J'|]. \tag{S89}$$

In the last line, we observed that the right-hand side vanishes because $\overline{\eta}$ satisfies Eq. (S82). Equation (S88) thus shows that, if Eq. (S81) holds, then $\eta$ satisfies the stationarity condition (S63) for $\mathcal{R}$. Given that Section S12 shows that Eq. (S63) admits a unique solution, we obtain that $\eta$ is related to $\overline{\eta}$ by Eq. (S81).

Finally, given Eq. (S89), we may choose the scaling factor as

$$s = \mathbb{E}[|J'|] \tag{S90}$$

so as to make the expectation value $\overline{\mathbb{E}}[|J'|]$ in the modified rescaling procedure equal to unity.

### S15.2 Decimation

Here, we will show how to perform the decimation procedure by using the scaled distribution of Section S15.1, and obtain the distribution $p'$ of the decimated model in terms of its quantiles.

We restore the subscript $k+1$ and substitute Eq. (S87) in the last two lines of Eq. (S78):

$$\int dJ\, \overline{p}_{k+1}(J)\frac{1}{2}\mathbb{I}(-\mathcal{J}_\varsigma(sJ,T) < L_i^{k+1}) = F_i. \tag{S91}$$

Equation (S91) is the analog of Eq. (S78): it involves only the modified distribution $\overline{p}_{k+1}$, and can be solved for the quantiles $L^{k+1}$ of the decimated distribution $p'_{k+1}$.

## S16 Numerical solution with stochastic-approximation methods

In this Section, we will briefly discuss the numerical procedure used to solve numerically the RG equations.

To illustrate this procedure, we will focus on Eq. (S63) and show how, in the rescaling procedure, it can be solved for $\eta$ by means of stochastic-approximation methods. The solution of Eq. (S78) for $L^{k+1}$ in the decimation procedure can be worked out along the same lines [11].

Following the approach by Robbins and Monro [12], we observe that Eq. (S63) can be rewritten as

$$\mathbb{E}_\eta[\mathfrak{J}] = \mathbb{E}[|J'|], \tag{S92}$$

where $\mathfrak{J}$ is a random variable distributed according to the probability density function (PDF)

$$\rho_\eta(\mathfrak{J}) \equiv \int dJ\, p_{k+1}(J)\delta\left(\mathcal{J}_\varsigma(J,0) - \mathfrak{J}\right), \tag{S93}$$

and we indicated the dependence on $\eta$ explicitly in Eqs. (S92) and (S93), where $\rho_\eta$ depends on $\eta$ through $p_{k+1}$, see Eq. (S67).

In what follows, we will shortly illustrate a stochastic procedure to solve Eq. (S92):

1. Given a tentative solution $\eta_n$ of Eq. (S92)

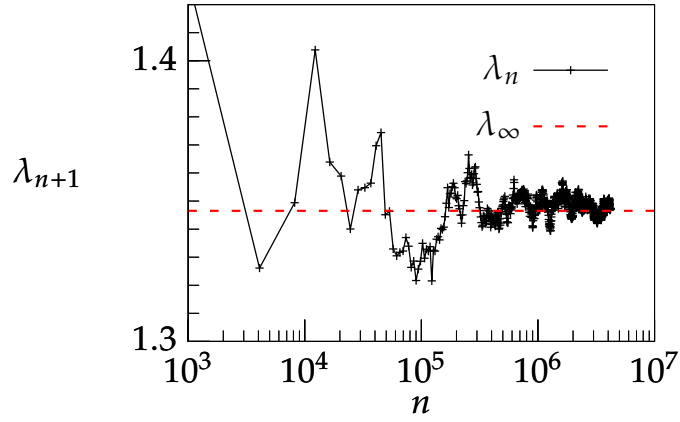

**Figure S6:** **Solution of the renormalization-group (RG) equations with stochastic-approximation methods.** Semi-logarithmic plot of the Lagrange multiplier $\eta_{n+1}$ as a function of the number $n$ of Robbins-Monro iterations, see Eq. (S97), for $\varsigma = 0.7$, $N = 2^6$ discretization bins, $S = 2^{12}$ samples and a total number of $M = 2^{22}$ iterations (black points). Here, the RG steps are iterated by seeking for the critical point, see Section 3, and the data shown is for one given RG step. The estimate of the solution $\eta_\infty$ is also shown (red dashed line).

2. Draw a random variable $\mathfrak{J}_n$ according to the PDF $\rho_{\eta_n}$ as follows [13]:

    (a) Draw $S$ samples
    $$J_1^*, \cdots, J_S^* \tag{S94}$$
    according to $p_{k+1}^*$, see Eq. (S67), where $S$ is a large enough integer.

    (b) Compute the weight $w_s$ of each sample given by the exponential factor and by the condition $J \in \mathcal{E}$ of Eq. (S67):
    $$w_s = \begin{cases} \exp\left[\eta \mathcal{J}_\varsigma(J_s, 0)\right] & \text{if } J \in \mathcal{E} \\ 0 & \text{otherwise} \end{cases}. \tag{S95}$$

    (c) Reweigh the population (S94) according to the weights (S95), and obtain a population
    $$J_1, \cdots, J_S \tag{S96}$$
    distributed according to $p_{k+1}$.

    (d) Randomly draw one element $J_s$ in the population (S96) and obtain $\mathfrak{J}_n \equiv \mathcal{J}_\varsigma(J_s, 0)$, which is the desired random sample drawn from the distribution $\rho_{\eta_n}$.

3. Obtain an updated value of the tentative solution as
    $$\eta_{n+1} = \eta_n + \frac{C}{n^\alpha}(\mathbb{E}[|J'|] - \mathfrak{J}_n), \tag{S97}$$
    where $C$ is a positive constant and $1/2 < \alpha < 1$.

It can be proved that, under some regularity conditions, the procedure (S97) converges to a value $\eta_\infty$, which satisfies Eq. (S63) [12], no matter what the initial value $\eta_1$. An example of the numerical solution of the RG equations with this method is given in Fig. S6.

## S17 Numerical results for the fixed distributions

The high- and low-temperature fixed distributions can be characterized by rewriting the RG transformation in terms of a scaled coupling distribution $\overline{p}_k'(J')$, where the scaling factor $s$ is chosen according to Eq. (S90), see Section S15: Unlike $p_k'$, for large $k$ the width of $\overline{p}_k'$ stays finite, and $\overline{p}_k'$ converges to a fixed distribution.

Figures S7 and S8 show the scaled high- and low-temperature fixed distributions, respectively, where in Figure S7 we used a discretization with $N = 2^6$ quantiles, $S = 2^{10}$ samples to represent $\overline{p}$, and $M = 2^{22}$ iterations

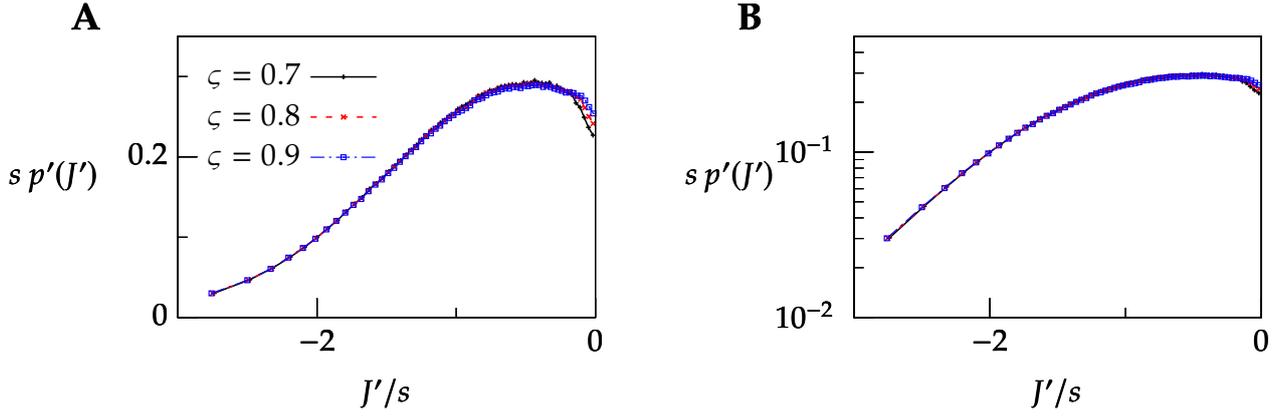

**Figure S7: High-temperature fixed distribution of spin couplings.** **A)** Fixed distribution $sp'(J')$ of the scaled spin-spin coupling $J'/s$ as a function of $J'/s$ for $\varsigma = 0.7$ (black), $\varsigma = 0.8$ (red) and $\varsigma = 0.9$ (blue), where the scaling factor is $s = \mathbb{E}[|J'|]$. The scaling factor $s$ has been chosen so as to keep the width of $p'$ finite, see Section S15. Given that $p'$ is an even function, only negative values of $J'$ are shown. **B)** Same as **A**, in semi-logarithmic scale

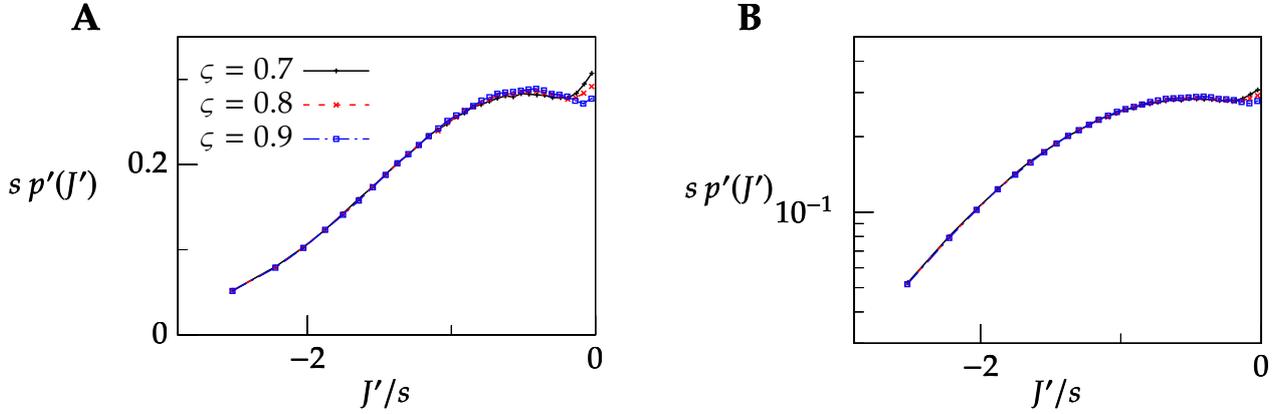

**Figure S8: Low-temperature fixed distribution of spin couplings.** Same as Fig. S7, for the low-temperature fixed distribution.

to solve the RG equations with stochastic-approximation methods [11, 12], while in Figure S8 we have $N = 2^5$, $S = 2^{12}$ and $M = 2^{22}$, see Sections S15 and S16. Finally, Fig. S9 shows the distribution of the dimensionless coupling $\beta_c J'$ [1] at the critical fixed distribution, where $N = 2^7$, $S = 2^{12}$, and we made $k = 2^{18}$ iterations for the solution of the RG equations at the critical point, see Eqs. (S63) and (S78) and Section 3.

## S18 Linearization of the renormalization-group transformation

In what follows, we will work out an analytical expression for $\mathcal{K}$. We derive the last two lines of Eq. (S78) as well as Eq. (S63), where we recall that $\varsigma = \varsigma_{\text{low}}$, with respect to $L_j^k$, and we obtain

$$\sum_{m<n} \mathbb{E}\left[ \frac{\partial \log p'_k(J_{mn})}{\partial L_j^k} \left( \mathcal{J}_{\varsigma_{\text{low}}}(J,0) - \mathbb{E}[|J'|] \right) \right] + \mathbb{E}\left[ \left( \mathcal{J}_{\varsigma_{\text{low}}}(J,0) - \mathbb{E}[|J'|] \right)^2 \right] \frac{\partial \eta}{\partial L_j^k} - \frac{\partial \mathbb{E}[|J'|]}{\partial L_j^k} = 0, \quad \text{(S98)}$$

$$\sum_{m<n} \mathbb{E}\left[ \frac{\partial \log p'_k(J_{mn})}{\partial L_j^k} \left[ \frac{1}{2}\mathbb{I}\left(-\mathcal{J}_\varsigma(J,T) < L_i^{k+1}\right) - F_i \right] \right]$$
$$+ \mathbb{E}\left[ \left( \mathcal{J}_{\varsigma_{\text{low}}}(J,0) - \mathbb{E}[|J'|] \right) \left[ \frac{1}{2}\mathbb{I}\left(-\mathcal{J}_\varsigma(J,T) < L_i^{k+1}\right) - F_i \right] \right] \frac{\partial \eta}{\partial L_j^k} + \frac{1}{2}\mathbb{E}\left[ \delta\left( \mathcal{J}_\varsigma(J,T) + L_i^{k+1} \right) \right] \mathcal{K}_{ij} = 0, \quad \text{(S99)}$$

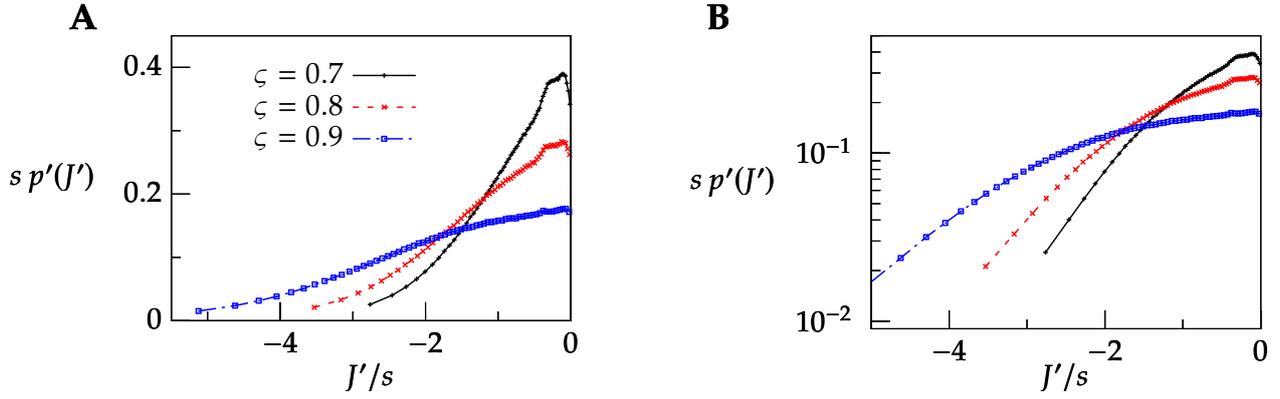

**Figure S9: Critical fixed distribution of spin couplings**. **A**) Fixed distribution $sp'(J')$ of the scaled spin-spin coupling $J'/s$ as a function of $J'/s$ for $\varsigma = 0.7$ (black), $\varsigma = 0.8$ (red) and $\varsigma = 0.9$ (blue), where the scaling factor is $s = 1/\beta_c$, see Section S15. Given that $p'$ is an even function, only negative values of $J'$ are shown. **B**) Same as **A**, in semi-logarithmic scale.

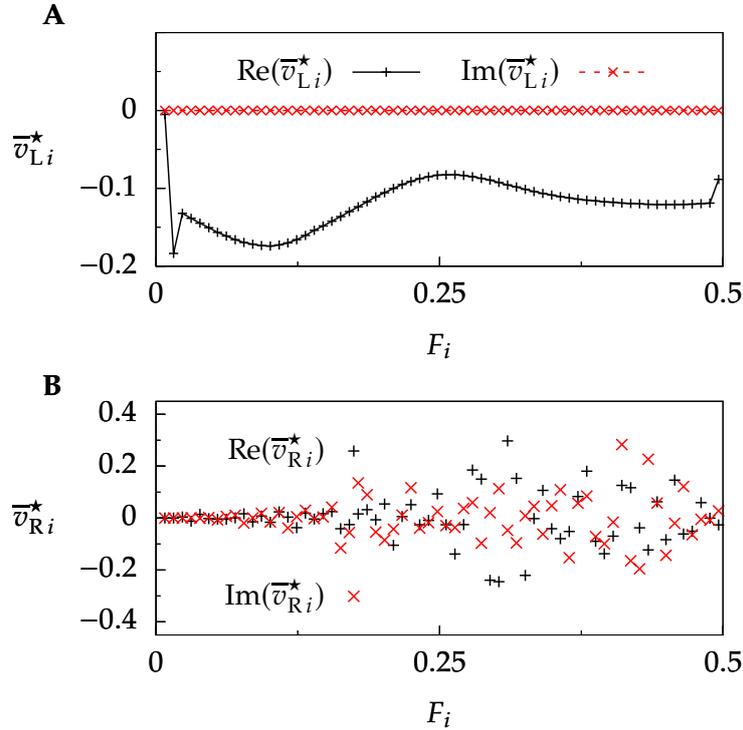

**Figure S10: Eigenvectors of the linearized renormalization group (RG) transformation at the critical fixed distribution**. Components of the left (**A**) and right (**B**) eigenvector relative to the largest eigenvalue $\overline{\lambda}_\star$ of the Jacobian $\mathcal{J}$ of the linearized RG transformation, as functions of the cumulative distribution function $F$. The left eigenvector characterizes the instability of the fixed distribution, as shown in Section S23. The fixed distribution and parameters are those of Fig. 2A.

where $i, j = 1, \cdots, N$, we used Eq. (S67), the definition of $p^*_{k+1}$ in Eq. (S50), and that of the Jacobian

$$\mathcal{K}_{ij} \equiv \frac{\partial L_i^{k+1}}{\partial L_j^k}. \tag{S100}$$

Here, in contrast to Eq. (S63), we now write explicitly the subscript in $\varsigma_{\text{low}}$.

By solving Eq. (S98) for $\partial \eta / \partial L_j^k$, substituting the solution in Eq. (S99), and solving Eq. (S99) for $\mathcal{K}_{ij}$, we obtain the desired expression

$$\mathcal{K}_{ij} = \frac{1}{\mathbb{E}\left[\delta\left(\mathcal{J}_\varsigma(J,T) + L_i^{k+1}\right)\right]} \left\{ - \sum_{m<n} \mathbb{E}\left[ \frac{\partial \log p'_k(J_{mn})}{\partial L_j^k} \left[ \mathbb{I}\left(-\mathcal{J}_\varsigma(J,T) < L_i^{k+1}\right) - 2 F_i \right] \right] \right.$$

$$+ \frac{\mathbb{E}\left[ \left(\mathcal{J}_{\varsigma_{\text{low}}}(J,0) - \mathbb{E}[|J'|]\right) \left[\mathbb{I}\left(-\mathcal{J}_\varsigma(J,T) < L_i^{k+1}\right) - 2 F_i\right] \right]}{\mathbb{E}\left[ \left(\mathcal{J}_{\varsigma_{\text{low}}}(J,0) - \mathbb{E}[|J'|]\right)^2 \right]} \left[ \sum_{m<n} \mathbb{E}\left[ \frac{\partial \log p'_k(J_{mn})}{\partial L_j^k} \left(\mathcal{J}_{\varsigma_{\text{low}}}(J,0) - \mathbb{E}[|J'|]\right) \right] \right.$$

$$\left. \left. - \frac{\partial \mathbb{E}[|J'|]}{\partial L_j^k} \right] \right\}. \tag{S101}$$

## S19 Numerical evaluation of the Jacobian

We evaluated numerically the Jacobian (S129) as follows: We substituted in the right-hand side of Eq. (S129) the explicit expression for the derivatives of $\overline{p}'_k$ with respect to $\boldsymbol{K}^k$ given by Eqs. (S73) and (S79).

We then computed the expectation values $\overline{\mathbb{E}}[]$ by randomly drawing a population of samples $J$ from $\overline{p}_k$, proceeding along the lines of Items 2a–2d of Section S16.

Finally, the term

$$\overline{\mathbb{E}}[\delta(\mathcal{K}_\varsigma(J) + K_i^{k+1})] \tag{S102}$$

has been evaluated by replacing the Dirac delta function with a piecewise constant function with width $\Delta$:

$$\delta(\mathcal{K}_\varsigma(J) + K_i^{k+1}) \to \begin{cases} \frac{1}{\Delta} & \text{if } -K_i^{k+1} - \frac{\Delta}{2} < \mathcal{K}_\varsigma(J) < -K_i^{k+1} + \frac{\Delta}{2} \\ 0 & \text{otherwise} \end{cases}. \tag{S103}$$

As shown in Fig. S11, we computed the term (S102) with the substitution (S103) for multiple values of $\Delta$: The numerical estimate of (S102) is then given by the value of $\Delta$ at which this quantity plateaus in the plot.

## S20 Limits

In what follows, we will discuss some specific limits of the RG transformation proposed in Section 2, and relate them to results in the literature.

### S20.1 Lower-critical-dimension limit

In this Section we will show that the RG transformation that we proposed satisfies the limit [2]

$$T_c \stackrel{\varsigma \to \varsigma_{\text{low}}}{\to} 0. \tag{S104}$$

By iterating the transformation of Section 3

$$p'_k(J') \to p'_{k+1}(J'). \tag{S105}$$

at $\varsigma = \varsigma_{\text{low}}$ and $T = 0$, the condition (12) in the rescaling procedure ensures that, when the rescaled distribution $p_{k+1}$ is decimated to obtain $p'_{k+1}$, the expectation value of $|J'|$ taken with the latter equals the one taken with $p'_k$. This implies that $p'_{k+1}$ and $p'_k$ have the same width, and the RG transformation reaches a finite, critical fixed distribution for large $k$.

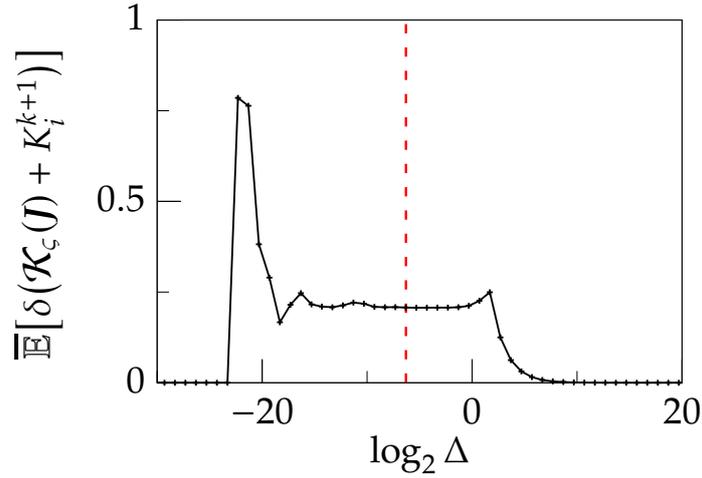

**Figure S11: Numerical evaluation of the Jacobian term involving a Dirac delta function.** The term (S102) approximated by replacing the delta function with a piecewise constant function of width $\Delta$, Eq. (S103), is shown as a function of $\log_2 \Delta$ (black solid lines), for $\varsigma = 0.75$, $N = 2^6$, and a given $i$. The value of $\log_2 \Delta$ at which this term, as well as all the other terms for $i = 1, \cdots, N$, plateau, is also shown (red dashed line).

### S20.2 Ferromagnetic limit

In what follows we will discuss how the RG transformation that we propose reproduces the known RG iteration for the ferromagnetic version of the HEA, known as Dyson's hierarchical model [14], in the limit where the spin-spin couplings are all ferromagnetic. We will consider Dyson's hierarchical model for

$$\varsigma_\infty^{\text{FM}} < \varsigma \leq \varsigma_{\text{low}}^{\text{FM}}, \tag{S106}$$

where [14]

$$\varsigma_\infty^{\text{FM}} = 1, \varsigma_{\text{low}}^{\text{FM}} = 2, \tag{S107}$$

are the values of the interaction parameter below which the thermodynamic limit is defined, and the one corresponding to the lower critical dimension, respectively.

#### S20.2.1 Decimation

In this Section we will start with a model $\mathcal{M}$ with coupling distribution

$$p_k(J) = \delta(J - J_{\text{FM}}^{k-1}), \tag{S108}$$

where, for all $k \geq 1$,

$$J_{\text{FM}}^{k-1} \equiv \{J'_{k-1}, \cdots, J'_{k-1}\} \tag{S109}$$

is a coupling configuration where all spin-spin interactions are equal to the ferromagnetic coupling $J'_{k-1} \geq 0$, and $\delta$ is the multi-dimensional Dirac delta function. We will go through the decimation procedure $\mathcal{M} \to \mathcal{M}'$ of Section 2.1 in the ferromagnetic limit, and obtain the coupling distribution of $\mathcal{M}'$.

**Energy excitations** First, let us discuss the structure of the energy excitations of Section 2.1 of $\mathcal{M}$ for

$$J = J_{\text{FM}}^{k-1}. \tag{S110}$$

The GS is

$$\sigma = \{+, +, +, +\}. \tag{S111}$$

The excited state $S_{\uparrow\uparrow\downarrow\downarrow}$ of group i) in Section 2.1, has an energy gap $2^{3-\varsigma}J'_{k-1}$ with respect to the GS, while those of groups ii) and iii) have a gap of $2(1 + 2^{1-\varsigma})J'_{k-1}$ and $2(2 + 2^{1-\varsigma})J'_{k-1}$, respectively. Since $\varsigma$ lies in the interval given by Eqs. (S106) and (S107), it is easy to show that the gap of $S_{\uparrow\uparrow\downarrow\downarrow}$ is always the smallest one, i.e., the first excited state is

$$S_2 = S_{\uparrow\uparrow\downarrow\downarrow}. \tag{S112}$$

**Form of the order parameter**  Here, we will show that the order parameter discussed in Section 2.1 reduces to the ferromagnetic order parameter in the ferromagnetic limit (S110). In fact, by using Eq. (S111), Eq. (8) reduces to

$$\Phi_{\rm L}[S] = \frac{S_1 + S_2}{2}, \quad \Phi_{\rm R}[S] = \frac{S_2 + S_3}{2}, \tag{S113}$$

and similarly for $\mathcal{M}'$. The order parameter in Eq. (S113) is the local magnetization, and spins are blocked according to the block-spin majority rule: the decimated spin points either up or down if the majority of the spins in the block points up or down, respectively [15]. In the general case where the $J_{ij}$s are either positive or negative, the order parameter in Eqs. (8) and (9) constitutes a generalization of the majority rule above: Spins are no longer decimated according to a ferromagnetic majority rule, but with a majority rule relative to the structure of the GS of the system, see Section 2.1.

By imposing that the decimation procedure conserves the ferromagnetic form of the coupling distribution, i.e., that $p'(J')$ is different from zero for non-negative values of $J'$ only, in the ferromagnetic limit Eq. (S49) becomes $J' = +\mathcal{J}_\varsigma(J, T)$, and Eq. (10)

$$\begin{aligned} p'_k(J') &= \int dJ\, p_k(J)\delta\left(J' - \mathcal{J}_\varsigma(J, T)\right) \\ &= \delta(J' - J'_k), \end{aligned} \tag{S114}$$

where we have set

$$J'_k \equiv \mathcal{J}_\varsigma(J_{\rm FM}^{k-1}, T). \tag{S115}$$

### S20.2.2  Rescaling

We will now show that, if we start with a model $\mathcal{M}'$ with coupling distribution (S114), then, in the ferromagnetic limit, the solution of the rescaling procedure of Section 2.2 is

$$p_{k+1}(J) = p^*_{k+1}(J), \tag{S116}$$

where $p^*_{k+1}$ is defined by the equality in Eq. (S50). To prove this, first we observe that, by definition, Eq. (S116) realizes the absolute minimum of the Kullback-Leibler divergence (13). Second, we will show that Eq. (S116) satisfies the constraints (11), (12) and the normalization condition in Eq. (13).

Let us focus on constraint (11) first: By substituting the definition in Eq. (S50), Eqs. (S114) and (S116) into Eq. (10), we obtain

$$p'_{k+1}(J') = \delta(J' - \mathcal{J}_\varsigma(J_{\rm FM}^k, T)). \tag{S117}$$

By combining Eqs. (3), (S55), (S107), (S112) and (S117), the LHS of Eq. (11) reads

$$\begin{aligned} \mathbb{E}[H'^{k+1}[S'_2] - H'^{k+1}[S'_1]]\big|_{\varsigma=\varsigma_{\rm low}^{\rm FM}, T=0} &= 2\mathcal{J}_{\varsigma_{\rm low}^{\rm FM}}(J_{\rm FM}^k, 0) \\ &= 2J'_k. \end{aligned} \tag{S118}$$

In addition, by substituting Eqs. (4) and (S114) in the right-hand side of Eq. (11), we obtain

$$\mathbb{E}[H'^k[S'_2] - H'^k[S'_1]] = 2J'_k. \tag{S119}$$

As a result, Eqs. (S118) and (S119) show that constraint (11) is satisfied.

Second, let us consider constraint (12), which is equivalent to Eq. (S64). By substituting the definition in Eq. (S50), Eqs. (S114) and (S116) into (S64), Eq. (12) can be rewritten as

$$\mathbb{I}(S_2 \neq S_{\uparrow\uparrow\downarrow\downarrow})\big|_{\varsigma=\varsigma_{\rm low}, J=J_{\rm FM}^k} = 0 \tag{S120}$$

which, according to the discussion of Section S20.2.1, is satisfied.

Finally, Eqs. (S114) and (S116) and the definition in Eq. (S50) show that the normalization condition in Eq. (13) is satisfied.

We have thus shown that in the ferromagnetic limit the simple, factorized solution (S116) realizes the absolute minimum of the Kullback-Leibler divergence in the optimization problem, Eq. (13), and that it satisfies all of its constraints. As a result, (S116) is the solution of the rescaling problem in this limit.

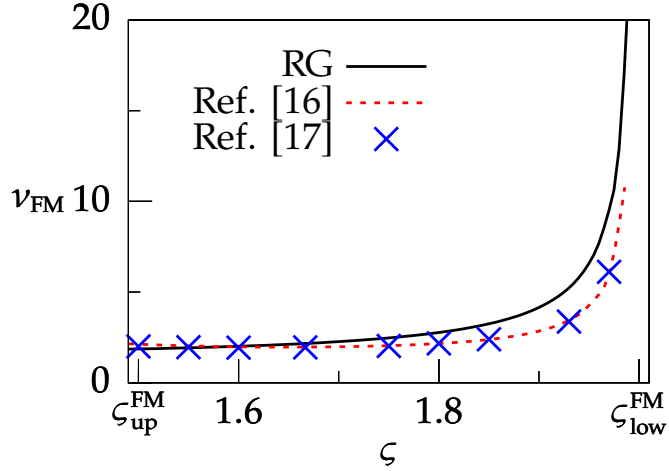

**Figure S12: Critical exponent $\nu$ in the ferromagnetic limit**. The exponent $\nu_{\text{FM}}$ describing the divergence of the correlation length is shown as a function of the coupling-range exponent $\varsigma$ from the renormalization group (RG) approach in the ferromagnetic (FM) limit (black solid curve), from the numerically exact study of [16] and [17] (red dashed curved and blue points, respectively). Here $\nu_{\text{FM}}$ corresponds to the critical exponent $\nu$ for Dyson's hierarchical model [14]—the ferromagnetic version of the hierarchical Edwards-Anderson model—and this plot is the ferromagnetic analog of Fig. 2B.

### S20.2.3 Renormalization-group transformation

The discussion of Sections S20.2.1 and S20.2.2 shows that, in the ferromagnetic limit, the RG transformation—the combination of decimation and rescaling—maps $p_k(J) = \delta(J - J_{\text{FM}}^{k-1})$ into $p_{k+1}(J) = \delta(J - J_{\text{FM}}^k)$, see Eqs. (S114) and (S116) and the definition in Eq. (S50). As a result, the RG functional flow of Eq. (S105) reduces to the flow of a single quantity $J'_k$, i.e.,

$$J'_{k-1} \to J'_k, \tag{S121}$$

and is given by Eqs. (S109) and (S115).

The RG iteration relation (S121) is similar, but not identical, to the real-space RG flow equation for Dyson's hierarchical model in the limit where a four-spin model is mapped into a two-spin model [6, 16]. In fact, both Eqs. (S115) and (S121) and the RG iteration relation of [16], i.e. Eq. (3) in there, result from imposing that the two-point correlation function of $\mathcal{M}$ and $\mathcal{M}'$ are equal, cf. Eq. (5). However, Eq. (3) in [16] includes an additional normalization factor, which represents the magnitude of the square magnetization in each spin block—see the the two equations preceding Eq. (3) in [16]. Given such difference between the two transformation, in what follows we will study the critical exponents resulting from the RG transformation (S121): This will allow us to compare with the exact values the predictions for the critical exponents of our RG framework in the ferromagnetic limit, mirroring the analysis of Section 4 and Fig. 2 for the HEA, and thus illustrating further the predictive capabilities of our method.

### S20.2.4 Critical exponents

The critical exponent related to the divergence of the correlation length in the ferromagnetic limit, $\nu_{\text{FM}}$, is obtained from Eq. (S121) as [1]

$$2^{1/\nu_{\text{FM}}} = \frac{\partial \overline{J}'_{k+1}}{\partial \overline{J}'_k}, \tag{S122}$$

where $\overline{J}'_k \equiv \beta J'_k$ and Eq. (S122) is evaluated at the critical fixed point given by $\overline{J}'_k = \overline{J}'_{k+1}$. The exponent $\nu_{\text{FM}}$ is shown in Fig. S12 as a function of $\varsigma$, together with numerically exact results from the literature.

## S20.3 Zero temperature, first-level couplings only

The decimation procedure of our RG approach reproduces some features of a recently proposed decimation procedure [18] in the limit where the temperature is zero and couplings on the second hierarchical level vanish.

At null temperature the ensemble averages in Eq. (2) are dominated by the GS, i.e., $\boldsymbol{S} = \boldsymbol{\sigma}$. Also, in this case the GS is determined by the first-hierarchical-level couplings only, i.e., $J_{12}$ and $J_{34}$: as a result, the GS satisfies $\sigma_2 = \text{sgn}(J_{12})\sigma_1$, and $\sigma_4 = \text{sgn}(J_{34})\sigma_3$. It follows that the ensemble averages are dominated by the terms with

$$\begin{aligned} S_2 &= \text{sgn}(J_{12})S_1, \\ S_4 &= \text{sgn}(J_{34})S_3. \end{aligned} \quad \text{(S123)}$$

Equation (S123) reproduces the decimation criterion of an RG approach for one-dimensional Ising spin glass (SG) models with long-range interactions [18]. It is clear that Eq. (S123) takes into account the short-range couplings, and neglects the long-range ones; as a result, it constitutes an approximation. On the other hand, our decimation rule (2) incorporates the alignment of the spin configuration $\boldsymbol{S}$ with $\boldsymbol{\sigma}$, i.e., the GS of the Hamiltonian (3), which is determined by both short- and long-range couplings.

## S21  Scaled renormalization-group transformation

In order to linearize the RG transformation at the critical fixed distribution in terms of dimensionless quantities only [1], we rewrite the RG iteration in terms of the coupling $\beta J$. To achieve this, we consider the rescaling and decimation procedure for a modified probability distribution of Section S15, where we choose the scaling factor

$$s = 1/\beta. \quad \text{(S124)}$$

As specified in Section S15, quantities which are scaled according to $s$ will be denoted by a $\bar{\phantom{x}}$, see for example Eq. (S80). The scaled transformation will then be given by Eq. (S82) and Eq. (S127) below.

First, by combining Eqs. (S73) and (S79), we observe that the modified coupling distribution $\overline{p}'_k$ depends on the quantiles $\boldsymbol{L}^k$ and $\beta$ through the dimensionless quantiles

$$\boldsymbol{K}^k \equiv \beta \boldsymbol{L}^k. \quad \text{(S125)}$$

By combining this observation with Eqs. (S55) and (S86), we obtain that Eq. (S82) depends on $\boldsymbol{L}^k$, $\beta$ and $\eta$ only through the scaled quantiles and Lagrange multiplier (S125) and $\overline{\eta}$, respectively.

Second, by using Eq. (S12), we observe that $\beta \mathcal{J}_\varsigma(\boldsymbol{J}, T)$ depends on $\beta$ and $\boldsymbol{J}$ through the combination $\beta \boldsymbol{J}$, and thus can be rewritten in terms of an additional function $\mathcal{K}_\varsigma$ as

$$\beta \mathcal{J}_\varsigma(\boldsymbol{J}, T) \equiv \mathcal{K}_\varsigma(\beta \boldsymbol{J}). \quad \text{(S126)}$$

We then substitute Eq. (S126) in Eq. (S91), multiply by $\beta$ both sides of the equality in the indicator function, and obtain the second equation for the linearization

$$\int d\boldsymbol{J}\, \overline{p}_{k+1}(\boldsymbol{J}) \left[ \mathbb{I}\big( -\mathcal{K}_\varsigma(\boldsymbol{J}) < K_i^{k+1} \big) - 2F_i \right] = 0, \quad \text{(S127)}$$

which depends on $\boldsymbol{L}^k$, $\boldsymbol{L}^{k+1}$, $\beta$ and $\eta$ through the combinations $\boldsymbol{K}^k$, $\boldsymbol{K}^{k+1}$ and $\overline{\eta}$ only.

## S22  Jacobian

The Jacobian of the scaled RG transformation

$$\mathcal{J}_{ij} \equiv \frac{\partial K_i^{k+1}}{\partial K_j^k}, \quad \text{(S128)}$$

is obtained by proceeding along the lines of Section 3. We derive both sides of Eqs. (S82) and (S127) with respect to $K^k$, and obtain

$$\mathcal{J}_{ij} = \frac{1}{\overline{\mathbb{E}}\left[\delta\left(\mathcal{K}_\varsigma(J) + K_i^{k+1}\right)\right]} \left\{ -\sum_{m<n} \overline{\mathbb{E}}\left[\frac{\partial \log \overline{p}'_k(J_{mn})}{\partial K_j^k}\left[\mathbb{I}\left(-\mathcal{K}_\varsigma(J) < K_i^{k+1}\right) - 2 F_i\right]\right] \right.$$

$$+ \frac{\overline{\mathbb{E}}\left[\left(\mathcal{J}_{\varsigma_{\text{low}}}(J, 0) - \overline{\mathbb{E}}[|J'|]\right)\left[\mathbb{I}\left(-\mathcal{K}_\varsigma(J) < K_i^{k+1}\right) - 2 F_i\right]\right]}{\overline{\mathbb{E}}\left[\left(\mathcal{J}_{\varsigma_{\text{low}}}(J, 0) - \overline{\mathbb{E}}[|J'|]\right)^2\right]} \left[\sum_{m<n} \overline{\mathbb{E}}\left[\frac{\partial \log \overline{p}'_k(J_{mn})}{\partial K_j^k}\left(\mathcal{J}_{\varsigma_{\text{low}}}(J, 0) - \overline{\mathbb{E}}[|J'|]\right)\right]\right]$$

$$\left. - \frac{\partial \overline{\mathbb{E}}[|J'|]}{\partial K_j^k} \right\}. \qquad \text{(S129)}$$

where, for the sake of clarity, we recall that the expectation value $\overline{\mathbb{E}}[]$ of a quantity which depends on $J$ is taken with respect to $\overline{p}_{k+1}(J)$, and $\overline{\mathbb{E}}[|J'|]$ denotes the average with respect to $\overline{p}'_k$, and it is thus a function of $K^k$. Finally, proceeding along the lines of Section 3, we write $\mathcal{J}$ in terms of its eigenvalues $\overline{\lambda}_n$ and its left and right eigenvectors $\overline{v}_L^n, \overline{v}_R^n$ as

$$\mathcal{J}_{ij} = \sum_n \overline{\lambda}_n \overline{v}_{R\,i}^n \overline{v}_{L\,j}^n. \qquad \text{(S130)}$$

## S23 Characterization of fixed-point instability

Here, we recall how the eigenvectors of the Jacobian matrix characterize the instability of fixed distributions [1].

Given a fixed distribution $p'_*$ with one eigenvalue, $\lambda_\star$, with norm larger than unity, let us denote by $v_L^\star$ its left eigenvector. Denoting the quantiles of $p'_*$ by $L_*$ and the deviation with respect to them by $\delta L^k \equiv L^k - L_*$, then for small $\delta L$ and $l \geq 0$ we have

$$\delta L_i^{k+l} = \sum_n \lambda_n^l v_{R\,i}^n \left(v_L^n \cdot \delta L^k\right). \qquad \text{(S131)}$$

As a result, if the distribution is perturbed by altering its quantiles by $\delta L^k \propto v_L^\star$, the RG iteration will flow away from $p'_*$. The eigenvector $v_L^\star$ thus denotes the unstable direction of the fixed distribution.

## S24 Numerical simulations

In what follows, we will compare the predictions for the critical exponent $\nu$ related to the divergence of the correlation length [19] from the RG method, with that from numerical simulations.

### S24.1 Models

We aim at testing how the RG method handles coupling-coupling correlations, in the region $\varsigma \lesssim \varsigma_{\text{low}}$ where such correlations are expected to be most important, see Section S7: in such region, the critical temperature tends to zero, thus implying long equilibration times for numerical simulations in the critical region [20, 21]. We thus considered the asynchronous multispin-coding method, which allows one to simulate simultaneously multiple disorder samples by writing the values of the spin-spin couplings into a the bits of an integer [22] with a significant computational gain, and two variants of the HEA which are fit for this technique. Both these variants are systems of $2^k$ Ising spins with Hamiltonian

$$H_k^{\text{d}}[S] \equiv -\sum_{i<j=1}^{2^k} J_{ij}^{\text{d}} S_i S_j, \qquad \text{(S132)}$$

where $\{J_{ij}^{\text{d}}\}$ are independent and identically distributed random variables, and the superscript d stands for 'diluted.' Given two sites $i$ and $j$, we denote by $d_{ij}$ the number of hierarchical levels that we need to ascend in the hierarchical tree starting from spins $i$ and $j$, to find a root common to such spins [23], see Fig. S2. We then

choose $J_{ij}^\mathrm{d}$ to be nonzero with probability $\mathfrak{p}_{ij}$ and zero otherwise, where a nonzero $J_{ij}^\mathrm{d}$ is equal to ±1 with equal probability. The two variants of the HEA are

- The HEA with power-law interaction decay (hierarchical Edwards-Anderson model with power-law interaction decay (pHEA)) [23, 24], where

$$\mathfrak{p}_{ij} = 2^{-2\varsigma(d_{ij}-1)}. \tag{S133}$$

- The HEA with fixed average coordination number (hierarchical Edwards-Anderson model with fixed average coordination number (cHEA)) [4, 25], where

$$\mathfrak{p}_{ij} = 1 - \exp\left(-A\, 2^{-2\varsigma(d_{ij}-1)}\right). \tag{S134}$$

The coefficient $A$ is set by imposing that the coordination number of any spin $i$—the number of spins $S_j$ such that $J_{ij}^\mathrm{d} \neq 0$—is equal, on average, to a given value $z$, which plays the role of a model parameter. In particular, $A$ is determined from the relation

$$\begin{aligned}
z &= \mathbb{E}\Big[\sum_{j \neq i} \mathbb{I}(J_{ij}^\mathrm{d} \neq 0)\Big] \\
&= \sum_{l=1}^{k} \sum_{j \in B_l} \mathfrak{p}_{1j} \\
&= \sum_{l=1}^{k} 2^{l-1}[1 - \exp\left(-A\, 2^{-2\varsigma(l-1)}\right)],
\end{aligned} \tag{S135}$$

where in the first line the sum runs on $j$ but not on $i$, which is fixed, and we equated $z$ to the expression of the average coordination number of site $i$. In the second line we observed that the average coordination number is the same for all spins and thus replaced $i$ by 1, we rewrote the sum as a sum over blocks $B_l$ of $2^{l-1}$ spins which lie at hierarchical distance $d_{1j} = l$ from $S_1$, and we used the relation $\mathbb{E}[\mathbb{I}(J_{1j}^\mathrm{d} \neq 0)] = \mathfrak{p}_{1j}$. Finally, in the third line we used the fact that $\mathfrak{p}_{1j}$ in Eq. (S134) depends through $j$ through $d_{1j}$ only. As a result, Eq. (S135) has been substantially simplified and can be readily solved numerically for $A$.

As we discussed in Section S7, the limit of the critical temperature at the lower critical dimension, Eq. (S104), holds for thMonte Carlo (MC)e HEA defined in Eq. (1). Given that the couplings in the pHEA have the same scaling as in the HEA of Eq. (1), Eq. (S104) holds for the pHEA as well [24]. In fact, the larger $\varsigma$, the lower the overall number of spin-spin interactions, and thus the lower the temperature $T_c$ below which spin-glass order appears. Given that the equilibration times of MC simulations grow dramatically at low temperatures, for $\varsigma$ close to one, MC simulations for the pHEA in the critical region are computationally unfeasible [26]. On the other hand, for the cHEA the condition (S104) ensures that the average coordination number is fixed and independent of $\varsigma$: As a result, for $\varsigma \to 1$, the decrease of $T_c$ is hindered, and simulations in the critical region around the lower critical dimension are feasible.

By making the hypothesis the critical exponents of the model depend on the long-range scaling of its spin-spin couplings only, and not on the full coupling distribution [27, 28], we will assume that the critical exponents of the cHEA and pHEA are the same: as a result, we will use the cHEA to extract the critical exponents for $\varsigma \lesssim 1$, and the pHEA for other values of $\varsigma$.

In what follows, we will evaluate numerically the finite-size critical temperature and the correction-to-scaling exponent; these results will then be used in a finite-size scaling analysis at the critical point to estimate $\nu$.

### S24.2  Finite-size critical temperature

In order to estimate the critical temperature, we introduce the reduced temperature

$$t \equiv \frac{T - T_c}{T_c}, \tag{S136}$$

and consider a thermodynamic observable $f_k(t)$ for a HEA with $2^k$ spins which, in the critical region, scales with the system size and temperature as

$$f_k(t) = f^L(2^{\phi k}t) + \frac{1}{2^{k\omega}}f^S(2^{\phi k}t), \tag{S137}$$

where $f^L$ and $f^S$ are the leading and subleading terms, $\omega$ the correction-to-scaling exponent [4], and [24, 29, 30]

$$\phi = \begin{cases} 1/3 & \text{if } \varsigma \leq \varsigma_{\text{up}} \\ 1/\nu & \text{if } \varsigma > \varsigma_{\text{up}} \end{cases}, \tag{S138}$$

where

$$\varsigma_{\text{up}} \equiv \frac{2}{3} \tag{S139}$$

is the upper critical dimension of the model [23]. We introduce a finite-size critical temperature $T_c^k$ and its reduced value

$$t_c^k \equiv \frac{T_c^k - T_c}{T_c}, \tag{S140}$$

where $t_c^k$ is defined as the value of $t$ at which $f_k$ and $f_{k+1}$ are equal:

$$f_k(t_c^k) = f_{k+1}(t_c^k). \tag{S141}$$

Substituting Eq. (S137) into Eq. (S141), we obtain

$$\begin{aligned}
f^L(2^{\phi k}t_c^k) + \frac{1}{2^{k\omega}}f^S(2^{\phi k}t_c^k) &= f^L(2^{\phi(k+1)}t_c^k) + \frac{1}{2^{(k+1)\omega}}f^S(2^{\phi(k+1)}t_c^k) \\
&= f^L(0) + \left.\frac{df^L}{dt}\right|_0 2^{\phi k}t_c^k + \frac{1}{2^{k\omega}}f^S(0) + \mathcal{O}((2^{\phi k}t_c^k)^2) + \mathcal{O}\left(\frac{1}{2^{k\omega}}2^{\phi k}t_c^k\right) \\
&= f^L(0) + \left.\frac{df^L}{dt}\right|_0 2^{\phi(k+1)}t_c^k + \frac{1}{2^{(k+1)\omega}}f^S(0) + \mathcal{O}((2^{\phi(k+1)}t_c^k)^2) + \mathcal{O}\left(\frac{1}{2^{(k+1)\omega}}2^{\phi(k+1)}t_c^k\right).
\end{aligned} \tag{S142}$$

By neglecting the $\mathcal{O}$ terms in the last two lines of Eq. (S142), and setting $\beta_c^k \equiv 1/T_c^k$, we obtain

$$\beta_c^k = \beta_c + \frac{A_f}{2^{k(\omega+\phi)}}, \tag{S143}$$

where $A_f$ is a constant which depends on $f$ and which is independent of $k$.

We considered the following choices for $f_k$:

1. Given the overlap $q \equiv \frac{1}{2^k}\sum_{i=1}^{2^k} S_i S'_i$ between spin configurations $S$ and $S'$, we write $f_k$ in terms of the spin-glass susceptibility $\chi_k$ as

$$f_k = \log[2^{k(1-2\varsigma)}\chi_k], \tag{S144}$$

   where

$$\chi_k \equiv 2^k \mathbb{E}[\langle q^2 \rangle], \tag{S145}$$

   and $\langle \rangle$ is the Boltzmann average with respect to $S$ and $S'$ [31]. Here and in Items 2 and 3, the Boltzmann average is taken for a model with $2^k$ spins.

2. We consider the fourth-moment ratio

$$U_k \equiv \frac{\mathbb{E}[\langle q^4 \rangle]}{(\mathbb{E}[\langle q^2 \rangle])^2} \tag{S146}$$

   and write $f_k$ in terms of the Binder cumulant [32]

$$B_k \equiv \frac{3 - U_k}{2} \tag{S147}$$

   as

$$f_k = \log B_k. \tag{S148}$$

3. We write $f_k$ in terms of the correlation length $\xi_k$ as

$$f_k = \log \frac{\xi_k}{2^k}, \tag{S149}$$

where $\xi_k$ is defined as follows. Given four spin replicas $S^a$, $a = 1, \cdots, 4$, denoting by $\langle \rangle$ the Boltzmann average over the four replicas, their overlaps in the left and right half of the model are defined as

$$q^{\text{L}}_{ab} \equiv \frac{1}{2^{k-1}} \sum_{i=1}^{2^{k-1}} S_i^a S_i^b, \quad q^{\text{R}}_{ab} \equiv \frac{1}{2^{k-1}} \sum_{i=2^{k-1}+1}^{2^k} S_i^a S_i^b, \tag{S150}$$

respectively, and the correlation length is given by [24, 28]

$$\frac{\xi_k}{2^k} = \frac{1}{2} \left( \frac{\mathbb{E}[\langle (q^{\text{L}}_{12} + q^{\text{R}}_{12})^2 \rangle]}{\mathbb{E}[\langle (q^{\text{L}}_{12} - q^{\text{R}}_{12})^2 \rangle]} - 1 \right)^{\frac{1}{2\varsigma - 1}}. \tag{S151}$$

## S24.3 Correction to scaling

Given the presence of strong finite-size effects, in order to estimate the critical exponents, we need to estimate the correction-to-scaling exponent $\omega$ [4].

To achieve this, let us consider a thermodynamic observable $g_k(t)$ for a HEA with $2^k$ spins which, in the critical region, scales with the system size and temperature as in Eq. (S137):

$$g_k(t) = g^{\text{L}}(2^{\phi k} t) + \frac{1}{2^{k\omega}} g^{\text{S}}(2^{\phi k} t). \tag{S152}$$

By defining the quotient of $g_k$ as

$$\mathcal{Q}_k[g] \equiv \frac{g_{k+1}(t_c^k)}{g_k(t_c^k)}, \tag{S153}$$

substituting Eqs. (S140) and (S143) into Eq. (S152), we obtain that $g_k$ computed at the finite-size critical temperature (S141) satisfies the simple scaling relation

$$\mathcal{Q}_k[g] = 1 + \frac{B_{f,g}}{2^{k\omega}}, \tag{S154}$$

where $\mathcal{Q}_k[g]$ denotes the quotient of $g_k$, and $B_{f,g}$ depends on the observables $f_k$ and $g_k$.

**Algorithm** We simulated both the pHEA and the cHEA with parallel tempering [33], and run the simulation with three independent replicas for each temperature. Observables which require only one pair of replicas only to be computed, e.g., (S144), were evaluated by considering all replica pairs among the three simulated replicas, and averaging the data across such pairs. The simulation parameters are shown in Table S1, where $N_S$ is the number of samples $\{J^{\text{d}}_{ij}\}$ in Eq. (S132), $N_{\text{sw}}$ the number of MC sweeps, $n_\beta$ the number of simulated temperatures, which lie between $\beta_{\min}$ and $\beta_{\max}$, and temperatures are swapped every $n_{\text{swap}}$ sweeps. We used the second half of the MC sweeps to compute the observables, and the equilibration for the SG susceptibility, Binder ratio and correlation length is illustrated in Fig. S19 where, for each value of $\varsigma$, we show only the largest volume and the smallest temperature that we simulated, in order to assess equilibration in the worst-case scenario. Finally, the error on the observables was estimated as the sum of the error resulting from the finite number of disorder samples, the one resulting from the finite number of MC steps, and the systematic error related to equilibration [34], see Fig. S19. The resulting MC estimates of the considered observables are shown in Figs. S13–S18.

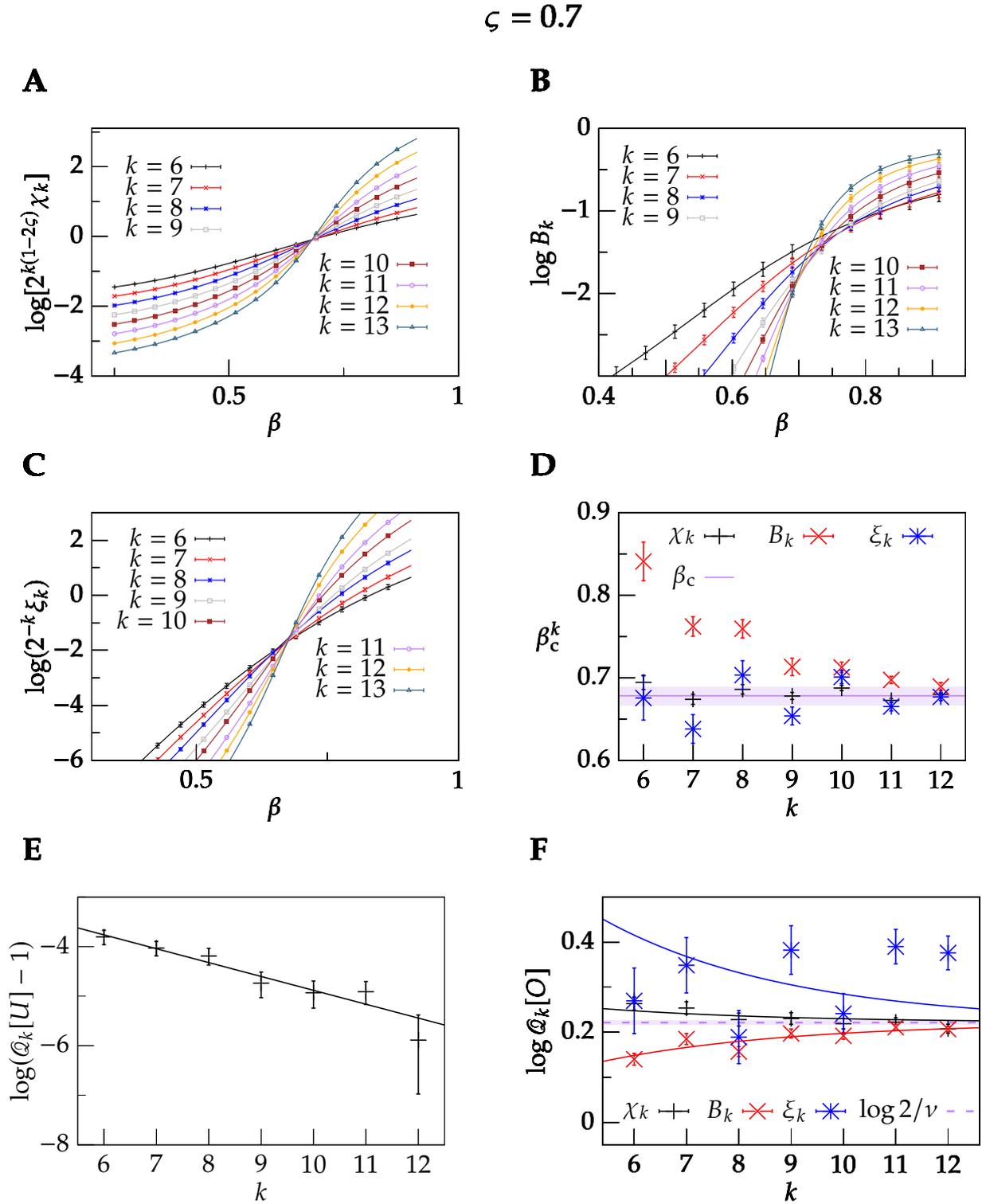

**Figure S13: Finite-size-scaling analysis for the hierarchical Edwards-Anderson model with power-law interaction decay and coupling-range exponent $\varsigma = 0.7$, from Monte Carlo simulations**. **A**) Logarithm of the scaled spin-glass susceptibility $\chi_k$ as a function of the inverse temperature, for different system sizes. **B**) Same as **A**, for the Binder cumulant $B_k$. **C**) Same as **A**, for the scaled correlation length $\xi_k/2^k$. **D**) Finite-size inverse critical temperatures $\beta_c^k$ determined from the data in **A**, **B** and **C** (black, red and blue points, respectively) and Eq. (S141). The infinite-volume inverse critical temperature $\beta_c = 0.678 \pm 0.011$ (purple line) with its error bar (light-purple band), has been determined with a combined fit of $\beta_c^k$ vs. $k$ for the three observables with Eq. (S143), where $\omega$ and $\nu$ have been taken from **E** and **F**, respectively. **E**) Logarithm of $\mathcal{Q}_k[U] - 1$ as a function of $k$ (points), where $\mathcal{Q}_k$ is the quotient given by Eq. (S153) and $U_k$ is the moment ratio. In addition, we show the fitting function $A - k\omega \log 2$ (line), where $\omega$ is the correction-to-scaling exponent. The fit yields $\omega = 0.404 \pm 0.056$. **F**) Logarithm of the quotient $\mathcal{Q}_k[O]$, for the three choices of $O_k$ given in Items (i)–(iii) of Section S24.4, denoted by $\chi$, $B$ and $\xi$ (black, red and blue points, respectively), and fitting function of Eq. (S156) (black, red and blue lines, respectively), where $\omega$ is taken from **E**. The result of the combined fit for the three observables is $\nu = 3.13 \pm 0.064$, and it is shown together with its error bar (purple dashed line and light-purple band, respectively). The simulation parameters are shown in Table S1.

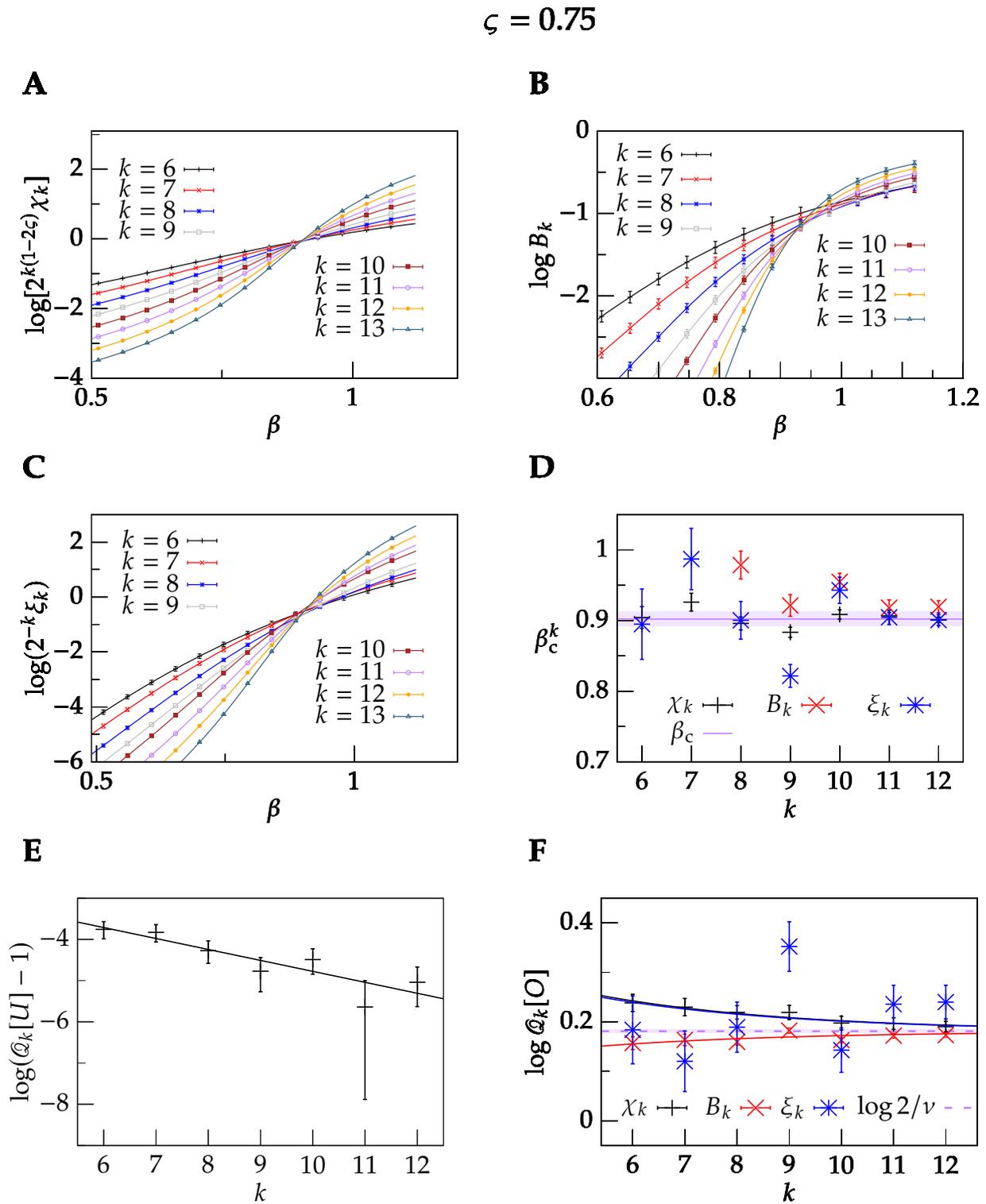

**Figure S14: Finite-size-scaling analysis for the hierarchical Edwards-Anderson model with power-law interaction decay and coupling-range exponent** $\varsigma = 0.75$, **from Monte Carlo simulations**. Same as Fig. S13, with $\beta_c = 0.902 \pm 0.010$, $\omega = 0.383 \pm 0.087$, $\nu = 3.830 \pm 0.079$. Here and in what follows, for the observables and system sizes where the curve crossing which determines $\beta_c^k$ does not occur, the finite-size critical temperature is not shown, nor it is included in the fit which determines $\beta_c$.

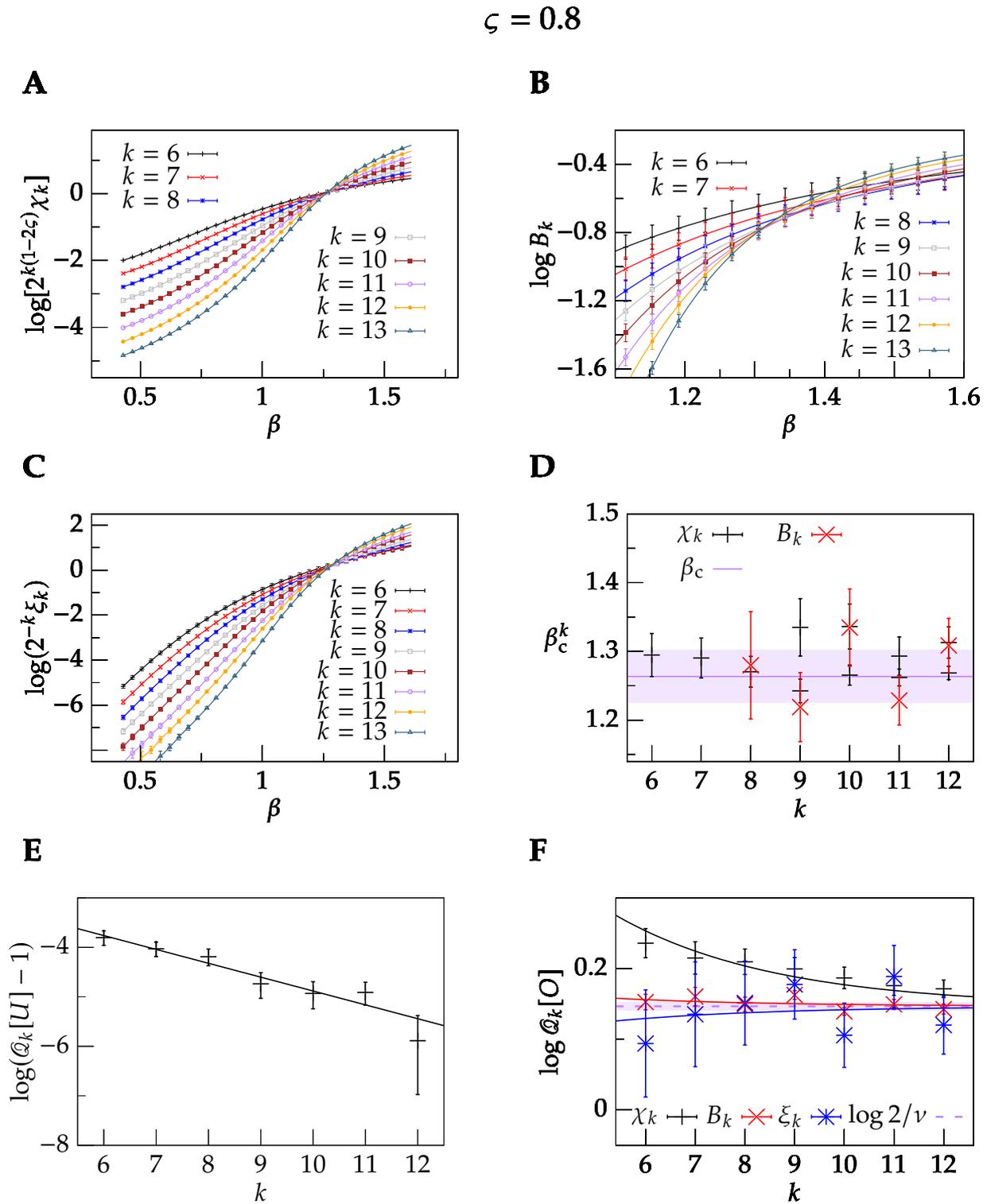

**Figure S15: Finite-size-scaling analysis for the hierarchical Edwards-Anderson model with power-law interaction decay and coupling-range exponent $\varsigma = 0.8$, from Monte Carlo simulations.** Same as Fig. S13, with $\beta_c = 1.263 \pm 0.038$, $\omega = 0.45 \pm 0.13$ and $\nu = 4.73 \pm 0.18$.

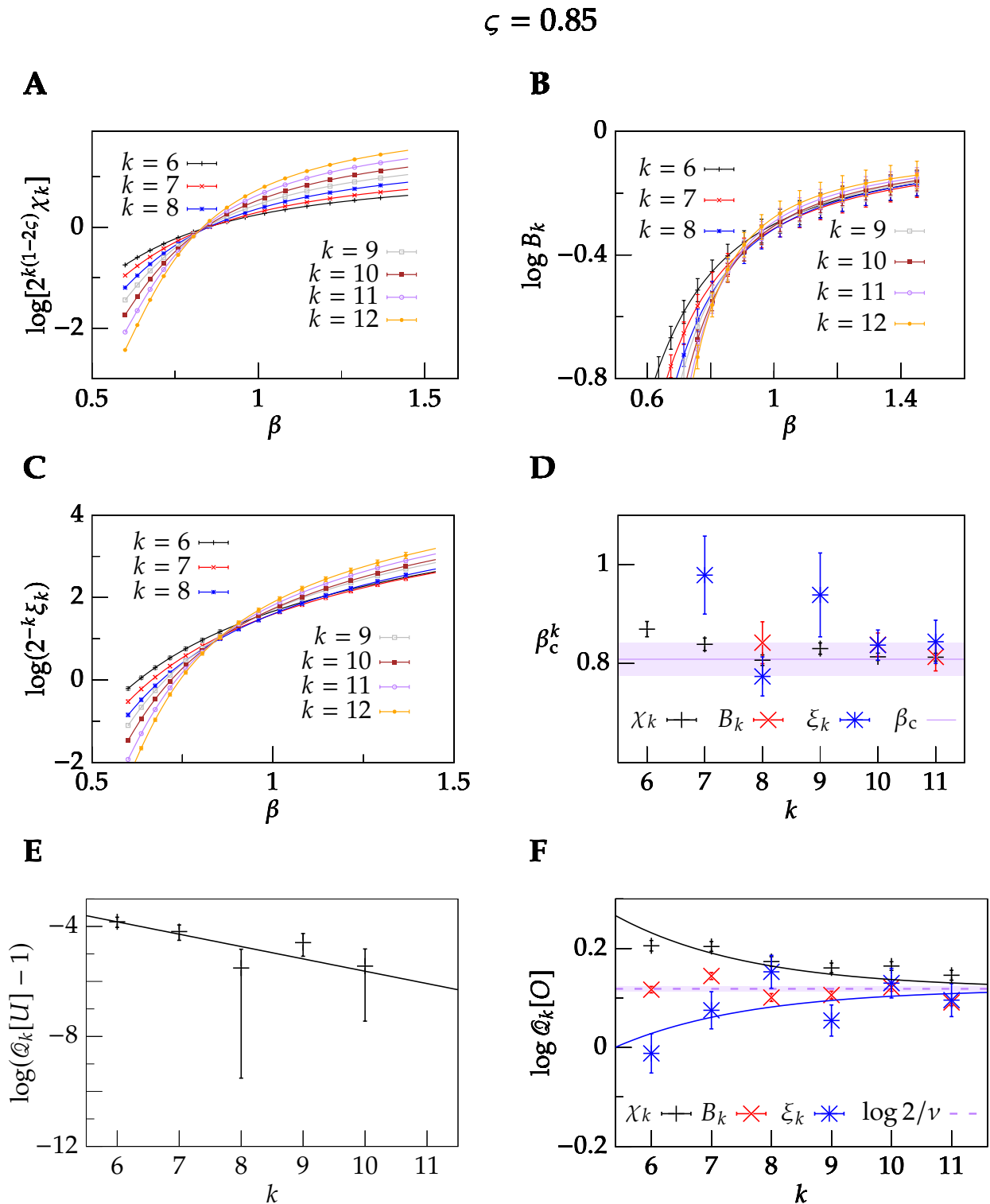

**Figure S16: Finite-size-scaling analysis for the hierarchical Edwards-Anderson model with fixed coordination number and coupling-range exponent $\varsigma = 0.85$, from Monte Carlo simulations.** Same as Fig. S13, with $\beta_c = 0.809 \pm 0.033$, $\omega = 0.65 \pm 0.21$ and $\nu = 5.86 \pm 0.26$. In **D**, the error bar of $\log(Q_k[U] - 1)$ goes to $-\infty$ because the error on $Q_k[U]$ is such that $Q_k[U] - 1$ may fluctuate to negative values. In **F**, the fitting function for $B_k$ is not shown for clarity.

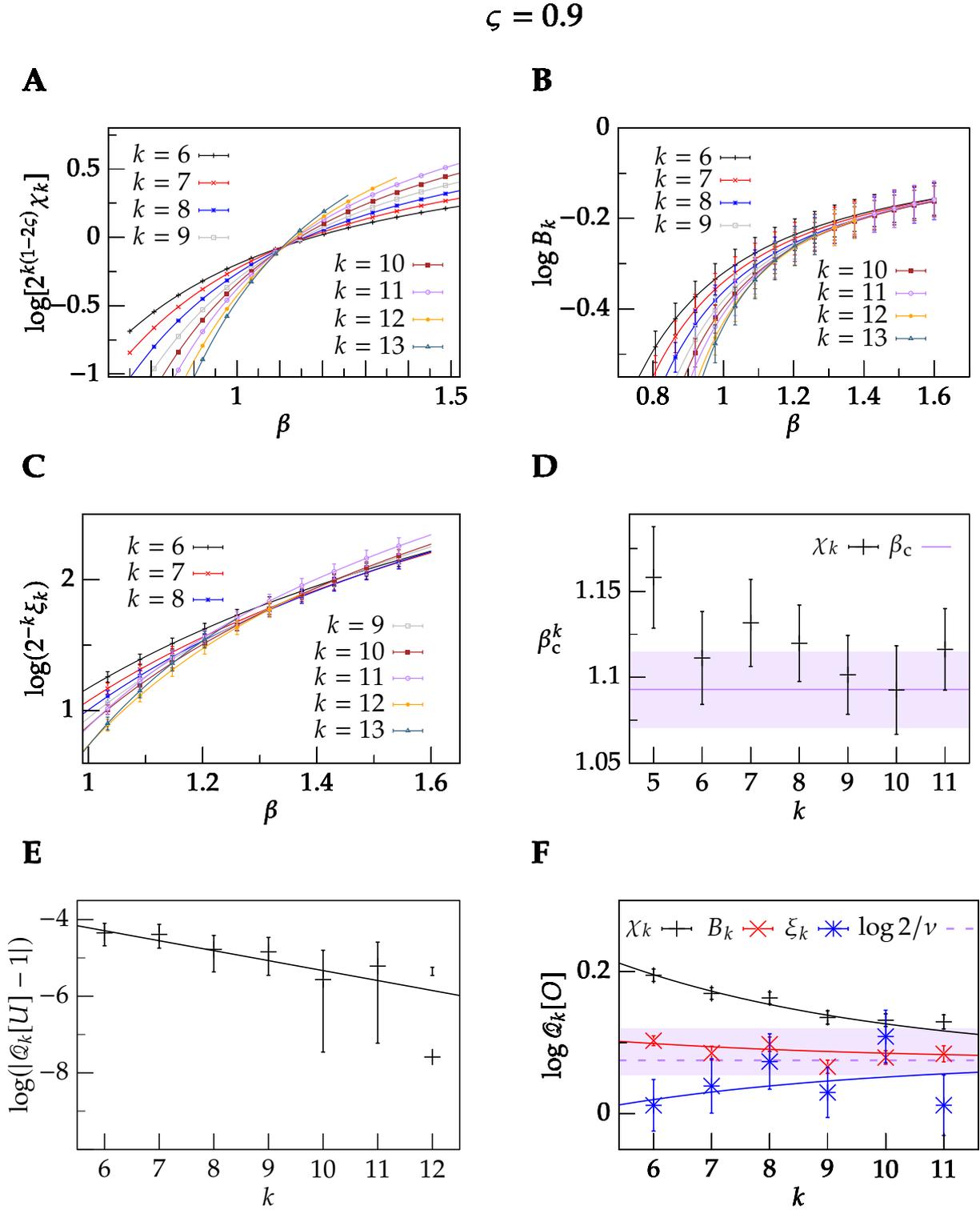

**Figure S17: Finite-size-scaling analysis for the hierarchical Edwards-Anderson model with fixed coordination number and coupling-range exponent $\varsigma = 0.9$, from Monte Carlo simulations.** Same as Fig. S13, with $\beta_c = 1.093 \pm 0.022$, $\omega = 0.31 \pm 0.17$ and $\nu = 9.2 \pm 3.4$. Unlike Fig. S13, here only the data in **A** display sensible crossings: as a result, in **D** we only plot $\beta_c^k$ for $\chi_k$. Given that the fluctuations of $\mathcal{Q}_k[U] - 1$ related to its error bar are such that $\mathcal{Q}_k[U] - 1$ may become negative, in **D** we replaced $\mathcal{Q}_k[U] - 1$ with its absolute value.

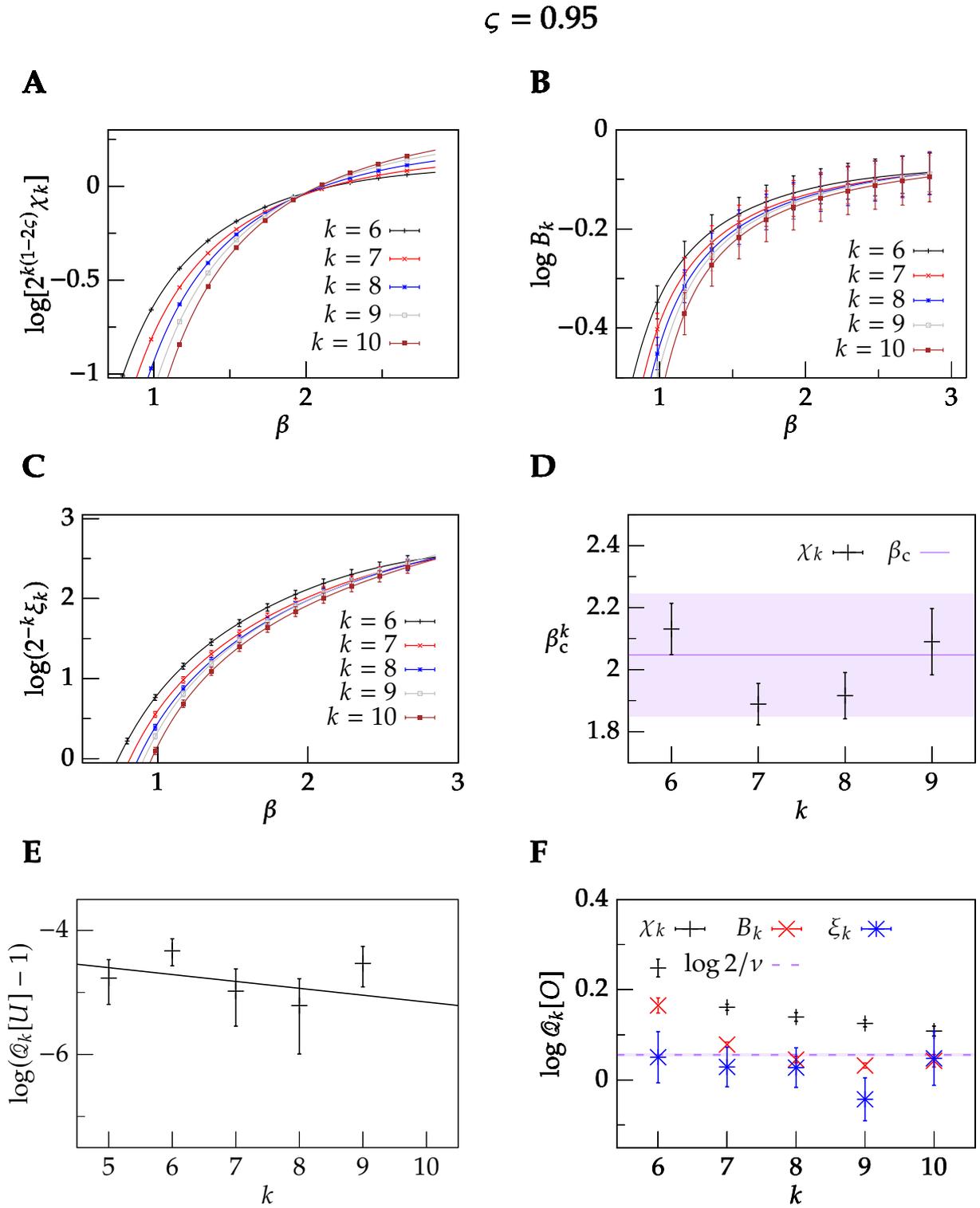

**Figure S18: Finite-size-scaling analysis for the hierarchical Edwards-Anderson model with fixed coordination number and coupling-range exponent** $\varsigma = 0.95$, **from Monte Carlo simulations**. Same as Fig. S13, with $\beta_c = 2.05 \pm 0.20$ and $\nu = 12.50 \pm 0.73$. Unlike Fig. S13, here only the data in **A** display sensible crossings: as a result, in **E** we only show $\beta_c^k$ for $\chi_k$. Given that $\log(\mathcal{Q}_k[U] - 1)$ vs. $k$ does not appear to have a nonzero slope, **E** suggests that corrections to scaling are negligible. As a a result, in the fit of **D** we dropped the correction-to-scaling term in Eq. (S143). Also, in the combined fit of **F**, we dropped the correction-to-scaling term in Eq. (S156), and considered only the data points which appear to plateau, i.e., the data for $B_k$ and $\xi_k$ with $k = 6, \cdots, 10$.

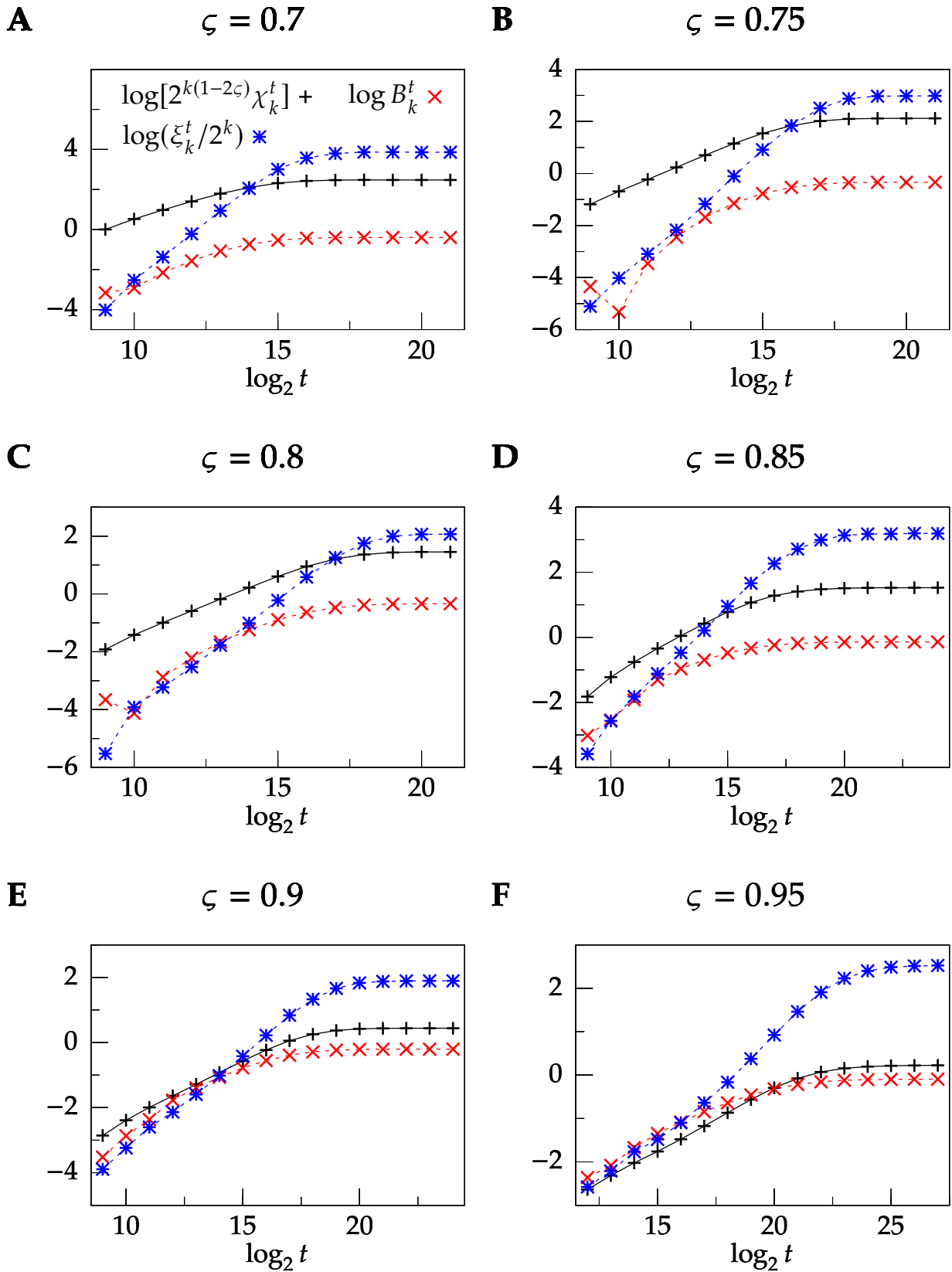

**Figure S19: Equilibration of Monte Carlo simulations** . Scaled spin-glass susceptibility (black), Binder ratio (red) and correlation length (blue) evaluated with $t$ Monte Carlo sweeps, and averaged across the second half of the sweeps, as functions of $\log_2 t$, for the hierarchical Edwards-Anderson model with power-law interaction decay (**A-C**) and the hierarchical Edwards-Anderson model with fixed coordination number (**D-F**), where each panel corresponds to a value of the coupling-range exponent $\varsigma$, and the color legend in panels **B-F** is the same as in panel **A**. For each value of $\varsigma$, we show the largest value of $k$ and the lowest temperature shown in Figs. S13–S18 and Table S1.

| Model | $\varsigma$ | $k$ | $N_S$ | $N_{sw}$ | $n_\beta$ | $\beta_{min}$ | $\beta_{max}$ | $n_{swap}$ |
|---|---|---|---|---|---|---|---|---|
| pHEA | 0.7 | 6 - 12 | 16384 | 1048576 | 16 | 0.25 | 0.91 | 16 |
| | | 13 | 8192 | 2097152 | 19 | 0.03 | 0.91 | 16 |
| | 0.75 | 6 - 13 | 16384 | 1048576 | 16 | 0.42 | 1.12 | 16 |
| | | 14 | 16384 | 2097152 | 16 | 0.42 | 1.12 | 16 |
| | 0.8 | 6 - 13 | 16384 | 2097152 | 32 | 0.43 | 1.61 | 16 |
| cHEA | 0.85 | 6 - 10 | 16384 | 2097152 | 16 | 0.6 | 1.45 | 16 |
| | | 11 | 16384 | 8388608 | 16 | 0.6 | 1.45 | 16 |
| | | 12 | 8192 | 16777216 | 16 | 0.6 | 1.45 | 16 |
| | 0.9 | 5 - 9 | 16384 | 2097152 | 16 | 0.75 | 1.6 | 16 |
| | | 10 | 16384 | 8388608 | 16 | 0.75 | 1.6 | 16 |
| | | 11 | 8192 | 8388608 | 16 | 0.75 | 1.6 | 16 |
| | | 12 | 8192 | 16777216 | 16 | 0.52 | 1.37 | 16 |
| | 0.95 | 5 - 8 | 16384 | 1048576 | 16 | 0.05 | 2.85 | 16 |
| | | 9 | 16384 | 4194304 | 16 | 0.05 | 2.85 | 16 |
| | | 10 | 8192 | 16777216 | 16 | 0.05 | 2.85 | 16 |
| | | 11 | 8192 | 134217728 | 16 | 0.05 | 2.85 | 16 |

**Table S1: Parameters of Monte Carlo simulations.** The parameters are shown for both the hierarchical Edwards-Anderson model with power-law interaction decay (pHEA) and hierarchical Edwards-Anderson model with fixed average coordination number (cHEA), for different values of the coupling-range exponent $\varsigma$.

In particular, panels A-D of Figs. S13–S18 show the MC results for the finite-size critical temperatures $\beta_c^k$ with each of the three choices in Section S24.2. To obtain $\omega$, we determined $t_c^k$ from the crossings of observable (S144), which yields the cleanest data to determine the finite-size critical temperature. We then considered the moment ratio (S146): given that $U_k$ satisfies Eq. (S152), a fit of $\mathcal{Q}_k[U_k]$ vs. $k$ from Eq. (S154) allowed us to estimate $B_{f,g}$ and $\omega$ as fitting parameters, see panels E of Figs. S13–S18 and Fig. S20. In particular, Fig. S20 is compatible with $\omega \to 0$ for $\varsigma \to \varsigma_{low}$, which is consistent with the fact that the HEA behaves as a system of independent spins at its lower critical dimension.

## S24.4 Critical exponents

Now that we estimated the correction-to-scaling exponent, we can evaluate the critical exponent $\nu$. To achieve this, let us consider an observable $O_k$ which, unlike $f_k$ and $g_k$ above, diverges with the system size at the critical point as follows:

$$O_k(t) = 2^{k/\nu} \left[ O^L(2^{\phi k} t) + \frac{1}{2^{k\omega}} O^S(2^{\phi k} t) \right]. \tag{S155}$$

Substituting Eqs. (S140), (S143) and (S155) into Eq. (S153) and keeping only the first subleading term in $k$, we obtain that the quotient of $O_k$ scales as follows:

$$\log \mathcal{Q}_k[O] = \frac{\log 2}{\nu} + \frac{D_{f,O}}{2^{\omega k}}, \tag{S156}$$

where $D_{f,O}$ is a constant.

Proceeding along the lines of Section S24.2, we considered the three following choices for $O_k$:

(i)
$$O_k = \left| \frac{d[2^{k(1-2\varsigma)} \chi_k]}{dT} \right|, \tag{S157}$$

(ii)
$$O_k = \left| \frac{dB_k}{dT} \right|, \tag{S158}$$

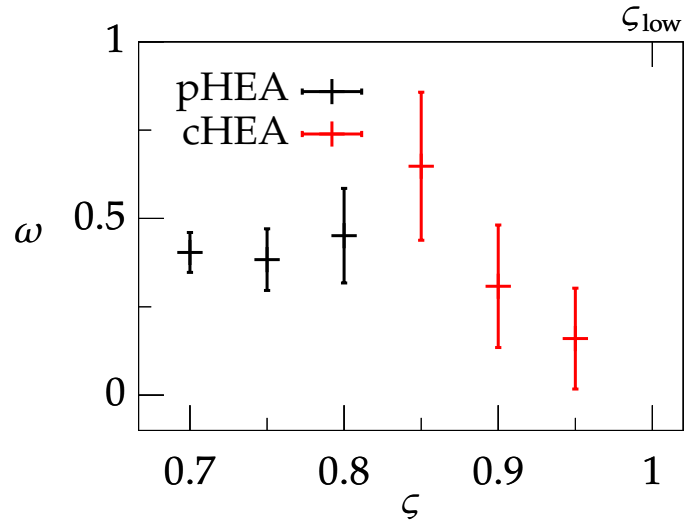

**Figure S20: Correction-to-scaling exponent** $\omega$. The exponent $\omega$ is shown as a function of the coupling-range exponent $\varsigma$, and it has been generated from Monte Carlo simulations the hierarchical Edwards-Anderson model with power-law interaction decay (pHEA) for $\varsigma \leq 0.8$ and from the hierarchical Edwards-Anderson model with fixed average coordination number (cHEA) for $\varsigma > 0.8$—see Figs. S13–S18. The lower critical dimension $\varsigma_{\text{low}}$ is also marked.

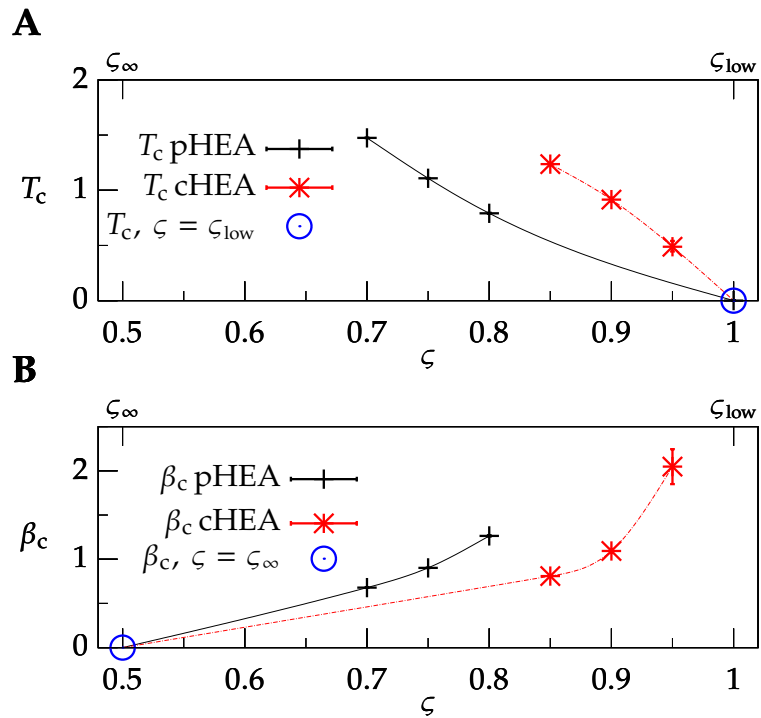

**Figure S21: Critical temperature from numerical simulations**. **A)** Critical temperature $T_c$ of the hierarchical Edwards-Anderson model (HEA) with power-law interaction decay (pHEA) and of the HEA with fixed coordination number (cHEA) (in black and red, respectively) as functions of the interaction exponent $\varsigma$, estimated as in Figs. S13–S18. The value of $\varsigma$ above which the thermodynamic limit of the model is defined and the lower critical dimension for the model, $\varsigma_\infty$ and $\varsigma_{\text{low}}$, respectively, are marked. We also show the critical temperature at the lower critical dimension (blue point). The curves connect the data points to the theoretical value of $T_c$ at $\varsigma = \varsigma_{\text{low}}$, and they are intended as guides for the eye. **B)** Same as **A**, for the inverse critical temperature, where the curves connect the data points to the theoretical value of $\beta_c$ at $\varsigma = \varsigma_\infty$.

(iii)
$$O_k = \left| \frac{\mathrm{d}[2^{-k}\xi_k]}{\mathrm{d}T} \right|. \tag{S159}$$

We computed $\mathcal{Q}_k[O]$ for choices (i), (ii) and (iii), with $t_c^k$ determined with observable (S144), which yields the cleanest data to determine the finite-size critical temperature. We then fixed $\omega$ in Eq. (S156) to the value determined in Section S24.3 and made a combined fit of the numerical data of $\mathcal{Q}_k[O]$ for the three choices above, with Eq. (S156). As a result, we determined $\nu$, $D_{\chi,O}$, $D_{B,O}$ and $D_{\xi,O}$ as fitting parameters. The results are shown in panels F of Figs. S13–S18 and in Fig. 2B.

### S24.5 Infinite-volume critical temperature

In order to test our numerical data and data-analysis procedure, we estimated the infinite-volume critical temperature $\beta_c$ from the data above.

We obtained $\beta_c$ by performing a combined fit of the three instances of Eq. (S143) discussed in Section S24.2 with fitting parameters $\beta_c$, $A_\chi$, $A_B$ and $A_\xi$, where the subscripts $\chi$, $B$ and $\xi$ denote the three observable choices above. The correction-to-scaling exponent $\omega$ has been obtained from Section S24.3. Given that here $\varsigma > \varsigma_{\mathrm{up}}$, Eq. (S138) implies that $\phi = 1/\nu$, and $\nu$ has been obtained from Section S24.4. The results of this analysis are shown in panels D of Figs. S13–S18. The resulting values of $T_c$ and $\beta_c$ as functions of $\varsigma$ are shown in Fig. S21, and, as a general test of our numerical simulations, we observe that they are compatible with the expected behaviors $\beta_c \to 0$ for $\varsigma \to \varsigma_\infty$ [16], and with Eq. (S104).

## S25 Supplementary discussion

A future direction of this work consists in studying the shape of the fixed distributions (FDs). In fact, while Fig. S9 shows that the critical FD depends on the coupling-range exponent $\varsigma$, the high- and low-temperature FDs appear to be nearly independent of $\varsigma$, see Figs. S7 and S8. It would be interesting to understand the physical meaning of this feature, in particular for the low-temperature FD, along with the potential implications on the structure of the low-temperature phase of the model.

In addition, further insights on this RG method may be obtained by studying its higher-order approximations where, for instance, an eight-spin model is decimated into a four-spin model [6, 16], see Section S26. This perspective presents a major challenge. In fact, in higher-order approximations, model $\mathcal{M}'$ would be composed of four or more spins, and the coupling distribution $p'_k$, which characterizes the RG flow, would be a multivariate distribution. It would thus be necessary to develop a suitable strategy to handle such a complex, multivariate structure in the discretization and linearization procedures of Sections 3 and S22.

Finally, one may consider the value of the critical temperature as a function of $\varsigma$ for $\varsigma \lesssim \varsigma_{\mathrm{low}}$. In this limit, analytical predictions exist for one-dimensional, long-range models with Gaussian-distributed couplings [2], which yield
$$T_c \sim \sqrt{\varsigma_{\mathrm{low}} - \varsigma}. \tag{S160}$$
Given that the critical temperature depends on the form of the coupling distribution, and it is thus not universal [27, 28], our RG prediction for $T_c$ cannot be quantitatively compared with Eq. (S160). However, it would be interesting to study whether our analysis reproduces some features of (S160), such as the scaling of $T_c$ with respect to $\varsigma_{\mathrm{low}} - \varsigma$.

## S26 Higher-order approximations

In higher-order approximations of our RG method, models $\mathcal{M}$ and $\mathcal{M}'$ contain $2^{k+1}$ and $2^k$ spins, respectively; the approximation presented here corresponds to $k = 1$, see Section 2.1.

The approximations with $k > 1$ present a major challenge: In fact, model $\mathcal{M}'$ would contain $2^k(2^k - 1)/2$ couplings $J'_{ij}$. The coupling distribution $p'_k$, which characterizes the RG flow, would thus be a multivariate distribution.

If the discrete approach used in the $k = 1$ case were used in such higher-order approximations, and the quantiles relative to each spin coupling were discretized into $N$ bins, then $p'_k$ would be represented by

$N^{2^k(2^k-1)/2}$ parameters, which would describe the RG flow. It would thus be necessary to develop a suitable strategy to handle such a complex, multivariate structure in the discretization and linearization procedures of Sections 3 and S22. A suitable candidate for such strategy could be Monte-Carlo methods, which allow to sample high-dimensional spaces [34], and could thus be used to sample the space of spin couplings.

# Acronyms

**CDF**  cumulative distribution function

**cHEA**  hierarchical Edwards-Anderson model with fixed average coordination number

**FD**  fixed distribution

**FM**  ferromagnetic

**GS**  ground state

**HEA**  hierarchical Edwards-Anderson model

**LHS**  left-hand side

**MC**  Monte Carlo

**PDF**  probability density function

**pHEA**  hierarchical Edwards-Anderson model with power-law interaction decay

**RG**  renormalization group

**SG**  spin glass

# Symbols

| | |
|---|---|
| $\mathcal{M}$ | Four-spin hierarchical Edwards-Anderson model (HEA) |
| $\mathcal{M}'$ | Two-spin HEA |
| $H$ | Hamiltonian |
| $\langle \rangle$ | Boltzmann average |
| $\epsilon_p$ | $p$th energy level |
| $J$ | Spin-spin coupling |
| $J^{\text{d}}$ | Spin-spin coupling for the diluted HEA |
| $\varsigma$ | Exponent for the coupling range |
| $\varsigma_\infty$ | Value of the coupling range $\varsigma = 1/2$, above which the thermodynamic limit is defined |
| $\varsigma_{\text{up}}$ | Value of the coupling range $\varsigma = 2/3$ corresponding to the upper critical dimension |
| $\varsigma_{\text{low}}$ | Value of the coupling range $\varsigma = 1$ corresponding to the lower critical dimension |
| $\nu$ | Critical exponent describing the divergence of the correlation length for the HEA |
| $\nu_{\text{FM}}$ | $\nu$ in the ferromagnetic limit |
| $\varsigma_\infty^{\text{FM}}$ | $\varsigma_\infty$ in the ferromagnetic limit |
| $\varsigma_{\text{up}}^{\text{FM}}$ | $\varsigma_{\text{up}}$ in the ferromagnetic limit |
| $\varsigma_{\text{low}}^{\text{FM}}$ | $\varsigma_{\text{low}}$ in the ferromagnetic limit |
| $Z$ | Partition function |
| $S_i$ | Ising spin |
| $\boldsymbol{S}$ | Spin configuration of the system |
| $\boldsymbol{\sigma}$ | Ground-state spin configuration |
| $\boldsymbol{S}_p$ | Spin configuration relative to the $p$th energy level |
| $\boldsymbol{S}_{\uparrow\uparrow\downarrow\downarrow}$ | Ground-state spin configuration with the right-half spins flipped |

| | |
|---|---|
| $\mathcal{E}$ | Set of couplings $J$ such that $S_{\uparrow\uparrow\downarrow\downarrow} = S_2$ |
| $\Phi$ | Order parameter |
| $k_B$ | Boltzmann constant |
| $T$ | Temperature |
| $T_c$ | Critical temperature |
| $\beta_c$ | Inverse critical temperature |
| $\xi$ | Correlation length |
| $\omega$ | Correction-to-scaling exponent |
| $\beta$ | $1/(k_B T)$ |
| $N$ | Number of bins in the discretization of the renormalization-group transformation |
| $S$ | Number of samples in the Robbins-Monro method |
| $M$ | Number of iterations in the Robbins-Monro method |
| $\mathcal{R}$ | Rescaling transformation |
| $\overline{\phantom{x}}$ | Label indicating that the quantity under the bar is scaled, or related to the scaled distribution $p'$ |
| $s$ | Scaling factor |
| $p^*$ | Factorized spin-coupling distribution |
| $\mathcal{Z}$ | Normalization factor |
| $F()$ | Cumulative distribution function |
| $L_i$ | $i$th quantile |
| $K_i$ | $i$th scaled quantile |
| $\mathcal{O}()$ | Big O notation for the limiting behavior of a quantity |
| $\mathbb{I}()$ | Indicator function |
| sgn() | Sign function |
| $\mathbb{E}[]$ | Expectation value taken with respect to the spin-spin couplings |
| $D[]$ | Kullback-Leibler divergence |
| $\mathcal{K}_{ij}$ | Jacobian of the renormalization group (RG) transformation |
| $\mathcal{J}_{ij}$ | Jacobian of the scaled RG transformation |
| $\mathcal{Q}_k[O]$ | Quotient of observable $O$ |
| $N_{sw}$ | Number of Monte Carlo (MC) sweeps |
| $N_S$ | Number of disorder samples in MC simulations |
| $n_\beta$ | Number of temperatures in MC simulations |
| $\beta_{\min (\max)}$ | Minimal (maximal) inverse temperature in MC simulations |
| $n_{swap}$ | Number of MC sweeps after which replicas are swapped |



\end{document}